\documentclass[12pt,a4paper, epsfig]{report}
\usepackage[centertags]{amsmath}
\usepackage{amsfonts}
\usepackage{amssymb}
\usepackage{amsthm}

\usepackage{xthesis} 
\usepackage{xtocinc} 

\usepackage[dvips]{graphicx}

\begin{document}
\thispagestyle{empty}

\title{\bf{Kondo Effect in Artificial and Real Molecules}}

\author{\bf{Thesis submitted in partial fulfillment}\\
        \bf{of the requirements for the degree of}\\
        \bf{DOCTOR OF PHILOSOPHY}\\
        \\
        \\
        \\
        {\Large\bf by}\\
        \\
        {\Large\bf Tetyana Kuzmenko}\\
        \\
        \\
        \\
        {\bf Submitted to the Senate of Ben-Gurion University}\\
        {\bf of the Negev}
        \\
        \\
        \\
        \date{\bf\today\\
        \vspace{10mm}
        {\bf Beer-Sheva}}}

\maketitle

\thispagestyle{empty}
\begin{center}
 {\Large Kondo Effect in Artificial and Real Molecules}
\end{center}

\vspace{36mm}

\begin{center}
 Thesis submitted in partial fulfillment\\
 of the requirements for the degree of\\
 DOCTOR OF PHILOSOPHY\\
 \vspace{10mm}
 {\large by}\\
 \vspace{10mm}
 {\large Tetyana Kuzmenko}\\
 \vspace{20mm}
 Submitted to the Senate of Ben-Gurion University\\
 of the Negev
\end{center}

 \vspace{15mm}

\begin{flushleft}
 Approved by the advisor\underline{\hspace{40mm}}

 \vspace{10mm}

 Approved by the Dean of the Kreitman School of Advanced Graduate
 Studies\underline{\hspace{20mm}}
\end{flushleft}

 \vspace{30mm}

\begin{center}
 \date{\today\\
 \vspace{15mm}
 Beer-Sheva}
\end{center}

\newpage

\thispagestyle{empty}
 \vspace*{5cm}

This work was carried out under the supervision of Prof. Yshai
Avishai\\

\vspace{35mm}

In the Department \underline{ of Physics }

\vspace{14mm}

Faculty \underline{ of Natural Sciences }

\newpage

 \thispagestyle{empty}
\thanks{\begin{center}{\Large\bf Acknowledgments}\end{center}
I am grateful to my supervisors Professor Yshai Avishai and
Professor Konstantin Kikoin for proposing this intriguing and
challenging subject, always having time for discussions, reading
my notes quickly and very carefully. \\
I am deeply indebted to the
{\it Clore Scholars Programme} for generous support. }


\begin{abstract}
\pagenumbering{roman}
 In this Thesis we develop a novel direction in the theory of nano-objects,
i.e., structures of nanometer size in a tunnel contact with
macroscopic electron reservoirs (metallic leads). This theory
arose and was developed rapidly during the two recent decades as a
response to challenging achievements of modern nanotechnology and
experimental techniques. Evolution of this technology enabled the
fabrication of various low-dimensional systems from semiconductor
heterostructures to quantum wires and constrictions, quantum dots,
molecular bridges and artificial structures constructed from large
molecules. This impressive experimental progress initiated the
development of a new direction in quantum physics, namely, the
physics of artificial nano-objects.

We focus in this work on a theoretical investigation of the Kondo
physics in quantum dots and molecules with strong correlations. An
exciting series of recent experiments on mesoscopic and nanoscale
systems has enabled a thorough and controlled study of basic
physical problems dealing with a local moment interacting with a
Fermi sea of conduction electrons. Scanning tunnel microscopy and
quantum--dot devices have provided new tools for studying the
Kondo effect in many new perspectives and with unprecedented
control. As experimental and theoretical investigations of
tunneling phenomena continue, it turns out that the physics of
tunneling spectroscopy of large molecules and complex quantum dots
have much in common.

In particular we elucidate the Kondo effect predicted in tunneling
through triple quantum dots and sandwich-type molecules adsorbed
on metallic substrate, which are referred to as {\it trimers}. The
unusual dynamical symmetry of nano-objects is one of the most
intriguing problems, which arise in the theory of these systems.
We demonstrate that trimers possess dynamical symmetries whose
realization in Kondo tunneling is experimentally tangible. Such
experimental tuning of dynamical symmetries is not possible in
conventional Kondo scattering. We develop the general approach to
the problem of dynamical symmetries in Kondo tunneling through
nano-objects and illustrate it by numerous examples of trimers in
various configurations, in parallel, in series and in ring
geometries.

In the first part of this Thesis, the evenly occupied trimer in a
{\it parallel} geometry is studied. We show that Kondo tunneling
through the trimer is controlled by a family of $SO(n)$ dynamical
symmetries. The most striking feature of this result is that the
value of the group index $n=3,4,5,7$ can be changed experimentally
by tuning the gate voltage applied to the trimer. Following the
construction of the corresponding $o_n$ algebras, the scaling
equations are derived and the Kondo temperatures are calculated.

In the second part, the Kondo physics of trimer with both even and
odd electron occupation in a {\it series} geometry is discussed.
We derive and solve the scaling equations for the evenly occupied
trimer in the cases of the $P\times SO(4)\times SO(4)$, $SO(5)$,
$SO(7)$ and $P\times SO(3)\times SO(3)$ dynamical symmetries. The
dynamical-symmetry phase diagram is displayed and the experimental
consequences are drawn. The map of Kondo temperature as a function
of gate voltages is constructed. In addition, the influence of
magnetic field on the dynamical symmetry and its role in the Kondo
tunneling through the trimer are studied. It is shown that the
anisotropic Kondo effect can be induced in the trimer by an
external magnetic field. The corresponding symmetry group is
$SU(3)$. In the case of odd electron occupation, the effective
spin Hamiltonian of the trimer manifests a two-channel Kondo
problem albeit only in the weak coupling regime (due to
unavoidable anisotropy).

In the third part, the point symmetry $C_{3v}$ of an artificial
trimer in a {\it ring} geometry and its interplay with the spin
rotation symmetry $SU(2)$ are studied. This nano-object is a
quantum dot analog of the Coqblin-Schrieffer model in which the
Kondo physics is governed by a subtle interplay between spin and
orbital degrees of freedom. The orbital degeneracy is tuned by a
magnetic field, which affects the electron phases thereby leading
to a peculiar Aharonov-Bohm effect.

The following novel results were obtained in the course of this
research:

\begin{itemize}
\item It is found that evenly occupied trimer manifests a new
type of Kondo effect that was not observed in conventional spin
$1/2$ quantum dots. The dynamical $SO(n)$ symmetries of Kondo
tunneling through evenly occupied trimer both in parallel and
series geometry are unravelled. These symmetries can be
experimentally realized and the specific value of $n=3,4,5,7$ can
be controlled by gate voltage and/or tunneling strength. The Kondo
temperature explicitly depends on the index $n$ and this
dependence may be traced experimentally by means of measuring the
variation of tunnel conductance as a function of gate voltages.
The hidden dynamical symmetry manifests itself, firstly in the
very existence of the Kondo effect in trimer with even occupation,
secondly in non-universal behavior of the Kondo temperature $T_K$.
In a singlet spin state the anisotropic Kondo effect can be
induced in the trimer by an external magnetic field.
\item It is shown that the effective spin
Hamiltonian of a trimer with odd electron occupation weakly
connected in series with left ($l$) and right ($r$) metal leads is
composed of two-channel exchange and co-tunneling terms.
Renormalization group equations for the corresponding three
exchange constants $J_{l}$, $J_{r}$ and $J_{lr}$ are solved in a
weak coupling limit (single loop approximation). Since $J_{lr}$ is
relevant, the system is mapped on an anisotropic two-channel Kondo
problem. The structure of the conductance as function of
temperature and gate voltage implies that in the weak and
intermediate coupling regimes, two-channel Kondo physics persists
at temperatures as low as several $T_{K}$. Analysis of the Kondo
effect in cases of higher spin degeneracy of the trimer ground
state is carried out in relation with dynamical symmetries. The
Kondo physics remains that of a fully screened impurity, and the
corresponding Kondo temperature is calculated.
\item  It is demonstrated that spin and orbital degrees
of freedom interlace in ring shaped artificial trimer thereby
establishing the analogy with the Coqblin-Schrieffer model in
magnetically doped metals. The orbital degrees of freedom are
tunable by an external magnetic field, and this implies a peculiar
Aharonov-Bohm effect, since the electron phase is also affected.
The conductance is calculated both in three- and two-terminal
geometries. It is shown that it can be sharply enhanced or
completely blocked at definite values of magnetic flux through the
triangular loop.

\end{itemize}

\vspace{5mm}

\begin{center}
 {\large\bf Key Words}
\end{center}
Kondo effect, sandwich-type molecule with strong correlations,
trimer, renormalization group procedure, Schrieffer-Wolff
transformation, effective spin Hamiltonian, dynamical symmetry,
group generators, scaling equations, Kondo temperature, orbital
effects.
\end{abstract}

\pagenumbering{arabic}

\tableofcontents

\chapter{Introduction}

{\underline{\bf Background.}} Recently, studies of the physical
properties of artificially fabricated nano-objects turn out to be
a rapidly developing branch of fundamental and applied physics.
Progress in these fields is stimulated both by the achievements of
nanotechnology and by the ambitious projects of information
processing, data storage, molecular electronics and spintronics.
The corresponding technological evolution enabled the fabrication
of various low-dimensional systems from semiconductor
heterostructures to quantum wires and constrictions, quantum dots
(QD), molecular bridges and artificial structures with large
molecules built in electric circuits \cite{Wire,KAT}. This
impressive experimental progress led to the development of
nanophysics, a new aspect and research direction in quantum
physics \cite{Nano}. Artificial nano-objects possess the familiar
features of quantum mechanical systems, but sometimes one may
create in artificially fabricated systems such conditions, which
are hardly observable "in natura". For example, one-dimensional to
two-dimensional $(1D \to 2D)$ crossover may be realized in quantum
networks \cite{Net,Net1,Net2} and constrictions
\cite{Wees,Thomas}. The Kondo effect may be observed in
non-equilibrium conditions \cite{Noneq,Noneq1,Noneq2}, at high
magnetic fields \cite{Magn,Magn1,Magn-v1,Magn-v2,Magn-v3}, and at
finite frequencies \cite{Freq,Freq1,Freq2,Freq3,Freq4,Freq5}.
Moreover, a quantum dot in the Kondo regime can be integrated into
a circuit exhibiting the Aharonov-Bohm effect
\cite{Moti,Moti1,GDCO}.

According to the theory of Kondo effect in QD \cite{Glaz,Ng}, the
spin degrees of the QD are involved in Kondo resonance, and Kondo
effect takes place only if the dot has a nonzero spin in the
ground state. Resonance Kondo tunneling was experimentally
observed in QD with {\it odd} electron occupation number and spin
one-half ground state \cite{KKK,KKK1,Sim,Wiel}. But real
nano-objects cannot be simply represented by a spin $1/2$ moment,
because low-lying spin excitations, which are always present in
few-electron systems, should be taken into account. Evenly
occupied dot usually has a singlet ground state. However, it was
predicted theoretically \cite{Magn} and confirmed experimentally
\cite{Magn1} that Kondo tunneling can be induced by an external
magnetic field in planar QD with {\it even} occupation. This
unconventional magnetic field induced Kondo effect arises because
the spectrum of the dot possesses a low-lying triplet excitation
above the singlet ground state. The Zeeman splitting energy of a
triplet in an external magnetic field may exactly compensate the
energy spacing between the two adjacent levels, and the lowest
spin excitation possesses an effective spin $1/2$, thus inducing a
Kondo-like zero-bias anomaly (ZBA) in the differential
conductance. A similar scenario may be realized in vertical QDs
\cite{Magn-v1,Magn-v2,Magn-v3,Magn-v} where the Larmor (instead of
the Zeeman) effect comes into play. The influence of an external
magnetic field on the orbital part of the wave functions of
electrons in vertical QDs is, in general, more pronounced than the
Zeeman effect. 
Hence, singlet-triplet level crossing are induced by this magnetic
field, causing the emergence of Kondo scattering \cite{Magn-v}.

Another device which manifests the Kondo effect in QDs with even
electron occupation is a double quantum dot. Double quantum dots
(DQDs), namely, quantum dots with two potential-wells, oriented
parallel to the lead surface were fabricated several years ago
\cite{Hoff95,Molen95}. The two wells in a DQD may be identical or
have different size; the DQD may be integrated within an electric
circuit either in series or in parallel; different gate voltages
may be applied to each well. Moreover, one of the two wells may be
disconnected from the leads (side geometry) \cite{KA01,KA02,KKAv}.
If the tunneling between the right and left wells of the DQD is
taken into account, DQD can be treated as an artificial molecule
where the interdot tunneling results in the formation of
complicated manifold of bonding and antibonding states which
modifies its degrees of freedom \cite{Tarucha,Rontani,Part,Yigal}.
The systematic treatment of the physics of DQD coupled to metallic
leads \cite{KA01,KA02} is based on the mechanism according to
which the transition from a singlet state in a weak coupling
regime to a triplet state in a strong coupling regime is an {\it
intrinsic} property of nano-objects with even occupation. It is
manifested in tunneling through real and artificial molecules in
which the electrons are spatially separated into two groups with
different degree of localization. Electrons in the first group are
responsible for strong correlation effects (Coulomb blockade),
whereas those in the second group are coupled to a metallic
reservoir. The necessary precondition under which the singlet
$S=0$ ground state changes into a partially screened triplet $S=1$
Kondo state due to hybridization with metallic leads is the
existence of charge-transfer exciton in the DQD. Unlike quantum
dot with odd occupation whose Hamiltonian is mapped on the
Kondo-type $sd$-exchange Hamiltonian with a localized spin $S=1/2$
obeying $SU(2)$ symmetry, DQD in contact with metallic leads can
be treated as a quantum spin rotator with $S=1$. The effective
Hamiltonian of DQD possesses the dynamical symmetry $SO(4)$ of a
spin rotator \cite{KA02}.

The Kondo physics seems to be richer in systems involving
tunneling through artificial molecules containing more than two
wells. Meanwhile, the Kondo effect was observed already in complex
molecules containing cages with magnetic ions \cite{Park2,Park3}.
Analogy between the Kondo effect in real and artificial molecules
in tunnel contact with metallic leads was noted some time ago in
Refs. \cite{KA01,Rontani,Part}. There exists a great variety of
molecules containing magnetic rare earth (RE) ions secluded in
carbon and nitrogen based cages. Endofullerenes REC$_{82}$ are the
most common among them \cite{endof,endof1,endof2,endof3}. In these
molecules magnetic ions are inserted in a nearly spherical carbon
cage. Lanthanocenes Ln(C$_8$H$_8$)$_2$ are sandwich-type molecules
formed by two rings of CH radicals and magnetic ion Ln=Ce, Yb in
between \cite{Dolg,Dolg1,Liu}. In these molecules the electrons in
a strongly correlated $f$~shell of Ln are coupled with loosely
bound $\pi$ electrons in the cage. The ground state of cerocene
molecule is a spin singlet combination $^1A_{1g}(f{\pi}^3)$ of an
$f$ electron and $\pi$~orbitals, and the energy of the first
excited triplet state $^3E_{2g}$ is rather small ($\sim0.5$~eV).
In the ytterbocene (hole counterpart of cerocene) the ground state
with one $f$~hole is a triplet, and the gap for a singlet
excitation is tiny, $\sim0.1$~eV. In all these systems there is no
direct overlap between the strongly correlated $f$ electrons and
the metallic reservoir. However, these electrons can influence the
tunnel properties of the molecule via covalent bonding with outer
$\pi$~electrons which are coupled to the metallic reservoir. Other
examples of molecules secluding magnetic ions may be found, e.g.,
in \cite{Park2,Park3}.

{\underline{\bf Objectives.}} In this thesis we develop a theory
of Kondo tunneling through triple quantum dots and sandwich-type
molecules adsorbed on metallic substrate, which are referred to as
{\it trimers}. One of the most intriguing phenomenon which arise
in the theory of these systems is the emergence of an unusual {\it
dynamical symmetry of nano-objects}. Our main purpose is to
demonstrate that trimers possess dynamical symmetries whose
realization in Kondo tunneling is experimentally tangible. Such
experimental tuning of dynamical symmetries is not possible in
conventional Kondo scattering. In many cases even the very
existence of Kondo tunneling crucially depends on the dynamical
spin symmetry of trimer. We develop the general approach to the
problem of dynamical symmetries in Kondo tunneling through
nano-objects and illustrate it by numerous examples of trimer in
various configurations, in parallel, in series and in ring
geometries. We show that Kondo tunneling reveals hidden $SO(n)$
dynamical symmetries of evenly occupied trimers both in parallel
and series geometry. The possible values $n=3,4,5,7$ can be
controlled by gate voltages, indicating that abstract concepts
such as dynamical symmetry groups are experimentally realizable.
We construct the corresponding $o_n$ algebras, derive and solve
scaling equations and calculate the Kondo temperatures. We
elucidate the role of discrete and continuous symmetries exposed
by the Kondo effect in artificial trimer in ring geometry, i.e.,
triangular triple quantum dot (TTQD). In comparison with a linear
configuration of three quantum dots, a TTQD possesses additional
degrees of freedom, namely, discrete rotations. We show that the
Kondo physics of TTQD is determined by a subtle interplay between
continuous spin rotation symmetry $SU(2)$, and discrete point
symmetry $C_{3v}$. Moreover, such ring shaped nano-object can
serve both as a Kondo-scatterer and as a peculiar Aharonov-Bohm
(AB) interferometer, since the magnetic flux affects not only the
electron phase, but also the nature of the ground and excited
states of the trimer. The main lesson to be learned is that Kondo
physics in trimer suggests a novel and in some sense rather
appealing aspect of low-dimensional physics of interacting
electrons. It substantiates, in a systematic way, that dynamical
symmetry groups play an important role in mesoscopic physics. In
particular, we encounter here some "famous" groups which appear in
other branches of physics. Thus, the celebrated group $SU(3)$
enters also here when a trimer is subject to an external magnetic
field. And the group $SO(5)$ which plays a role in the theory of
superconductivity is found here when a certain tuning of the gate
voltages in trimer is exercised.

{\underline{\bf Structure.}} The structure of the Thesis is as
follows. In the second Chapter, the basic physics of Kondo effect
in quantum dots is briefly reviewed. First, the Kondo tunneling
through single quantum dot with odd electron occupation is
described. Next, quantum dot with even number of electrons,
subject to an external magnetic field is considered. It is shown
that the system exhibits Kondo effect in a finite magnetic field,
when the Zeeman energy is equal to the single-particle level
spacing in the dot. Then, the Kondo physics of evenly occupied
double quantum dot (DQD) is presented. Special attention is given
to the symmetry properties of the DQD. It is shown that the DQD
possesses the $SO(4)$ dynamical symmetry of a spin rotator.
Finally, we explain the concept of dynamical symmetry and its
realization in complex quantum dots.

In the third Chapter, the special case of trimer with even
electron occupation in the {\it parallel} geometry is studied. We
discuss the energy spectrum of the isolated trimer, derive
renormalization group equations and demonstrate possible
singlet-triplet level crossing due to tunneling. We show that the
trimer manifests $SO(n)$ dynamical symmetry in the Kondo tunneling
regime. We expose the effective spin Hamiltonian of the trimer and
construct the corresponding $o_n$ algebras for the $P\times
SO(4)\times SO(4)$, $SO(5)$ and $SO(7)$ dynamical symmetries. We
derive scaling equations and calculate the Kondo temperatures for
the cases of $P\times SO(4)\times SO(4)$ and $SO(5)$ symmetries.

In the fourth Chapter we discuss the physics of trimer in a {\it
series} geometry and point out similarities and differences
between Kondo physics in the parallel and series geometries. The
scaling equations are derived and the Kondo temperatures are
calculated for the evenly occupied trimer in the cases of the
$P\times SO(4)\times SO(4)$, $SO(5)$, $SO(7)$ and $P\times
SO(3)\times SO(3)$ dynamical symmetries. The dynamical symmetries
of the trimer are summarized by a phase diagram which can be
scanned {\it experimentally} by appropriate variations of gate
voltages. We discuss a novel phenomena, namely, {\it a Kondo
effect without a localized spin}. The anisotropic exchange
interaction occurs between the metal electron spin and the trimer
Runge-Lenz operator alone in an external magnetic field. The
symmetry group for such magnetic field induced anisotropic Kondo
tunneling is $SU(3)$. We show that in the case of odd occupation,
the effective Hamiltonian of the trimer manifests generic futures
of a two-channel Kondo problem at least in the weak coupling
regime.

In the fifth Chapter we concentrate on the {\it point symmetry}
$C_{3v}$ of artificial trimer in a {\it ring} geometry and its
interplay with the {\it spin rotation symmetry} $SU(2)$ in the
context of Kondo tunneling through triangular artificial molecule.
The underlying Kondo physics is determined by the product of a
discrete rotation symmetry group in real space and a continuous
rotation symmetry in spin space. These symmetries are reflected in
the resulting exchange Hamiltonian which naturally involves spin
and orbital degrees of freedom thereby establishing the analogy
between the Coqblin-Schrieffer model in real metals and the
physics of transport in complex quantum dots. The ensuing poor-man
scaling equations are solved and the Kondo temperature is
calculated. We show that the trimer also reveals a peculiar
Aharonov-Bohm effect where the magnetic field affects not only the
electron phase but also controls the underlying dynamical symmetry
group.

 The derivation of the pertinent effective spin
Hamiltonians and the establishment of group properties (in
particular identification of the group generators and checking the
corresponding commutation relations) sometimes require lengthy
mathematical expressions. These are collected in the appendices.

This work was partially presented by posters and lectures in
scientific conferences and schools (see List of Presentations).
The first part of the results was published in Refs. 1, 2 (see
List of Publications). The second part was published in Refs. 3-5.
The third part was published in Refs. 6, 7.

\newpage

\begin{center}
 {\LARGE\bf List of Presentations}
\end{center}
\begin{enumerate}
\item \underline{T. Kuzmenko}. {\sl Kondo Effect in Artificial Molecules}
      (lecture). {\it Condensed Matter
      Seminar}, Department of Physics, Ben-Gurion University of the
      Negev, Beer Sheva, Israel, June 20, 2005.
\item  \underline{T. Kuzmenko}, K. Kikoin, and Y. Avishai.
       {\sl Kondo Effect in Molecules with Strong Correlations} (poster).
       {\it The International Conference on Strongly Correlated
       Electron Systems SCES'04}. Karlsruhe, Germany, July 26 - August 30, 2004.
\item \underline{T. Kuzmenko}, K. Kikoin, and Y. Avishai.
     {\sl $SO(n)$ Symmetries in Kondo Tunneling through Evenly
     Occupied Triple Quantum Dots} (lecture). {\it International School and
     Workshop on Nanotubes $\&$ Nanostructures}. Frascati, Italy, September 15-19, 2003.
\item \underline{T. Kuzmenko}, K. Kikoin, and Y. Avishai. {\sl Kondo Effect in
      Evenly Occupied Triple Quantum Dot} (poster). {\it International Seminar
      and Workshop on Quantum Transport and Correlations
      in Mesoscopic Systems and Quantum Hall Effect}.
      Dresden, Germany, July 28 - August 22, 2003.
\item \underline{T. Kuzmenko}, K. Kikoin, and Y. Avishai.
      {\sl Dynamical Symmetries in Kondo Tunneling Through Complex
      Quantum Dots} (poster). {\it International School of Physics
      "Enrico Fermi"}, Varenna, Italy, July 9--19, 2002.
\item \underline{T. Kuzmenko}, K. Kikoin, and Y. Avishai.
      {\sl $SO(5)$ Symmetry in Kondo Tunneling Through a Triple
      Quantum Dot} (lecture). {\it Correlated Electrons Day}, Institute
      for Theoretical Physics, Haifa, May 2, 2002.
\end{enumerate}

\newpage

\begin{center}
 {\LARGE\bf List of Publications}
\end{center}
\begin{enumerate}
\item T. Kuzmenko, K. Kikoin, and Y. Avishai.
      {\it Dynamical Symmetries in Kondo Tunneling through Complex
      Quantum Dots}. Phys. Rev. Lett. {\bf 89}, 156602 (2002);
      {\tt cond-mat/0206050}.
\item  K. Kikoin, T. Kuzmenko, and Y. Avishai.
      {\it Unconventional Mechanism of Resonance
      Tunneling through Complex Quantum Dots}.
      Physica E {\bf 17}, 149 (2003).
\item T. Kuzmenko, K. Kikoin, and Y. Avishai.
      {\it Towards Two-Channel Kondo Effect in Triple Quantum Dot}.
      EuroPhys. Lett. {\bf 64}, 218 (2003); {\tt cond-mat/0211281}.
\item T. Kuzmenko, K. Kikoin, and Y. Avishai.
      {\it Kondo Effect in Systems with Dynamical Symmetries}.
       Phys. Rev. B {\bf 69}, 195109 (2004); {\tt cond-mat/0306670}.
\item  T. Kuzmenko, K. Kikoin, and Y. Avishai.
      {\it Kondo Effect in Molecules with Strong Correlations}.
      Physica B {\bf 359-361}, 1460 (2005).
\item  Y. Avishai, T. Kuzmenko, and K. Kikoin.
      {\it Dynamical and Point Symmetry of the Kondo Effect in
      Triangular Quantum Dot}. To appear in Physica E;
      {\tt cond-mat/0412527}.
\item T. Kuzmenko, K. Kikoin, and Y. Avishai,
      {\it Magnetically Tunable Spin and Orbital Kondo Effect in
      Triangular Quantum Dot}. Submitted to Phys. Rev. Lett.;
      {\tt cond-mat/0507488}.
\end{enumerate}

\chapter{Kondo Physics in Artificial Nano-objects}\label{chap.2}

{\it In this Chapter we review the basic physics of Kondo effect
in single and double quantum dots. Quantum dot with odd electron
occupation in tunnel contact with metallic leads is considered in
Section \ref{sec.2.1}. The Hamiltonian of the system is written
down within the framework of the Anderson model. The Haldane
renormalization group procedure is described and the effective
spin Hamiltonian is obtained by means of the Schrieffer-Wolff
transformation. The expressions for the Kondo temperature and
zero-bias conductance of the dot are then derived. In Section
\ref{sec.2.2} the possibility of Kondo effect in evenly occupied
single quantum dots subject to an external magnetic field is
discussed. The Kondo effect in evenly occupied double quantum dot
(DQD) is studied in Section \ref{sec.2.3}. It is shown that DQD
possesses $SO(4)$ dynamical symmetry of a spin rotator. Finally in
Section \ref{sec.2.4} we introduce the concept of dynamical
symmetry and its emergence in complex quantum dots.}
\section{Kondo Effect in Single Quantum Dot (QD)}\label{sec.2.1}
\begin{figure}[htb]
\centering
\includegraphics[width=100mm,height=25mm,angle=0]{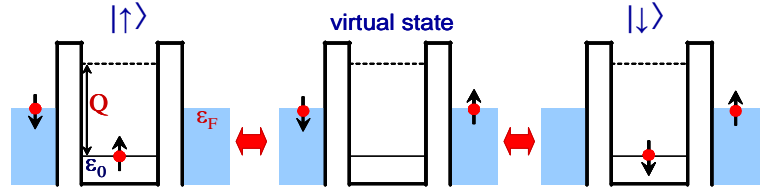}\label{spin-fl}
\caption{Spin-flip cotunneling process in a quantum dot with odd
occupation. The left and right panels refer to spin-up
$|\uparrow\rangle$ and spin-down $|\downarrow\rangle$ ground
states, which are coupled by a cotunneling process. The middle
panel corresponds to a high-energy virtual state.}
\end{figure}

The conventional Kondo effect appears in scattering of conduction
electrons by localized magnetic impurity \cite{Kondo}. The latter
is represented by its spin ${\bf S}$, which possesses the $SU(2)$
symmetry of rotationally invariant moment. Spin scattering of
conduction electrons dynamically screens this moment, and the
system transforms into a local Fermi liquid with separate branches
of charge and spin excitations \cite{Tsvelik,Andrei}. According to
Refs. \cite{Glaz, Ng}, the problem of tunneling through a
nano-object with odd electron occupation and strong Coulomb
blockade suppressing charge fluctuations can be mapped on the
Kondo scattering problem. The Kondo effect emerges in a quantum
dot occupied by an odd number of electrons at temperatures below
the mean level spacing in the dot. Under such conditions, the
highest occupied level $\varepsilon_0$ filled by a single electron
produces the Kondo effect. The other levels, occupied by pairs of
electrons with opposite spins, don't contribute to the Kondo
screening. Therefore, a dot attached to two metallic leads can be
described in the framework of the Anderson {\it single-level}
impurity model. Figure \ref{spin-fl} illustrates schematically one
of the spin-flip co-tunneling processes. The spin-up electron
tunnels out of the dot, and then it is replaced by the spin-down
electron. At low temperature, the coherent superposition of all
possible co-tunneling processes involving spin-flip results in the
screening of the local spin.

Let us consider a quantum dot (QD) occupied by one electron with
energy $\varepsilon_0$ in the ground state. The dot in tunneling
contact with the source and drain leads is described by the
Anderson Hamiltonian
\begin{eqnarray}
 H_{A} &=& H_{d}+H_{lead}+H_{t}.
 \label{And-ham}
 \end{eqnarray}
Here the first term, $H_d$, is the Hamiltonian of the isolated
dot,
\begin{eqnarray}
 H_{d} &=&
 \sum_{\sigma}\varepsilon_0 d_{\sigma}^{\dagger}d_{\sigma}+
 Qn_{\uparrow}n_{\downarrow},
 \end{eqnarray}
where ${\sigma}=\uparrow,\downarrow$ is the spin index, $Q>0$ is
the Coulomb blockade energy, and
$n_{\sigma}=d_{\sigma}^{\dagger}d_{\sigma}$.
 The second term, $H_{lead}$,
describes the electrons in the source $(s)$ and drain $(d)$
electrodes,
\begin{eqnarray}
 H_{lead} &=&\sum_{k\sigma\alpha}\epsilon_k
 c_{k\sigma \alpha}^{\dagger}c_{k\sigma \alpha}, \ \ \ \alpha=s,d.
\end{eqnarray}
The last term, $H_{t}$, is the tunneling Hamiltonian,
\begin{eqnarray}
 H_{t} &=&
 \sum_{k\sigma\alpha}
 \left(
      V_{\alpha}
      c_{k\sigma\alpha}^{\dagger}
      d_{\sigma}+H.c.
 \right),\label{tun-SQD}
\end{eqnarray}
where $V_{\alpha}$ $(\alpha=s,d)$ are tunneling matrix elements.
Here and below we assume that $V_{\alpha}$ are real and positive.
It is convenient to perform a canonical transformation \cite{Glaz}
\begin{eqnarray}
c_{k\sigma}=uc_{k\sigma s}+vc_{k\sigma d}, \ \ \ \ \ \ \
a_{k\sigma}=uc_{k\sigma d}-vc_{k\sigma s},\label{G-R-rot}
\end{eqnarray}
with
\begin{eqnarray}
&&u=\frac{V_s}{\sqrt{V_s^2+V_d^2}}, \ \ \
v=\frac{V_d}{\sqrt{V_s^2+V_d^2}}.\label{uv}
\end{eqnarray}
 As a result, only the fermions $c_{k\sigma}$ contribute to
 tunneling, and the tunneling Hamiltonian (\ref{tun-SQD}) takes
 the form,
\begin{eqnarray}
 H_{t} &=&
 V\sum_{k\sigma}
 \left(
      c_{k\sigma}^{\dagger}
      d_{\sigma}+H.c.
 \right),\label{tun-SQD-rot}
\end{eqnarray}
where $V=\sqrt{V_s^2+V_d^2}$.

\begin{figure}[htb]
\centering
\includegraphics[width=120mm,height=45mm,angle=0]{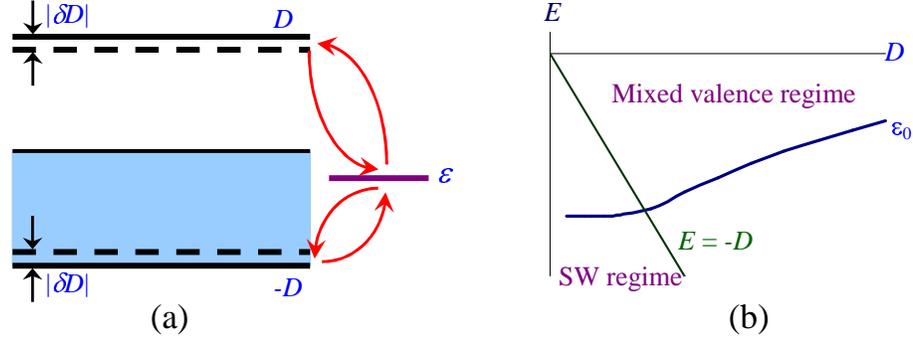}
\caption{Haldane renormalization group procedure. Reducing the
bandwidth by $2|\delta D|$ (panel (a)) results in the
renormalization of the energy level of the dot (panel (b)).}
\label{fig-Haldane}
\end{figure}

The spectrum of electrons in the leads forms a band with bandwidth
$2D_0$. In accordance with the Haldane renormalization group (RG)
procedure, the low energy physics can be exposed by integrating
out the high-energy charge excitations \cite{Hald}. This procedure
implies the renormalization of the energy level of the dot by
mapping the initial energy spectrum $-D_0<\epsilon_k<D_0$ onto a
reduced energy band $-D_0+|\delta D|<\epsilon_k<D_0-|\delta D|$
(Fig.\ref{fig-Haldane}):
\begin{eqnarray}
\varepsilon&=&\varepsilon_0-\frac{\Gamma |\delta
D|}{D},\label{en-ren}
\end{eqnarray}
where $\Gamma=\pi \rho_0 V^2$ is the tunneling rate, $\rho_0$ is
the density of electron states in the leads, which is assumed to
be constant. Iterating the renormalization procedure
(\ref{en-ren}), one obtains the scaling equation, which describes
the evolution of the one-electron energy state of the dot with
reducing the energy scale of the band continuum,
\begin{eqnarray}
\frac{d\varepsilon}{d\ln D}&=&\frac{\Gamma}{\pi}.\label{en-sc}
\end{eqnarray}
The processes involving charge scattering to the band edges lead
to renormalization of ${\Gamma}$ only in higher order in $V$:
$$\frac{d\Gamma}{d\ln D}=O\left(\frac{\Gamma}{D}\right),$$
and hence for $\Gamma\ll D$ there is no significant
renormalization of $\Gamma$.

The scaling invariant for equation (\ref{en-sc}),
\begin{equation}
\varepsilon^{\ast }=\varepsilon(D)-\frac{\Gamma}{\pi }\ln \left(
\frac{\pi D}{\Gamma }\right) . \label{inv-SQD}
\end{equation}
is tuned to satisfy the initial condition
$\varepsilon(D_{0})=\varepsilon_0$.

The above Haldane RG procedure brings us to the Schrieffer-Wolf
(SW) limit \cite{SW} ${\bar D}\sim |\varepsilon ({\bar D})|$,
where all charge degrees of freedom are quenched for excitation
energies within the interval $-{\bar D}<\epsilon_k<{\bar D}$ and
scaling terminates. The excited states with two or zero electrons
are higher in energy by $Q+\varepsilon_0$ and $|\varepsilon_0|$
$(Q+\varepsilon_0, |\varepsilon_0|\gg V)$, respectively, and are
not important for the low-energy dynamics of the dot. The
effective spin Hamiltonian with the two- and zero-electron states
frozen out can be obtained by means of the Schrieffer-Wolff
unitary-transformation \cite{SW} applied to the Hamiltonian
(\ref{And-ham}),
\begin{eqnarray}
H&=&e^{i{\cal S}}H_{A}e^{-i{\cal S}},\label{SW-tran}
\end{eqnarray}
where the operator ${\cal S}$ is found from the condition
\begin{eqnarray}
H_{t}+i[{\cal S}, H_{d}+H_{lead}]=0.\label{S-cond}
\end{eqnarray}
The condition (\ref{S-cond}) means that the effective Hamiltonian
(\ref{SW-tran}) does not contain linear in $V$ terms, which allow
the variation in the number of electrons in the dot.

Retaining the terms to order $O(V^2)$ on the right-hand side of
Eq.(\ref{SW-tran}), one comes to the Kondo Hamiltonian
\begin{eqnarray}
 H_{K} &=&
 \sum_{k \sigma}\epsilon_{k}
 c_{k\sigma }^{\dagger}c_{k\sigma }+
 \frac{J}{4}\sum_{k k'\sigma}
 c_{k\sigma}^{\dagger}c_{k'\sigma}
 +J{\bf{S}}\cdot{\bf{s}}.
 \label{H-cot-one-el}
\end{eqnarray}
Here $\bf S$ is the spin one-half operator of the dot,
\begin{eqnarray}
{\bf{S}}&=&
 \frac{1}{2}\sum_{\sigma\sigma'}
 d_{\sigma}^{\dag}
 {\mbox{$\boldsymbol\tau$}}_{\sigma\sigma'}
 d_{\sigma'},
 \end{eqnarray}
${\bf s}$ represents the spin states of the conduction electrons,
\begin{eqnarray}
 {\bf{s}}&=&
 \frac{1}{2}\sum_{kk'\sigma\sigma'}
 c_{k\sigma}^{\dagger}
 {\mbox{$\boldsymbol\tau$}}_{\sigma\sigma'}
 c_{k'\sigma'},
 \label{sp-l}
\end{eqnarray}
and $\boldsymbol\tau$ is the vector of Pauli matrices. The
antiferromagnetic coupling constant is,
\begin{eqnarray}
 J &=& \frac{V^2}{2}
 \left(
      \frac{1}{\epsilon_F-\varepsilon_0}+
      \frac{1}{\varepsilon_0+Q-\epsilon_F}
 \right).\label{J-1dot}
\end{eqnarray}

\begin{figure}[htb]
\centering
\includegraphics[width=120mm,height=40mm,angle=0]{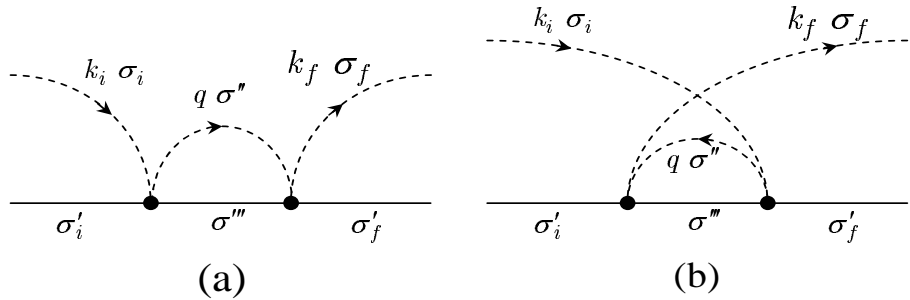}
\caption{Second order diagrams describing the scattering of a
conduction electron from the state $k_i\sigma_i$ into an
intermediate particle (panel (a)) or hole (panel (b)) state
$q\sigma''$ and then to a final state $k_f\sigma_f$. The dashed
lines represent the conduction electron, whereas the solid lines
correspond to the localized spin of the dot.}
\label{fig-PoorManSc}
\end{figure}

Scaling equation for the coupling constant (\ref{J-1dot}) can be
derived by the poor-man's scaling method \cite{Anderson}. The
essence of the scaling approach is that the higher energy
excitations can be absorbed as a renormalization of the coupling
constant $J$. To carry out the scaling we divide the conduction
band into states, $-{\bar D}+|\delta D|<\epsilon_k<{\bar
D}-|\delta D|$, which are retained, and states within $|\delta D|$
of the band edge which are to be eliminated provided the effective
exchange Hamiltonian, $J{\bf{S}}\cdot{\bf{s}}$, (the last term of
the Kondo Hamiltonian (\ref{H-cot-one-el})) is perturbatively
renormalized by changing the coupling constant $J\to J+\delta J$.
The lowest order corrections to $J$ due to virtual scattering of
conduction electrons into the band edge can be represented by the
second order diagrams (Fig. \ref{fig-PoorManSc}). Calculating the
contribution of these diagrams one obtains,
\begin{eqnarray}\label{delta-J}
\delta J&=&-\rho_0 J^2\frac{|\delta D|}{D}.
\end{eqnarray}
Eq. (\ref{delta-J}) leads to the scaling equation,
\begin{eqnarray}
  \frac{d J}{d\ln (\rho_0 D)}&=&-\rho_0 J^2.
    \label{scaling-one-el}
\end{eqnarray}
Integrating Eq. (\ref{scaling-one-el}) from the initial band width
${\bar D}$ and coupling constant $J$ (\ref{J-1dot}) to a new band
width ${\widetilde D}$ and renormalized coupling constant
$\widetilde J$ yields,
\begin{eqnarray}\label{J-tildeJ}
{\bar D}\exp \left(-\frac{1}{\rho_0 J}\right)&=&{\widetilde D}\exp
\left(-\frac{1}{\rho_0 \widetilde J}\right).
\end{eqnarray}
Eq. (\ref{J-tildeJ}) shows that the solution of the scaling
equation (\ref{scaling-one-el}) is characterized by a single
parameter which plays the role of a {\it scaling invariant}
\cite{Hewson}. This scaling invariant is called Kondo temperature,
\begin{eqnarray}
  T_K &=& {\bar D}\exp
  \left(
       -\frac{1}{\rho_0 J}
  \right).
    \label{TK-one-el}
\end{eqnarray}

The Kondo effect can be observed by measuring the dc current
induced by a direct bias voltage $V_{dc}$ applied across the dot.
The corresponding differential conductance
$G(V_{dc})={dI}/{dV_{dc}}$ exhibits a sharp peak at $V_{dc}=0$,
which is called zero-bias anomaly (ZBA) \cite{KKK,KKK1,Sim,Wiel}.
At finite value of the bias $eV_{dc}\gg T_K$ the Kondo effect is
suppressed \cite{Noneq1}. Therefore, the width of the peak of the
differential conductance at zero bias is of the order of $T_K$. In
the weak coupling regime $T\gg T_K$ the Kondo contribution $G_K$
to the differential conductance can be calculated by means of
Keldysh technique \cite{kng},
\begin{eqnarray}
 G_K &=& \frac{3\pi^2 }{16}
 \frac{1}{[\ln(T/T_K)]^2}G_U, \ \ \
 G_U=\frac{e^2}{\pi\hbar}\frac{4V_{s}^2V_{d}^2}{(V_s^2+V_d^2)^2}.
 \label{Cond-one-el}
\end{eqnarray}
The Kondo temperature $T_K$ is the only energy scale which
controls all low-energy properties of the quantum dot. Eq.
(\ref{Cond-one-el}) shows that the ratio $G_K/G_U$ depends only on
the dimensionless variable $T/T_K$ \cite{kng}. In the strong
coupling regime $T\ll T_K$ the spin-flip scattering is suppressed,
and the system allows an effective Fermi liquid description
\cite{Noz}. The zero-bias conductance then follows from the
Landauer formula, $G_K=G_U$.

If a magnetic field is applied to the system, the zero-bias peak
splits into two peaks at $eV_{dc}=\pm E_Z$, where $E_Z$ is the
Zeeman energy. These peaks are observable even at $eV_{dc},\pm E_Z
\gg T_K$ \cite{Noneq1,KKK1,Klitz}. However, due to a
nonequilibrium-induced decoherence, these peaks are wider than
$T_K$, and the value of the conductance at the peaks never reaches
the unitary limit $G_U$ \cite{Noneq1,win-meir,kng-prl}. In the
next section we demonstrate that quantum dots with even electron
occupation may exhibit a generic Kondo effect at certain value of
the Zeeman energy $E_Z\gg T_K$.

\section{Magnetic Field Induced Kondo Tunneling through Evenly Occupied QD}
\label{sec.2.2}

The Kondo effect discussed in the previous section takes place
only if the dot has a nonzero spin in the ground state. This is
always the case for odd electron occupation ${\cal N}$. When
${\cal N}$ is even, the ground state of the spin-degenerate QD is
a singlet ($S=0$) since all single-particle levels are occupied by
pairs of electrons with opposite spins. According to Hund's rule,
the lowest excited state of the dot is a triplet ($S=1$) at a
distance $\delta$ above the ground state. The spacing $\delta$ can
be tuned by means of a magnetic field. Application of a magnetic
field results in a singlet-triplet transition in the ground state
of the dot, leading to a Kondo effect. This {\it magnetic field
induced Kondo effect} occurs both in vertical QDs with shell-like
structure of electronic states \cite{Magn-v1,Magn-v2,Magn-v3} and
in planar (lateral) QDs formed by orbitally non-degenerate
electron states \cite{Magn}. In the former case the Zeeman effect,
which lifts the spin degeneracy, is negligibly  small in
comparison with diamagnetic shift because of a small value of the
effective $g$-factor in semiconductor heterostructures. Magnetic
field affects mostly the orbital states, leading to
singlet-triplet level crossing. In the latter case, the Zeeman
contribution dominates, and twofold degeneracy of the ground state
of an isolated dot appears only if the Zeeman energy $E_Z$ is
equal to the single-particle level spacing $\delta$ in the dot.
Both types of field--induced Kondo tunneling were observed
\cite{Magn1,Magn-v}.

\begin{figure}[htb]
\centering
\includegraphics[width=65mm,height=55mm,angle=0]{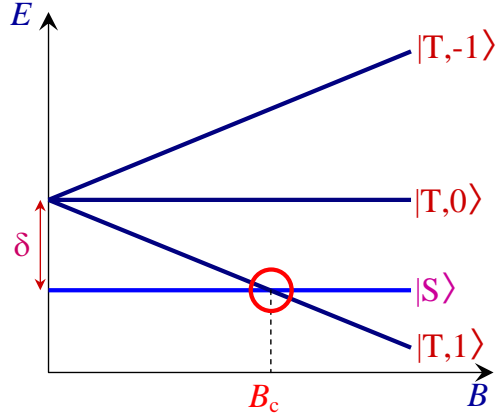}
\caption{Low-energy states of an evenly occupied QD in magnetic
field. The spin-up projection $|T,1\rangle$ of the triplet becomes
degenerate with the singlet ground state at
$B_c=\delta/g{\mu}_B$.} \label{magn-ind-th}
\end{figure}
\begin{figure}[htb]
\centering
\includegraphics[width=120mm,height=40mm,angle=0]{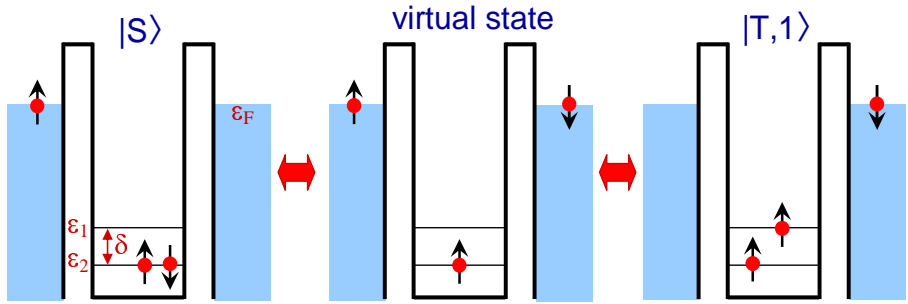}
\caption{Magnetic field induced Kondo effect in a QD with even
occupation. Spin-flip transitions connecting the singlet
$|S\rangle$ (left panel) and triplet $|T,1\rangle$ (right panel)
states. The intermediate high-energy virtual state is shown in the
middle panel.} \label{spin-flip-M}
\end{figure}
\begin{figure}[htb]
\centering
\includegraphics[width=45mm,height=60mm,angle=0]{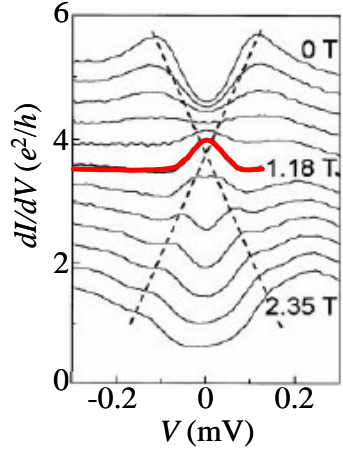}
\caption{Differential conductance $dI/dV$ as a function of bias
voltage $V$ at different values of magnetic field $B$
\cite{Magn1}. At $B_c=1.18$ T, the differential conductance has a
peak at zero voltage.} \label{magn-ind-exp}
\end{figure}

Let us consider a planar QD with even number of electrons, weakly
connected to the metallic electrodes, and subject to an external
magnetic field. In QDs, charge and spin excitations are controlled
by two energy scales, charging energy $Q$ and single-particle
level spacing $\delta$ respectively, which typically differ by an
order of magnitude: $Q\sim 1$meV, $\delta\sim 0.1$meV
\cite{KKK,KKK1,Sim,Wiel}. This separation of energy scales allows
one to change the spin state of the dot, without changing its
charge. In an applied field $B_c=\delta/g{\mu}_B$ \cite{Magn}, the
spin-up projection of the triplet becomes degenerate with the
singlet ground state (Fig. \ref{magn-ind-th}). At this point, the
spin-flip transitions shown in Fig. \ref{spin-flip-M} become
possible, leading to a new type of Kondo resonance. It was found
\cite{Magn1} that at a certain magnetic field the differential
conductance has a peak at zero bias voltage (Fig.
\ref{magn-ind-exp}).


Initially, the QD in Fig. \ref{spin-flip-M}  is treated within an
Anderson-type model with bare level operators $d_{\sigma i}$,
energies $\varepsilon_{i}$ and tunneling matrix elements $V_{i}$
with $i=1,2$ (we consider the symmetric case $V_{is}=V_{id}\equiv
V_i$). Next, the isolated dot Hamiltonian is diagonalized in the
Hilbert space which is a direct sum of two ($|\Lambda\rangle$),
one and three ($|\lambda\rangle$) electron states, using Hubbard
operators $X^{\gamma \gamma}=|\gamma\rangle\langle\gamma|$
($\gamma=\lambda, \Lambda$) \cite{Hub,Hub1}. The two particle
states $|\Lambda\rangle$ exhaust the lowest part of the spectrum
consisting of a singlet $|S\rangle$ and triplet $|T,\mu\rangle \
(\mu=1,0,-1)$. The corresponding energies are,
\begin{eqnarray}
E_{S} &=& 2{\varepsilon}_2,\ \ \ \ \  E_{{T\mu}} = {\varepsilon}_1
+{\varepsilon}_{2}+g\mu_B \mu B,
 \label{En-K}
\end{eqnarray}
with $\varepsilon_1-\varepsilon_2=\delta$, $\mu_B$ is the Bohr
magneton and $g\approx 2$ is the free electron $g$-factor. The
energies of one and three electron states are of order of the
charging energy, $E_{\lambda}\sim Q$. Finally, tunneling operators
in the bare Anderson Hamiltonian are replaced by a product of
number changing Hubbard operators $X^{\lambda \Lambda}$ and a
combination $c_{k\sigma}=2^{-1/2}(c_{k\sigma s}+c_{k\sigma d})$ of
lead electron operators, ($k=$momentum, $\sigma=$ spin projection
and $s,d$ stand for source and drain).

With these preliminaries, the starting point is a generalized
Anderson Hamiltonian describing the QD in tunneling contact with
the leads,
\begin{eqnarray}
H_{A}= \sum_{\alpha=s,d}\sum_{k\sigma}\epsilon_{k}
c^{\dagger}_{k\sigma \alpha}c_{k\sigma \alpha} +
\sum_{\gamma=\Lambda \lambda} {E}_\gamma
X^{\gamma\gamma} 
+\left(\sum_{\Lambda\lambda} \sum_{k
\sigma}V^{\lambda\Lambda}_{\sigma i}
c^{\dagger}_{k\sigma}X^{\lambda \Lambda}+ H.c.\right), \label{H-K}
\end{eqnarray}
with dispersion $\epsilon_{k}$ of electrons in the leads and
$V^{\lambda\Lambda}_{\sigma i}\equiv V_{i}\langle\lambda|d_{\sigma
i}|\Lambda\rangle$. The two states of the dot which become
degenerate at $B_c=\delta/g{\mu}_B$, are
\begin{eqnarray}\label{s-1}
|S\rangle=d^{\dag}_{2\uparrow}d^{\dag}_{2\downarrow}|0\rangle, \ \
\ \ \ \ \
|T,1\rangle=d^{\dag}_{1\uparrow}d^{\dag}_{2\uparrow}|0\rangle.
\end{eqnarray}

Since $Q\gg \delta$, one- and three-electron states can be
integrated out by means of the SW transformation \cite{SW}. The
resulting exchange Hamiltonian has a form of {\it anisotropic}
Kondo Hamiltonian,
\begin{eqnarray}
 H_{ex} =J_{\parallel}S_zs_z+
               \frac{J_{\perp}}{2}
               \left(
                    S^{+}s^{-}+
                    S^{-}s^{+}
               \right).
 \label{int5m-K}
\end{eqnarray}
Here the effective exchange constants are
\begin{eqnarray}
  J_{\parallel}=
  \frac{2(V_{1}^2+V_{2}^2)}{Q},
  \ \ \ \ \
  J_{\perp}=
  \frac{4V_{1}V_{2}}{Q}.
  \label{J-K}
\end{eqnarray}
The spherical components of the dot spin operator ${\bf S}$ are
defined via Hubbard operators connecting the $|S\rangle$ and
$|T,1\rangle$ states of the dot,
\begin{eqnarray}
 S_z=\frac{1}{2}\left(X^{1,1}-X^{SS}\right),
 \ \ \ \ \
 S^{+}=X^{1,S},
 \ \ \ \ \
 S^{-}=X^{S,1}.
 \label{S-K}
\end{eqnarray}
The conduction electron spin operators are determined by Eq.
(\ref{sp-l}).

The scaling equations for dimensionless exchange constants
$j_{\nu}=\rho_0 J_{\nu}$ $(\nu=\parallel,\perp)$ read,
\begin{eqnarray}
\frac{dj_{\parallel}}{d\ln d} = -(j_\perp)^2,\ \ \ \ \ \ \ \
\frac{dj_{\perp}}{d\ln d} = -j_{\parallel}j_{\perp},
 \label{sc-m}
\end{eqnarray}
yielding the Kondo temperature,
\begin{eqnarray}
T_{K}&=&\bar{D}\exp\left(-\frac{A}{2j_{\parallel}}\right),
\label{T-m}
\end{eqnarray}
where $\bar D$ is the effective bandwidth in the SW limit, and
$$A=\frac{j_{\parallel}}{\sqrt{j_{\parallel}^2-j_{\perp}^2}}
\ln \left(\frac{j_{\parallel}-\sqrt{j_{\parallel}^2-j_{\perp}^2}}
{j_{\parallel}+\sqrt{j_{\parallel}^2-j_{\perp}^2}}\right).$$
 In the isotropic limit $j_{\parallel}-j_{\perp}\to 0$ one has
 $A\to 2$ and Eq.(\ref{T-m}) reduces to the usual expression
 $T_K={\bar D}\exp(-1/j_{\parallel})$.

\section{Singlet-Triplet Kondo Effect in Double Quantum
Dot}\label{sec.2.3}

The isolated quantum dots considered above are typical examples of
artificial atoms. A double valley quantum dot with weak capacitive
and/or tunnelling coupling between its two wells may be considered
as the simplest case of artificial two-atom molecule. Its closest
natural analog is the hydrogen molecule H$_2$ or the corresponding
molecular ions H$_2^\pm$ for the occupation number ${\cal
N}=2,1,3$, respectively \cite{Smit}.
\begin{figure}[htb]
\centering
\includegraphics[width=50mm,height=30mm,angle=0]{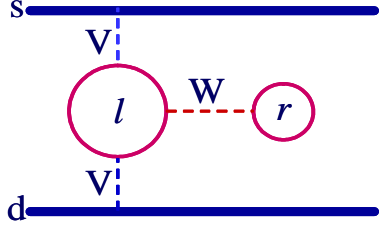}
\caption{Double Quantum Dot in side (T-shaped) geometry.}
\label{DQD}
\end{figure}

Let us consider the DQD with two electron occupation in a side
geometry (Fig. \ref{DQD}). Each valley is described by the
one-electron level $\varepsilon_a$, Coulomb blockade energy $Q_a$,
and bare level operators $d_{\sigma a}$ ($a=l,r$ for left and
right dot, respectively). The right dot is assumed to have a
smaller radius and, hence, larger capacitive energy than the left
dot, i.e., $Q_r{\gg}Q_l$. The left dot is coupled by tunneling $W$
to the right dot ($W{\ll}Q_{l,r}$) and by tunneling $V$ to the
source (s) and drain (d) leads. The spectrum of an isolated DQD
consists of the singlet ground state $|S\rangle$, the low-lying
triplet exciton $|T,\mu\rangle$ $(\mu=1,0,{\bar 1})$ and
high-energy charge transfer singlet exciton $|Ex\rangle$,
\begin{eqnarray}
  &&|S\rangle =
  \alpha |s\rangle
         -\sqrt{2}\beta |ex\rangle,
  \nonumber\\
  &&|T,1\rangle =
  d_{l\uparrow }^{\dag}d_{r\uparrow }^{\dag}
  |0\rangle, \ \ \ \ |T,{\bar 1}\rangle =
  d_{l\downarrow }^{\dag}d_{r\downarrow }^{\dag}
  |0\rangle,
 \nonumber \\
  &&|T,0\rangle =
  \frac{1}{\sqrt{2}}
  \left(
       d_{l\uparrow }^{\dag}d_{r\downarrow }^{\dag}+
       d_{l\downarrow }^{\dag}d_{r\uparrow }^{\dag}
  \right)|0\rangle,
  \nonumber\\
  &&|Ex\rangle =\alpha |ex\rangle +\sqrt{2}\beta |s\rangle,
  \label{STex-DQD}
\end{eqnarray}
where
$$|s\rangle =
  \frac{1}{\sqrt{2}}\left(d_{l\uparrow }^{\dag}d_{r\downarrow }^{\dag}-
       d_{l\downarrow }^{\dag}d_{r\uparrow }^{\dag}
  \right)|0\rangle, \ \ \
  |ex\rangle =d_{l\uparrow }^{\dag}d_{l\downarrow }^{\dag}
  |0\rangle .$$
  The corresponding energies are,
\begin{eqnarray}
  E_{S} = \epsilon_l +{\epsilon}_r-2W\beta,
  \ \ \ \ \ \
  E_{T} = \epsilon_l +{\epsilon}_r,
  \ \ \ \ \
  E_{Ex} = 2\epsilon_l+2W\beta,
  \label{En-DQDq}
\end{eqnarray}
where $\beta=W/{\Delta}_{lr}\ll 1$
$({\Delta}_{lr}=Q_l+\epsilon_l-\epsilon_r)$.

\begin{figure}[htb]
\centering
\includegraphics[width=60mm,height=40mm,angle=0]{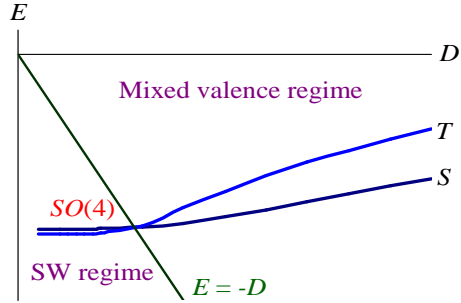}
\caption{Haldane renormalization group procedure. }
\label{DQD-Hald}
\end{figure}

The DQD in tunneling contact with the leads can be described by a
generalized Anderson Hamiltonian,
\begin{eqnarray}
H_{A}= \sum_{b=s,d}\sum_{k\sigma}\epsilon_{kb}
c^{\dagger}_{k\sigma b}c_{k\sigma b} + \sum_{\gamma=\Lambda
\lambda} {E}_\gamma
X^{\gamma\gamma} 
+\left(\sum_{\Lambda\lambda} \sum_{k \sigma
}V^{\lambda\Lambda}_{\sigma } c^{\dagger}_{k\sigma}X^{\lambda
\Lambda}+ H.c.\right). \label{H}
\end{eqnarray}
Here $|\Lambda\rangle$ are the two-electron eigenfunctions
(\ref{STex-DQD}), $|\lambda\rangle$ are the one- and
 three-electron eigenstates;
  $X^{\lambda\Lambda}=|\lambda\rangle\langle\Lambda|$ are
 number changing dot Hubbard
 operators; $V^{\lambda\Lambda}_{\sigma }\equiv
V\langle\lambda|d_{l\sigma }|\Lambda\rangle$. The Kondo effect at
$T>T_K$ is unravelled by employing a renormalization group (RG)
procedure \cite{Hald} in which the energies $E_{\gamma}$ are
renormalized as a result of rescaling high-energy charge
excitations (see Eqs.(\ref{en-sc}) and (\ref{inv-SQD})). Our
attention, though, is focused on renormalization of $E_{S},E_{T}$
(\ref{En-DQDq}). Since the tunnel constants are irrelevant
variables \cite{KA01,Hald}, the scaling equations are
\begin{eqnarray}
\frac{dE_\Lambda}{d\ln D}=\frac{\Gamma_\Lambda}{\pi}. \label{DE-2}
\end{eqnarray}
Here $2D$ is the conduction electron bandwidth, $\Gamma_\Lambda$
are the tunneling strengths,
\begin{eqnarray}
\Gamma_{T} = \pi\rho_0V^2,\ \ \ \  \Gamma_{S} =
{\alpha}^{2}\Gamma_{T},
 \label{GammaR-2}
\end{eqnarray}
with $\alpha=\sqrt{1-2\beta^2}<1$ and $\rho_0$ being the density
of states at $\varepsilon_F$. The scaling invariants for equations
(\ref{DE-2}),
\begin{eqnarray}
E_{\Lambda}^{\ast }=E_{\Lambda}(D)-\frac{\Gamma
_\Lambda}{\pi}\ln\left(\frac{\pi D}{\Gamma _\Lambda}\right),
\label{inv-2}
\end{eqnarray}
are tuned to satisfy the initial condition
$E_{\Lambda}(D_{0})=E_{\Lambda}^{(0)}$. Equations (\ref{DE-2})
determine two scaling trajectories $E_\Lambda(D)$ for singlet and
triplet states. Note that the level $E_{Ex}$ is irrelevant, but
admixture of the bare exciton $|ex\rangle$ in the singlet state is
crucial for the inequality of tunneling rates
$\Gamma_{T}>\Gamma_{S}$ \cite{KA01,KA02}. As a result, the energy
$E_{T}(D)$ decreases with $D$ faster than $E_{S}(D)$, so that the
trajectory $E_{T}(D, \Gamma_{T})$ usually intersects $E_{S}(D,
\Gamma_{S})$ at a certain point $D=D_{c}$. This level crossing may
occur either before or after reaching the Schrieffer-Wolff (SW)
limit ${\bar D}$ where $E_\Lambda({\bar D})\sim {\bar D}$ and
scaling terminates \cite{Hald}. When the scaling trajectories
cross near the SW boundary $D_c\sim {\bar D}$ the singlet ground
state becomes degenerate with a triplet one (Fig. \ref{DQD-Hald}).
As a result, the Kondo resonance may arise in spite of the even
electron occupation of the DQD.

The above Haldane RG procedure brings us to the SW limit
\cite{SW}, where all charge degrees of freedom are quenched. By
properly tuning the SW transformation $e^{iS}$ the effective
Hamiltonian $H=e^{iS}H_{A}e^{-iS}$ is of the $s-d$ type
\cite{Hewson}. However, unlike the conventional case \cite{SW} of
doublet spin $1/2$  we have here the degenerate singlet and
triplet states $\Lambda=\{S,T\}$, and the SW transformation
intermixes these states. To order $O(|V|^2)$, then, the tunneling
exchange Hamiltonian reads,
\begin{eqnarray}
&&H=\sum_{\Lambda}E_{\Lambda}X^{\Lambda\Lambda}+\sum_{k\sigma
b}\epsilon_{kb} c^{\dagger}_{k\sigma b}c_{k\sigma b}+
 J^T {\bf S}\cdot {\bf s}+ J^{ST} {\bf R}\cdot {\bf s}. \label{ex-2}
\end{eqnarray}
Here the (antiferromagnetic) coupling constants are
\begin{eqnarray}
J^T=\frac{V^2}{\varepsilon_{F}-\epsilon_l}, \ \ \ \ \ \
J^{ST}=\alpha J^T.\label{cons-2}
\end{eqnarray}
The conduction electron spin operator ${\bf s}$ is defined by Eq.
(\ref{sp-l}). ${\bf S}$ is the dot spin one operator with
projections ${\mu} =1,0,{\bar{1}}$, while ${\bf R}$ couples
singlet $|S\rangle$ with triplet $\langle T \mu|$. Their spherical
components are defined via Hubbard operators:
\begin{eqnarray}
&&S^{+}= \sqrt{2}( X^{10}+X^{0\bar{1}}),\ \ \ \ \ \; S^{-} =
(S^{+})^{\dagger}, \ \ \ \ \
S^z=X^{11}-X^{\bar{1}\bar{1}}, \nonumber\\
 &&R^{+}= \sqrt{2}(
X^{1S}-X^{S\bar{1}}),\ \ \ R^{-} = (R^{+})^{\dagger}, \ \ \
R^z=-(X^{0S}+X^{S0}).\label{comm-2}
\end{eqnarray}
The vector operators ${\bf R}$ and ${\bf S}$ obey the commutation
relations of $o_{4}$ Lie algebra,
\begin{eqnarray}
&&\lbrack S_{j},S_{k}]=ie_{jkm}S_{m},\ \ \
[R_{j},R_{k}]=ie_{jkm}S_{m}, \ \ \ [R_{j},S_{k}]=ie_{jkm}R_{m}.
\label{3.9e}
\end{eqnarray}
(here $j,k,m$ are Cartesian indices). Besides, ${\bf S}\cdot {\bf
R}=0,$ and the Casimir operator is ${\bf S}^{2}+{\bf R}^{2}=3.$
The operators ${\bf S}$ and ${\bf R}$ manifest the $SO(4)$
dynamical symmetry of the DQD. This justifies the qualification of
such DQD as a {\it spin rotator} \cite{KA01,KA02}.

Scaling equations for $J^T$ and $J^{ST}$ are,
\begin{eqnarray}
\frac{dj_1}{d\ln d} = -\left[(j_1)^2+(j_2)^2\right],\ \ \ \ \
\frac{dj_2}{d\ln d} = -2j_1j_2 , \label{dj-2}
\end{eqnarray}
with $j_1=\rho_0J^T, j_2=\rho_0J^{ST}, d=\rho_0D$. If
$\bar\delta=E_T({\bar D})-E_S({\bar D})$ is the smallest energy
scale, the energy spectrum of the DQD is quasi degenerate and the
system (\ref{dj-2}) is reduced to a single equation for the
effective integral $j_{+}=j_1+j_2$,
\begin{eqnarray}
 \frac{dj_{+}}{d\ln d}=-(j_{+})^2.
 \label{dj+2}
\end{eqnarray}
Then the RG flow diagram has an infinite fixed point, and the
solution of Eq. (\ref{dj+2}) gives the Kondo temperature
\begin{eqnarray}
  T_{K0}={\bar D}\exp\left(-\frac{1}{j_{+}}\right).
  \label{TK-2s}
\end{eqnarray}

In the general case, the scaling behavior is more complicated. The
flow diagram still has a fixed point at infinity, but the Kondo
temperature turns out to be a sharp function of $\bar\delta$. In
the case $\bar\delta<0$, $|\bar\delta|\gg T_K$, the scaling of
$J^{ST}$ terminates at $D\sim|\bar\delta|$
\cite{Magn-v1,Magn-v2,KA01,KA02}. Then one is left with the
familiar physics of an under-screened spin one Kondo model
\cite{NB}. The fixed point is still at infinite exchange coupling
$J^T$, but the Kondo temperature becomes a function of
$\bar\delta$,
\begin{eqnarray}
  \frac{T_K}{T_{K0}}\approx
  \left(\frac{T_{K0}}{|\bar\delta|}\right)^{\alpha},
  \label{TK-SO3-2}
\end{eqnarray}
where $\alpha<1$ is determined by the DQD parameters (see the text
after Eq. (\ref{GammaR-2})). The symmetry of the DQD in this case
is $SO(3)$.

\section{Dynamical Symmetry of Complex Quantum Dots} \label{sec.2.4}
In the previous sections an accidental level degeneracy is induced
by an external magnetic field (Sec. \ref{sec.2.2}) and the
dot-lead interaction (Sec. \ref{sec.2.3}). In both cases the
nano-object possesses dynamical symmetry. In this section we
present the concept of dynamical symmetry in more details, and
particularly, discuss its emergence in Complex Quantum Dots (CQD).

The term {\it Dynamical Symmetry} implies the symmetry of
eigenvectors of a quantum system forming an irreducible
representation of a certain Lie group. The main ideas and the
relevant mathematical tools can be found, e.g., in Refs.
\cite{Dynsym}. Here they are formulated in a form convenient for
our specific purposes without much mathematical rigor. We have in
mind a quantum system with Hamiltonian $H$ whose eigenstates
$|\Lambda \rangle = |M\mu\rangle$ form ( for a given $M$) a basis
to an irreducible representation of some Lie group ${\cal G}$. The
energies $E_{M}$ do not depend on the "magnetic" quantum number
$\mu$. For definiteness one may think of $M$ as an angular
momentum and of $\mu$ as its projection, so that ${\cal G}$ is
just $SU(2)$. Now let us look for operators which induce
transitions between different eigenstates. An economic way for
identifying them is through the Hubbard operators \cite{Hub}
\begin{equation}
X^{\Lambda\Lambda^\prime}=|\Lambda\rangle\langle \Lambda^\prime|.
\label{Hub}
\end{equation}
It is natural to divide this set of operators into two subsets.
The first one contains the operators $|M\mu\rangle \langle
\mu^{\prime}M| $ while the second one includes operators
$|M\mu\rangle \langle \mu^{\prime}M^ {\prime}|$ $(M\neq M^
{\prime})$ for which $|M \mu \rangle $ and $|M' \mu'\rangle$
belong to a {\it different} representation space of ${\cal G}$. A
central question at this stage is whether these operators (or
rather, certain linear combinations of them) form a close algebra.
In some particular cases it is possible to form linear
combinations within each set and obtain two new sets of operators
$\{S\}$ and $\{R\}$ with the following properties: 1) For a given
$M$ the operators $\{S\}$ generate the $M$ irreducible
representation of ${\cal G}$ and commute with $H$. 2) For a given
set $M_{i}$ the operators $\{S\}$ and $\{R\}$ form an algebra (the
{\it dynamic} algebra) and generate a non-compact Lie group ${\cal
A}$. The reason for the adjective {\it dynamic} is that,
originally,  the operators $\{R\}$ do not appear in the bare
Hamiltonian $H$ and emerge only when additional interactions
(e.g., dot-lead tunneling) are present. In the special case ${\cal
G}=SU(2)$ the operators in $\{S\}$ are the vector ${\bf S}$ of
spin operators determining the corresponding irreducible
representations,  while the operators in the set $\{R\}$ can be
grouped into a sequence ${\bf R}_{n}$ of vector operators
describing transitions between states belonging to different
representations of the $SU(2)$ group.

Strictly speaking, the group ${\cal A}$ is not a symmetry group of
the Hamiltonian $H$ since the operators $\{R\}$ do not commute
with $H$. Indeed, let us express $H$ in terms of diagonal Hubbard
operators,
\begin{equation} H=\sum_{\Lambda=M\mu}E_{\Lambda}
|\Lambda\rangle\langle \Lambda| =\sum_{\Lambda}E_{M} X^{\Lambda
\Lambda} ~, \label{HX}
\end{equation}
so that
\begin{equation}\label{comm}
[X^{\Lambda\Lambda^{\prime}},H]=-(E_{M}-E_{M^{\prime}})X^{\Lambda
\Lambda^{\prime}}.
\end{equation}
As we have mentioned above, the symmetry group ${\cal G}$ of the
Hamiltonian $H$, is generated by the operators
$X^{\Lambda=M\mu,\Lambda^{\prime}=M\mu^{\prime}} $. Remarkably,
however, the dynamics of CQD in contact with metallic leads and/or
an external magnetic field leads to renormalization of the
energies $\{E_{M}\}$ in such a way that a few levels at the bottom
of the spectrum become degenerate, $E_{M_{1}}=E_{M_{2}}=\ldots
E_{M_{n}}$. Hence, in this low energy subspace, the group ${\cal
A}$ which is generated by the operators $\{S\}$ and $\{R\}$ is a
symmetry group of $H$ referred to as the dynamical symmetry group.
The symbol $R$ is due to the analogy with the Runge-Lenz operator,
the hallmark of dynamical symmetry of the Kepler and Coulomb
problems. (The Coulomb potential possesses accidental degeneracy
of states with different angular momentum ${\bf l}$. Hence,
according to (\ref{comm}) the Runge-Lenz vector is an integral of
the motion. In this case one speaks about {\it hidden} symmetry of
the system.) Below we will use the term dynamical symmetry also in
cases where the levels are not strictly degenerate, but their
differences are bounded by a certain energy scale, which is the
Kondo energy in our special case. In that sense, the symmetry is
of course not exact, but rather, approximate.

Using the notions of dynamical symmetry, numerous familiar quantum
objects, such as hydrogen atom, quantum oscillator in
$d$-dimensions, quantum rotator, may be described in a compact and
elegant way. We are interested in a special application of this
theory, when the symmetry of the quantum system is approximate and
its violation may be treated as a perturbation. This aspect of
dynamical symmetry was first introduced in particle physics
\cite{DGN}, where the classification of hadron eigenstates is
given in terms of non-compact Lie groups. In our case, the
rotationally invariant object is an isolated quantum dot, whose
spin symmetry is violated by electron tunneling between the dot
and leads under the condition of strong Coulomb blockade.

The special cases ${\cal G}=SU(2)$ and ${\cal A}=SO(n)$ or $SU(n)$
is realizable in CQD. Let us first recall the manner in which the
spin vectors appear in the effective low energy Hamiltonian of the
QD in tunneling contact with metallic leads. When strong Coulomb
blockade completely suppresses charge fluctuations in QD, only
spin degrees of freedom are involved in tunneling via the Kondo
mechanism \cite{Glaz,Ng}. An {\it isolated} QD in this regime is
represented solely by its spin vector $\bf{S}$. This is a
manifestation of rotational symmetry which is of geometrical
origin. The exchange interaction $J{\bf S}\cdot{\bf s}$ (${\bf s}$
is the spin operator of the metallic electrons) induces
transitions between states belonging to the same spin (and breaks
$SU(2)$ invariance). On the other hand, the low energy spectrum of
spin excitations in CQD is not characterized solely by its spin
operator since there are states close in energy, which belong to
different representation spaces of $SU(2)$. Incidentally, these
might have either the same spin ${\bf S}$ (like, {\it e.g}, in two
different doublets) or a different spin (like, {\it e.g.}, in the
case of singlet-triplet or doublet-quartet transitions). The
exchange interaction must then contain also other operators ${\bf
R}_n$ (the R-operators mentioned above) inducing transitions
between states belonging to different representations. The
interesting physics occurs when the operators ${\bf R}_n$
``approximately'' commute with the Hamiltonian $H_{dot}$ of the
isolated dot. In accordance with our previous discussion, the
R-operators are expressible in terms of Hubbard operators and have
only non-diagonal matrix elements in the basis of the eigenstates
of $H_{dot}$. The spin algebra is then a subalgebra of a more
general non-compact Lie algebra formed by the whole set of vector
operators $\{{\bf S}, {\bf R}_n \}$. This algebra is characterized
by the commutation relations,
\begin{eqnarray}
&&[S_i,S_j]=it_{ijk}S_k,\ \ \ \ [S_i,R_{nj}] = i t_{ijk} R_{nk},\
\ \ \ [R_{ni},R_{nj}]  =  i t^n_{ijk} S_k, \label{1.1}
\end{eqnarray}
with structure constants $t_{ijk}$, $t^n_{ijk}$ (here $ijk$ are
Cartesian indices). The R-operators are orthogonal to $\bf S$,
\begin{equation}\label{orth}
{\bf S}\cdot {\bf R}_n=0.
\end{equation}
In the general case, CQDs possess also other symmetry elements
(permutations, reflections, finite rotations). Then, additional
scalar generators $A_p$ arise. These generators also may be
expressed via the bare Hubbard operators, and their commutation
relations with R-operators have the form
\begin{equation}
[R_{ni},R_{mj}] = i g_{ij}^{nmp} A_p,~~ [R_{ni},A_p] = i
f_{ij}^{nmp} R_{mj}, \label{1.2}
\end{equation}
with structure constants $g_{ij}^{nmp}$ and $f_{ij}^{nmp}$ ($n\neq
m$). The operators obeying the commutation relations (\ref{1.1})
and (\ref{1.2}) form an $o_n$ algebra. The Casimir operator for
this algebra is
\begin{equation}\label{1.K}
{\cal K}= {\bf S}^2 + \sum_n {\bf R}_n^2 + \sum_p A_p^2~.
\end{equation}
Various representations of all these operators via basic Hubbard
operators will be established in the following chapters, where the
properties of specific CQDs are studied.

 Next, we show how the dynamical symmetry of CQD is revealed in
the effective spin Hamiltonian describing Kondo tunneling. This
Hamiltonian is derived from the generalized Anderson Hamiltonian
\begin{equation}
H_A=H_{dot}+ H_{lead}+ H_{tun}. \label{1.3}
\end{equation}
The three terms on the right hand side are the dot, lead and
tunneling Hamiltonians, respectively. In the generic case, a
planar CQD is a confined region of a semiconductor secluded
between drain and source leads, with complicated multivalley
structure. The CQD contains several valleys numbered by index $a$.
Some of these valleys are connected with each other by tunnel
channels characterized by coupling constants $W_{aa'}$, and some
of them are connected with the leads by tunneling. The
corresponding tunneling matrix elements are $V_{ab}$ $(b=s,d$
stands for source and drain, respectively). The total number of
electrons $\cal N$ in a {\it neutral} CQD as well as the partial
occupation numbers ${\cal N}_a$ for the separate wells are
regulated by Coulomb blockade and gate voltages $v_{ga}$ applied
to these wells, with ${\cal N}=\sum_a {\cal N}_a$. It is assumed
that the capacitive energy for the whole CQD is strong enough to
suppress charged states with ${\cal N}'={\cal N}\pm1$, which may
arise in a process of lead-dot tunneling.

If the inter-well tunnel matrix elements $W_{aa'}$ are larger than
the dot-lead ones $V_{ab}$ (or if all tunneling strengths are
comparable), it is convenient first to diagonalize $H_{dot}$ and
then consider $H_{tun}$ as a perturbation. In this case $H_{dot}$
may be represented as
\begin{equation}\label{1.4}
H_{dot}=\sum_{\Lambda\in {\cal N}}E_{\Lambda}
|\Lambda\rangle\langle \Lambda| + \sum_{\lambda\in {\cal N}\pm
1}E_{\lambda} |\lambda\rangle\langle \lambda| .
\end{equation}
Here all intradot interactions are taken into account. The kets
$|\Lambda\rangle \equiv |{\cal N},q \rangle$ represent eigenstates
of $H_{dot}$ in the charge sector $\cal N$ and other quantum
numbers $q$, whereas the kets $|\lambda\rangle \equiv |{\cal N}\pm
1, p \rangle$ are eigenstates in the charge sectors ${\cal N}\pm
1$ with quantum numbers $p$.  All other charge states are
suppressed by Coulomb blockade. Usually, $q$ and $p$ refer to spin
quantum numbers but sometimes other specifications are required
(see below).

The lead Hamiltonian takes a form
\begin{equation}\label{1.5}
H_{lead}= \sum_{k,\alpha,\sigma} \varepsilon_{k
\alpha}c^\dagger_{\alpha k\sigma}c_{\alpha k\sigma}.
\end{equation}
In the general case, the individual dots composing the CQD are
spatially separated, so one should envisage the situation when
each dot is coupled by its own channel to the lead electron
states. So, the electrons in the leads are characterized by the
index $\alpha$, which specifies the lead (source and drain) and
the tunneling channel, as well as by the wave vector $k$ and spin
projection $\sigma$.

The tunnel Hamiltonian involves electron transfer between the
leads and the CQD, and thus couples states $|\Lambda \rangle$ of
the dot with occupation ${\cal N}$ and states  $|\lambda \rangle$
of the dot with occupation ${\cal N} \pm 1$. This is best encoded
in terms of non-diagonal dot Hubbard operators, which  intermix
the states from different {\it charge sectors}
\begin{equation}
 X^{\Lambda
\lambda}=|\Lambda \rangle \langle \lambda|, ~~~ X^{\lambda
\Lambda}=|\lambda \rangle \langle \Lambda|. \label{Xnondiag}
\end{equation}
Thus,
\begin{eqnarray}\label{1.6}
H_t  =\sum_{k\alpha a\sigma} \sum_{{\lambda\in {\cal N}+1}\atop
{\Lambda\in {\cal N}}}\left(V_{\alpha a
\sigma}^{\Lambda\lambda}c^\dagger_{\alpha k\sigma}|\Lambda\rangle
\langle \lambda| + H.c. \right )+\sum_{k\alpha
a\sigma}\sum_{{\lambda\in {\cal N}-1}\atop{\Lambda\in {\cal
N}}}\left( V_{\alpha a\sigma}^{\lambda\Lambda}c^\dagger_{\alpha
k\sigma}|\lambda\rangle \langle\Lambda| + H.c. \right ),
\end{eqnarray}
where $V_{\alpha a \sigma}^{\lambda\Lambda}=V_{\alpha }\langle
\lambda|d_{a\sigma}|\Lambda\rangle$.

Before turning to calculation of CQD conductance, the relevant
energy scales should be specified. First, we suppose that the
bandwidth of the continuum states in the leads, $D_\alpha$,
substantially exceeds the tunnel coupling constants, $D_\alpha \gg
W_{aa'},V_{\alpha }$ (actually, we consider leads made of the same
material with $D_{as}=D_{ad}=D_0$). Second, each well $a$ in the
CQD is characterized by the excitation energy defined as $\Delta_a
= E_\lambda({\cal N}_a-1)-E_\Lambda({\cal N}_a)$, i.e., the energy
necessary to extract one electron from the well containing ${\cal
N}_a$ electrons and move it to the Fermi level of the leads (from
now on the Fermi energy is used as the reference zero energy
level). Note that $\Delta_a$ is tunable by applying the
corresponding gate voltage $v_{ga}$. We are mainly interested in
situations where the condition
\begin{equation}\label{1.8} \Delta_c \sim D_0,~Q_c,
\end{equation}
is satisfied at least for one well labelled by the index $c$. Here
$Q_c$ is a capacitive energy, which is predetermined by the radius
of the well $c$. Eventually, this well with the largest charging
energy is responsible for Kondo-like effects in tunneling,
provided the occupation number ${\cal N}_c$ is odd. The third
condition assumed in most of our models is a weak enough Coulomb
blockade in all other wells except that with $a=c$, i.e., $Q_a\ll
Q_c$. Finally, we demand that
\begin{equation}\label{1.7}
b_{\alpha a}\equiv \frac{V_{\alpha }}{\Delta_a}\ll 1,
\end{equation}
for those wells, which are coupled with metallic leads, and
\begin{equation}\label{1.9}
\beta_{a}=\frac{W_{ac}}{E_{ac}}\ll 1.
\end{equation}
Here $E_{ac}$ are the charge transfer energies for electron
tunneling from the $c$-well to other wells in the CQD.

The interdot coupling under Coulomb blockade in each well
generates indirect exchange interactions between electrons
occupying different wells. Diagonalizing the dot Hamiltonian for a
\textit{given} ${\cal N}=\sum_a {\cal N}_a$, one easily finds that
the low-lying spin spectrum in the charge sectors with even
occupation ${\cal N}$ consists of singlet/triplet pairs (spin
$S=0$ or 1, respectively). In charge sectors with odd $\cal N$ the
manifold of spin states consists of doublets and quartets (spin
$S$=1/2 and 3/2, respectively).

The resonance Kondo tunneling is observed as a temperature
dependent zero bias anomaly in tunnel conductance \cite{KKK,KKK1}.
According to existing theoretical understanding, the quasielastic
cotunneling accompanied by the spin flip transitions in a quantum
dot is responsible for this anomaly. To describe the cotunneling
through a neutral CQD with given $\cal N$, one should integrate
out transitions involving high-energy states from charge sectors
with ${\cal N}'={\cal N}\pm 1$. In the weak coupling regime at
$T>T_K$ this procedure is done by means of perturbation theory
which can be employed in a compact form within the renormalization
group (RG) approach formulated in Refs. \cite{Hald,Anderson}.

As a result of the RG iteration procedure, the energy levels
$E_\Lambda$ in the Hamiltonian (\ref{1.4}) are renormalized and
indirect exchange interactions between the CQD and the leads
arise. The RG procedure is equivalent to summation of the
perturbation series at $T>T_K$, where $T_K$ is the Kondo energy
characterizing the crossover from a perturbative weak coupling
limit to a non-perturbative strong coupling regime. The leading
logarithmic approximation of perturbation theory corresponds to a
single-loop approximation of RG theory. Within this accuracy the
tunnel constants $W$ and $V$ are not renormalized, as well as the
charge transfer energy $\Delta_c$ (\ref{1.8}). Reduction of the
energy scale from the initial value $D_0$ to a lower scale $\sim
T$ results in renormalization of the energy levels $E_\Lambda \to
\bar{E}_\Lambda(D_0/T)$ and generates an indirect exchange
interaction between the dot and the leads with an
(antiferromagnetic) exchange constant ${\it J}$.

The rotational symmetry of a {\it simple} quantum dot is broken by
the spin-dependent interaction with the leads, which arises in
second order in the tunneling amplitude $V_\alpha$. In complete
analogy, the {\it dynamical symmetry} of a {\it composite} quantum
dot is exposed (broken) as encoded in the effective exchange
Hamiltonian. In a generic case, there are, in fact, several
exchange constants arranged within an exchange matrix $\textsf{J}$
which is non-diagonal both in dot and lead quantum numbers. The
corresponding exchange Hamiltonian is responsible for spin-flip
assisted cotunneling through the CQD as well as for
singlet-triplet transitions.

The precise manner in which these statements are quantified will
now be explained. After completing the RG procedure, one arrives
at an effective (or renormalized) Hamiltonian $\bar{H}$ in a
reduced energy scale $\bar{D}$,
\begin{equation}\label{1.12a}
\bar{H} =\bar{H}_{dot}+\bar{H}_{lead}+\bar{H}_{cotun},
\end{equation}
where the effective dot Hamiltonian (\ref{1.4}) is reduced to
\begin{equation}\label{1.12b}
\bar{H}_{dot}=\sum_{\Lambda\in {\cal N}} \bar{E}_\Lambda
X^{\Lambda\Lambda}
\end{equation}
written in terms of {\it diagonal} Hubbard operators,
\begin{eqnarray}
&& X^{\Lambda\Lambda}=|\Lambda\rangle\langle \Lambda|.
\label{Xdiag}
\end{eqnarray}
 At this stage, the manifold $\{\Lambda\}\in {\cal N}$
contains only the renormalized low-energy states within the energy
interval comparable with $T_K$ (to be defined below). Some of
these states may be quasi degenerate, with energy differences
$|\bar{E}_\Lambda- \bar{E}_{\Lambda'}|< T_K$. However, $T_K$
itself is a function of these energy distances (see, e.g.,
\cite{Magn,Pust,Eto02}), and all the levels, which influence
$T_K$, should be retained in (\ref{1.12b}).

The effective cotunneling Hamiltonian acquires the form
\begin{equation}\label{1.11}
H_{cot}= \sum_{\alpha\alpha'}\left( J_0^{\alpha\alpha'}{\bf
S}\cdot {\bf s}^{\alpha\alpha'} + \sum_n J_n^{\alpha\alpha'}{\bf
R}_n\cdot {\bf s}^{\alpha\alpha'}\right).
\end{equation}
Here ${\bf S}$ is the spin operator of CQD in its ground state,
the operators ${\bf s}^{\alpha\alpha'}$ represent the spin states
of lead electrons,
\begin{equation}\label{1.10}
{\bf s}^{\alpha\alpha'}=\frac{1}{2}\sum_{kk'}\sum_{\sigma \sigma'}
c^\dag_{\alpha k\sigma} \hat{\tau}_{\sigma \sigma'} c_{\alpha' k'
\sigma'}~,
\end{equation}
where $\hat{\tau}$ is the vector of Pauli matrices. In the
conventional Kondo effect the logarithmic divergent processes
develop due to spin reversals given by the first term containing
the operator ${\bf S}$. In CQD possessing dynamical symmetry, all
R-vectors are involved in Kondo tunneling. In the following
chapters we will show how these additional processes are
manifested in resonance Kondo tunneling through CQD. Note that the
elements of the matrix $\textsf{J}$ are also subject to
temperature dependent renormalization $J_{n}^{\alpha \alpha' } \to
J_{n}^{\alpha \alpha' }(D_0/T)$.

The cotunneling Hamiltonian (\ref{1.11}) is the natural
generalization of the conventional Kondo Hamiltonian $J {\bf S}
\cdot {\bf s}$ for CQDs possessing dynamical symmetries.  In many
cases there are several dot spin 1 operators depending on which
pair of electrons is ``active''. In this pair, one electron sits
in well $c$ and the other one sits in some well $a$. The other
${\cal N}-2$ electrons are paired in singlet states. This scenario
applies if $\cal N$ is even. The spin 1 operator for the active
pair is denoted as ${\bf S}_{a}$. (In some sense, the need to
specify which pair couples to $S=1$ while all other pairs are
coupled to $S=0$ is the analog of the seniority scheme in atomic
and nuclear physics (see, e.g., \cite{Talmi})). The cotunneling
Hamiltonian for CQD contains exchange terms
$J_0^{\alpha\alpha'}{\bf S}_{a}\cdot {\bf s}^{\alpha\alpha'}$.
Then, instead of a single exchange term (first term on the RHS of
Eq. (\ref{1.11})), one has a sum $\sum_a J_a^{\alpha \alpha'}{\bf
S}_a\cdot {\bf s}^{\alpha \alpha'}$. Additional symmetry elements
(finite rotations and reflections) turn the cotunneling
Hamiltonian even more complicated. In the following chapters we
will consider several examples of such CQDs. It is seen from
(\ref{1.11}), that in the generic case, both spin and R-vectors
may be the sources of anomalous Kondo resonances. The contribution
of these vectors depends on the hierarchy of the energy states in
the manifold. In principle, it may happen that the main
contribution to the Kondo tunneling is given not by the spin of
the dot, but by one of the R-vectors.

Thus, we arrive at the conclusion that the regular procedure of
reducing the full Hamiltonian of a quantum dot in junctions with
metallic leads to an effective Hamiltonian describing only spin
degrees of freedom of this system reveals a rich dynamical
symmetry of CQD. Strictly speaking, only an isolated QD with
${\cal N}=1$ is fully described by its spin 1/2 operator obeying
$SU(2)$ symmetry without dynamical degrees of freedom. Yet even
the doubly occupied dot with ${\cal N}=2$ possesses the dynamical
symmetry of a {\it spin rotator} because its spin spectrum
consists of a singlet ground state (S) and a triplet excitation
(T). Therefore, an R-vector describing S/T transitions may be
introduced, and the Kondo tunneling through a dot of this kind may
involve spin excitation under definite physical conditions, e.g.,
in an external magnetic field \cite{Magn}. A two-electron quantum
dot under Coulomb blockade constitutes apparently the simplest
non-trivial example of a nano-object with dynamical symmetry of a
spin rotator possessing an $SO(4)$ symmetry.

Dynamical symmetries $SO(n)$ of CQDs are described by non-compact
semi-simple algebras \cite{Cahn}. This non-compactness implies
that the corresponding algebra $o_n$ may be presented as a direct
sum of subalgebras, e.g., $o_4 = o_3 \oplus o_3$. Therefore, the
dynamical symmetry group may be represented as a direct product of
two groups of lower rank. In case of spin rotator the product is
$SO(4)=SU(2)\otimes SU(2)$. Generators of these subgroups may be
constructed from those of the original group. The $SO(4)$ group
possesses a single R-operator ${\bf R}$, and the direct product is
realized by means of the transformation
\begin{equation}\label{1.14}
{\bf K}= \frac{{\bf S}+\bf R}{2},~~~{\bf N}= \frac{{\bf S}-\bf
R}{2}.
\end{equation}
 Both vectors ${\bf K}$ and
${\bf N}$ generate $SU(2)$ symmetry and may be treated as
fictitious S=1/2 spins \cite{Pust}. In some situations these
vectors are real spins localized in different valleys of CQD. In
particular, the transformation (\ref{1.14}) maps a single site
Kondo problem for a DQD possessing $SO(4)$ symmetry to a two-site
Kondo problem for spin 1/2 centers with an $SU(2)$ symmetry (see
discussion in Refs. \cite{KA01,KA02}). For groups of higher
dimensionality ($n\geq 4)$ one can use many different ways of
factorization, which may be represented by means of different
Young tableaux (see Appendix D).

Even in the case $n=4,$ the transformation (\ref{1.14}) is not the
only possible "two-spin" representation. An alternative
representation is realized in an external magnetic field
\cite{KA02}. When the ground state of S/T manifold is a singlet
(the energy $\delta=E_T-E_S>0)$, the Zeeman splitting energy of a
triplet in an external magnetic field may exactly compensate the
exchange splitting $\delta$. This accidental degeneracy is
described by the pseudospin 1/2 formed by the singlet and the up
projection of spin 1 triplet. Two other projections of the triplet
form the second pseudospin 1/2. The Kondo effect induced by
external magnetic field observed in several nano-objects
\cite{Magn1,Magn-v}, was the first experimental manifestation of
dynamical symmetry in quantum dots.

In conclusion, we outlined in this section the novel features
which appear in effective Kondo Hamiltonians due to the dynamical
symmetry of CQD exhibiting Kondo tunneling. In the following
chapters we will see how the additional terms in the Hamiltonian
(\ref{1.11}) influence the properties of Kondo resonance in
various structures of CQDs.

\chapter{Trimer in Parallel Geometry} \label{III}
{\it This Chapter is devoted to a systematic exposure of the Kondo
physics in evenly occupied artificial trimer, i.e., triple quantum
dot (TQD) in parallel geometry. We are interested in answering
questions pertaining to the nature of the underlying symmetry of
the trimer Hamiltonian and the algebra of operators appearing in
the exchange Hamiltonian. The energy spectrum of the isolated TQD
is discussed in Section \ref{sec.3.1}. In Section \ref{sec.3.2},
the renormalization group equations are derived and various cases
of accidental degeneracy arising due to dot-lead interaction are
discussed. It is shown that the TQD manifests $SO(n)$ dynamical
symmetry in Kondo tunneling regime. The effective spin
Hamiltonians are written down and the corresponding $o_n$ algebras
are constracted for the $P\times SO(4)\times SO(4)$, $SO(5)$ and
$SO(7)$ dynamical symmetries in Subsections \ref{subsec.3.2.1},
\ref{subsec.3.2.2} and \ref{subsec.3.2.3}, respectively. The
scaling equations are derived and the Kondo temperatures are
calculated for the cases of $P\times SO(4)\times SO(4)$ and
$SO(5)$ symmetries. The results are summarized in the
Conclusions.}

\section{Energy Spectrum}\label{sec.3.1}
\begin{figure}[htb]
\centering
\includegraphics[width=80mm,height=80mm,angle=0]{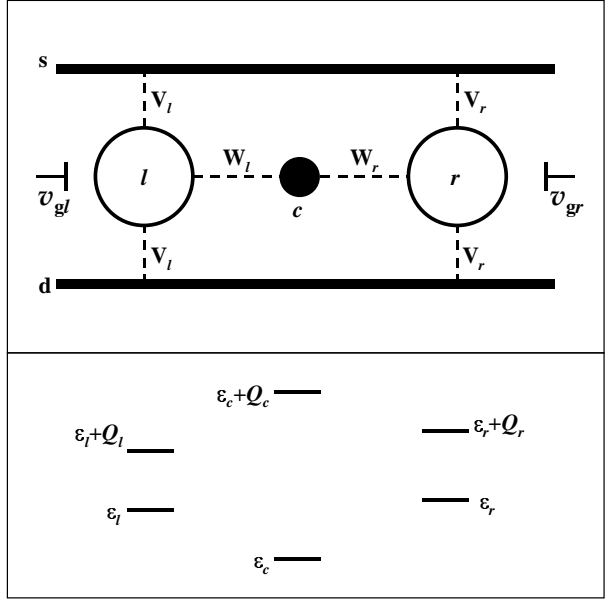}
\caption{Triple quantum dot in parallel geometry and energy levels
of each dot $\varepsilon_{a}=\epsilon_{a}-v_{ga}$ \label{TQD-par}
(bare energy minus gate voltage).}
\end{figure}

Double quantum dot with occupation ${\cal N}=2$ discussed in Sec.
\ref{sec.2.3} is the analog of a hydrogen molecule in the
Heitler-London limit \cite{KA01,KA02}, and its $SO(4)$ symmetry
reflects the spin properties of ortho/parahydrogen. A much richer
artificial object is a triple quantum dot (TQD), which can be
considered as an analog of a {\it linear molecule} RH$_2$. The
central $(c)$ dot is assumed to have a smaller radius (and, hence,
larger capacitive energy $Q_c$) than the left $(l)$  and right
$(r)$ dots, i.e., $Q_c \gg Q_{l,r}$. Fig. \ref{TQD-par}
illustrates this configuration in a parallel geometry, where the
"left-right" ($l-r$) reflection plane of the TQD is perpendicular
to the "source-drain" ($s-d$) reflection plane of metallic
electrodes.

To regulate the occupation of TQD as a whole and its constituents
in particular, there is a couple of gates $v_{gl},v_{gr}$ applied
to the $l,r$ dots. The energy levels of single- and two-electron
states in each one of the three constituent dots are shown in the
lower panel of Fig. \ref{TQD-par}. Here the gate voltages
$v_{gl,r}$ are applied in such a way that the one-electron level
$\epsilon_c$ of a $c$-dot is essentially deeper than those of the
$l,r$-dots, so that the condition (\ref{1.8}) is satisfied for the
$c$ dot, whereas the inequalities (\ref{1.7}) and (\ref{1.9}) are
satisfied for the "active" $l$ and $r$ dots. Tunneling between the
side dots $l,r$ and the central dot $c$ with amplitudes  $W_{l,r}$
determines the low energy spin spectrum of the isolated TQD once
its occupation $\cal N$ is given. This system enables the exposure
of much richer possibilities for additional degeneracy relative to
the DQD setup mentioned above due to the presence of two channels
$(l,r)$.

The full diagonalization procedure of the Hamiltonian $H_{dot}$
for the TQD is presented in Appendix \ref{diag}. When the
condition (\ref{1.9}) is valid, the low-energy manifold for ${\cal
N}=4$ is composed of two singlets $|S_l\rangle,|S_r\rangle$, two
triplets $|T_a\rangle=|\mu_a\rangle$ ($a=l,r,
\mu_a=1_a,0_a,\bar{1}_a$) and a charge transfer singlet exciton
$|Ex\rangle$ with an electron removed from the $c$-well to the
"outer" wells. Within first order in $\beta_a\ll1$ the
corresponding energies are,
\begin{eqnarray}
E_{S_{a}} &=& {\epsilon}_c +{\epsilon}_{a}+2{\epsilon}_{\bar a}
+Q_{\bar a} -2W_{a}\beta_{a},\nonumber\\
E_{{T_a}} &=& {\epsilon}_c +{\epsilon}_{a}+2{\epsilon}_{\bar
a}+Q_{\bar a} ,
\label{En} \\
E_{Ex} &=& 2{\epsilon}_{l} +2{\epsilon}_{r}+Q_l+Q_r
+2W_{l}\beta_{l}+ 2W_{r}\beta_{r}, \nonumber
\end{eqnarray}
where the charge transfer energies in Eq.(\ref{1.9}) (for
determining $\beta_a$) are $E_{ac}=Q_a+\epsilon_a-\epsilon_c$; the
notation $a=l,r$ and ${\bar a}=r,l$ is used ubiquitously
hereafter.

The completely symmetric configuration,
$\varepsilon_l=\varepsilon_{r}\equiv \varepsilon,~ Q_l=Q_{r}\equiv
Q ,~ W_l=W_{r}\equiv W,$ should be considered separately. In this
case the singlet states form even and odd combinations in close
analogy with the molecular states $\Sigma^\pm$ in axisymmetric
molecules. The odd state $S_-$ and two triplet states are
degenerate:
\begin{eqnarray}
&&E_{S^+}=\varepsilon_c+3\varepsilon+Q -4W\beta ,\nonumber
\\
 &&E_{S^-}=E_{T_{a}}=\varepsilon_c+3\varepsilon+Q, \label{degen}
 \\
&& E_{Ex} = 4{\epsilon} +2Q +4W\beta. \nonumber
\end{eqnarray}
Consideration of these two examples provide us with an opportunity
to investigate the dynamical symmetry of CQD.

\section{Derivation and Solution of Scaling Equations}\label{sec.3.2}
We consider the case of TQD with even occupation ${\cal N}=4$
discussed in Refs. \cite{KKA02,KKA04}. This configuration is a
direct generalization of an asymmetric spin rotator, i.e., the
double quantum dot in a side-bound geometry \cite{KA01}. Compared
with the asymmetric DQD, this composite dot possesses one more
symmetry element, i.e., the $l-r$ permutation, which, as will be
seen below, enriches the dynamical properties of CQD.

Following a glance at the energy level scheme (\ref{En}), one is
tempted to conclude outright that for finite $W,$ the ground state
of this TQD configuration is a singlet and consequently there is
no room for the Kondo effect to take place. A more attentive study
of the tunneling problem, however, shows that tunneling between
the TQD and the leads opens the way for a rich Kondo physics
accompanied by numerous dynamical symmetries.

Indeed, inspecting the expressions for the energy levels, one
notices that the singlet states $E_{S_a}$ are modified due to
inter-well tunneling, whereas the triplet states $E_{T_a}$ are
left intact. This difference is due to the admixture of the
singlet states with the charge transfer singlet exciton (see
Appendix \ref{diag}). As was mentioned in the previous chapter,
the Kondo cotunneling in the perturbative weak coupling regime at
$T,\varepsilon > T_K$ is excellently described within RG formalism
\cite{Hald,Anderson}. According to general prescriptions of this
theory, the renormalizable parameters of the effective low-energy
Hamiltonian in a one-loop approximation are the energy levels
$E_\Lambda$ and the effective indirect exchange vertices
$J_{\Lambda\Lambda'}^{\alpha\alpha'}$.

\subsection{$ {P\times SO(4)\times SO(4)}$ Symmetry}\label{subsec.3.2.1}
To apply the RG procedure to the Kondo tunneling through TQD, let
us first specify the terms $H_{lead}$ and $H_{tun}$ in the
Anderson Hamiltonian (\ref{1.3}). The most interesting for us are
situations where the accidental degeneracy of spin states is
realized. So we consider geometries where the device {\it as a
whole} possesses either complete or slightly violated $l-r$ axial
symmetry. Then the quantum number $\alpha$ in $H_{lead}$
(\ref{1.5}) contains the lead index ($s,d$) and the channel index
($l,r$). The two tunneling channels are not independent because of
weak interchannel hybridization in the leads. This hybridization
is characterized by a constant $t_{lr}\ll D_{0},$ which is small
first due to the angular symmetry, and second due to significant
spatial separation between the two channels. The wave vector $k$
is assumed to remain a good quantum number. Then, having in mind
that in our model $\epsilon_{kas}=\epsilon_{kad}\equiv
\epsilon_{ka}$, the generalized Hamiltonian (\ref{1.5}) acquires
the form
\begin{equation}\label{tlead}
H_{lead}=\sum_{ k
\sigma}\sum_{b=s,d}\sum_{a=l,r}\left(\epsilon_{ka} n_{abk\sigma} +
t_{lr} c^{\dag}_{abk\sigma}c_{\bar{a}bk\sigma}\right).
\end{equation}
The tunneling Hamiltonian (\ref{1.6}) is written as
\begin{equation}\label{hyb}
H_{tun}=\sum_{\Lambda\lambda}\sum_{k\sigma}\sum_{ab}(V^{\lambda\Lambda}_{ab\sigma}
c^{\dag}_{abk\sigma} X^{\lambda\Lambda}+H.c.).
\end{equation}
We assume below $V_{as}=V_{ad}\equiv V_a$ (see Fig
.\ref{TQD-par}).

\begin{figure}[htb]
\centering
\includegraphics[width=70mm,height=95mm,angle=0,]{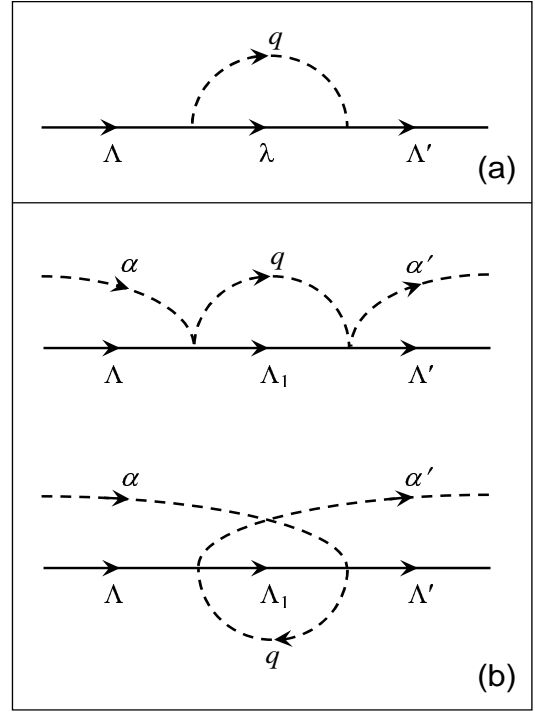}
\caption{RG diagrams for the energy levels $E_\Lambda$ $(a)$ and
the effective exchange vertices
$J_{\Lambda\Lambda'}^{\alpha\alpha'}$ $(b)$ (see text for further
explanations).} \label{RGfig}
\end{figure}
The iteration processes, which  characterize the two-step RG
procedure contributing to these parameters are illustrated in Fig.
\ref{RGfig}. The intermediate states in these diagrams are the
high-energy states $|q\rangle$ near the ultraviolet cut-off energy
$D$ of the band continuum in the leads (dashed lines) and the
states $|\lambda\rangle \in {\cal N}-1$ from adjacent charge
sectors, which are admixed with the low-energy states
$|\Lambda\rangle \in {\cal N}$ by the tunneling Hamiltonian $H_t$
(\ref{1.6}) (full lines). For the sake of simplicity we confine
ourselves with three-electron states in the charge sector.

In the upper panel, the diagrams contributing to the
renormalization of $H_{dot}$ are shown. In comparison with the
original theory \cite{Hald}, this procedure not only results in
renormalization of the energy levels but also an additional
hybridization of the states $|\Lambda_a\rangle$ via channel mixing
terms in the Hamiltonian (\ref{hyb}). Due to the condition
(\ref{1.8}), the central dot $c$ remains "passive" throughout the
RG procedure.

The mathematical realization of the diagrams displayed in Fig.
3.2a is encoded in the scaling equations for the energy levels
$E_{\Lambda}$,
\begin{equation}
\frac{\pi dE_\Lambda}{dD}=\sum_\lambda
\frac{\Gamma_{\Lambda}}{D-E_{\Lambda\lambda}}. \label{DE}
\end{equation}
Here $E_{\Lambda\lambda}=E_\Lambda-E_\lambda$, $\Gamma_\Lambda$
are the tunnel coupling constants which are different for
different $\Lambda$,
\begin{equation}
\Gamma_{T_a}=\pi \rho_0(V_a^2+2 V_{\bar a}^2),\ \ \ \ \
\Gamma_{S_a}=\alpha_a^2 \Gamma_{T_a}.\label{gam-4}
\end{equation}
Here $\alpha_a=\sqrt{1-2\beta_a^2}$, and $\rho_0$ is the density
of electron states in the leads, which is supposed to be energy
independent. These scaling equations should be solved at some
initial conditions
\begin{equation}\label{init}
E_{\Lambda}(D_{0})=E_{\Lambda}^{(0)},
\end{equation}
where the index $(0)$ marks the bare values of the model
parameters entering the Hamiltonian $H_A$ (\ref{1.3}).

Besides, the diagram of Fig. 3.2a generates a new vertex
$M_{lr}^{\Lambda\Lambda'}$, where the states $\Lambda,\Lambda'$
are either two singlets $S_l,S_{r}$ or two triplets $T_l,T_{r}$.
The third order Haldane iteration procedure results in a scaling
equation,
\begin{equation}
\frac{d M_{lr}}{dD} = - \frac{\gamma}{D^2} \label{sclr}
\end{equation}
with an initial condition $M_{lr}(D_0)=0$ and a flow rate
$\gamma=\rho_0 V_l V_r t_{lr}.$ After performing the Haldane
procedure we formally come to the scaled dot Hamiltonian
\begin{equation}\label{scah}
H_{dot}=\sum_{\Lambda_a}{E}_{\Lambda_a}
X^{\Lambda_a\Lambda_a}+\sum_{\Lambda_{a} \Lambda_{\bar a}}M_{lr}
X^{\Lambda_a \Lambda_{\bar a}}
\end{equation}
with the parameters ${E}_{\Lambda_a}$ and $M_{lr}$ depending on
the running variable $D$.

 Due to the above mentioned dependence of tunneling rates on
the index $\Lambda$, namely the possibility of $\Gamma_T>\Gamma_S$
and $\Gamma_{S_-}>\Gamma_{S_+}$, the scaling trajectories
$E_\Lambda (D)$ may cross at some value of the monotonically
decreasing energy parameter $D$. The nature of level crossing is
predetermined by the initial conditions (\ref{init}) and the
ratios between the tunneling rates $\Gamma_\Lambda$. As long as
the inequality $|E_{\Lambda\lambda}|\ll D$ is effective and all
levels are non-degenerate, the scaling equations (\ref{DE}) may be
approximated by Eqs. (\ref{DE-2}).
The scaling trajectories are determined by the scaling invariants
(\ref{inv-2}) for equations (\ref{DE}),
tuned to satisfy the initial conditions. With decreasing energy
scale $D$ these trajectories flatten and become $D$-independent in
the so called Schrieffer-Wolff (SW) limit, which is reached when
the excitation energies $\Delta_a$ become comparable with $D$. The
corresponding effective bandwidth is denoted as $\bar{D}$ (we
suppose, for the sake of simplicity, that $\Delta_a< Q_a$, so that
only the states $|\lambda\rangle$ with ${\cal N}'={\cal N}-1$ are
relevant). The simultaneous evolution of interchannel
hybridization parameter is described by the solution of scaling
equation (\ref{sclr}),
\begin{equation}\label{log}
M_{lr}(\bar{D})=\gamma \left(\frac{1}{{\bar
D}}-\frac{1}{D_0}\right).
\end{equation}

If this remarkable level crossing occurs at $D>\bar{D}$, we arrive
at the situation where {\it adding an indirect exchange
interaction between the TQD and the leads changes the magnetic
state of the TQD from singlet to triplet}. Those states
$E_\Lambda$, which remain close enough to the new ground state are
involved in Kondo tunneling. As a result, the TQD acquires a rich
dynamical symmetry structure instead of the trivial symmetry of
spin singlet predetermined by the initial energy level scheme
(\ref{En}). Appearance of the enhancement of the hybridization
parameter $M_{lr}$ (\ref{log}) does not radically influence the
general picture, provided the flow trajectories cross far from the
SW line , due to a very small hybridization $\gamma\ll
\Gamma_\Lambda\ll D$. However, we are interested just in cases
when the accidental degeneracy occurs at the SW line. Various
possibilities of this degeneracy are considered below.

\begin{figure}[htb]
\centering
\includegraphics[width=80mm,height=60mm,angle=0,]{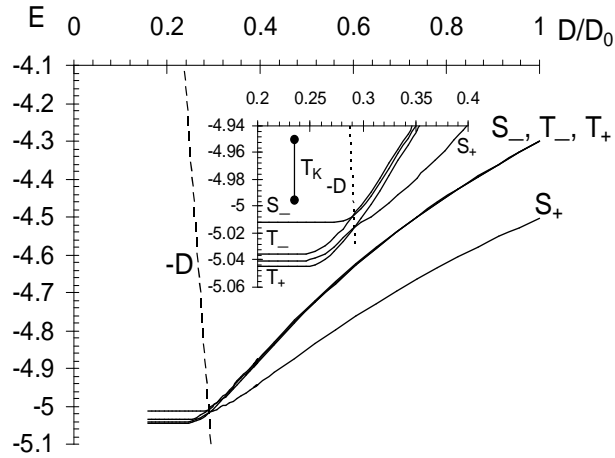}
\caption{Scaling trajectories for $P\times SO(4)\times SO(4)$
symmetry in the SW regime. Inset: Zoomed in avoided level crossing
pattern near the SW line.} \label{pdotso4}
\end{figure}

The flow diagrams leading to a non-trivial dynamical symmetry of
TQD with ${\cal N}=4$ are presented in Figs. \ref{pdotso4},
\ref{SO5}, \ref{SO7}. The horizontal axis on these diagrams
corresponds to the dimensionless energy scale $D/D_0$ for lead
electrons, where the vertical axes represent the energy levels
$E_\Lambda(D)$. The dashed line $E=-D$ establishes the SW boundary
for these levels.

Before turning to highly degenerate situations, where the system
possesses specific $SO(n)$ symmetry, it is instructive to consider
the general case, where all flow trajectories $E_\Lambda(\bar D)$
are involved in Kondo tunneling in the SW limit. This happens when
the whole octet of spin singlets and triplets forming the manifold
(\ref{En}) remains within the energy interval $\sim T_K$ in the SW
limit. The level repulsion effect does not prevent the formation
of such multiplet, provided $t_{lr}$ is small enough and the
inequality
\begin{equation}\label{ineq1}
M_{lr}(\bar D)<T_K
\end{equation}
is valid. At this stage, the SW procedure for constructing the
effective spin Hamiltonian in the subspace $\mathbb{R}_8=\{T_l,
S_l, T_r, S_r\}$ should be applied. This procedure excludes the
charged states generated by $H_t$ to second order in perturbation
theory (see, e.g., \cite{Hewson}).

The effective cotunneling Hamiltonian can be derived using
Schrieffer-Wolf procedure \cite{SW} (see Appendix \ref{H-spin}).
To simplify the SW transformation, one should first rationalize
the tunneling matrix ${\sf V}$ in the Hamiltonian (\ref{hyb}).
This $4\times 4$ matrix is diagonalized in the $s-d, l-r$ space by
means of the transformation to even/odd combinations of lead
electron $k$-states and similar symmetric/antisymmetric
combinations of $l,r$ electrons in the dots. The form of this
transformation for symmetric TQD can be found in Appendix
\ref{GR-trans}. Like in the case of conventional QD \cite{Glaz},
this transformation eliminates the odd combination of $s-d$
electron wave functions from tunneling Hamiltonian.

It should be emphasized that this transformation does  not exclude
the odd component from $H_{tun}$ in case of TQD in a series
geometry \cite{EPL}. The same is valid for the Hamiltonians
(\ref{tlead}), (\ref{hyb}) with $t_{lr}=0$: in this case the
rotation in $s-d$ space conformally maps the Hamiltonian $H_A$
(\ref{1.3}) for TQD in parallel geometry onto that for TQD in
series. Both these cases will be considered in Chapter \ref{IV}.

 Unlike the case of DQD studied in Refs.
\cite{KA01,KA02}, where the spin operators are the total spin
${\bf S}$ and a single R-operator, describing S/T transitions, the
TQD is represented by several spin operators corresponding to
different Young tableaux (see Appendix D). To order $O(|V|^2)$,
then,
\begin{eqnarray}
H&=&\sum_{\Lambda_a}\bar {E}_{\Lambda_a}X^{{\Lambda_a}{\Lambda_a}}
+\sum_{\Lambda_a \Lambda_{\bar a}}\bar
{M}_{{lr}}X^{{\Lambda_a}{\Lambda_{\bar a}}}
 +\sum_{ k
\sigma}\sum_{b=s,d}\sum_{a=l,r}\left(\epsilon_{ka} n_{abk\sigma} +
t_{lr} c^{+}_{abk\sigma}c_{\bar{a}bk\sigma}\right)\nonumber\\
&+&
 \sum_{a=l,r}J^T_{a} {\bf S}_{a}\cdot {\bf s}_a+ J_{lr}{\hat P}
\sum_{a=l,r}{\bf S}_{a}\cdot {\bf s}_{{\bar a}a}
+\sum_{a=l,r}J^{ST}_a {\bf R}_{a}\cdot {\bf s}_a
+J_{lr}\sum_{a=l,r}\tilde{\bf R}_{a}\cdot {\bf s}_{a{\bar
a}}\label{generic}.
\end{eqnarray}
Here we recall that $\bar {E}_{\Lambda_a}=E_{\Lambda_a}(\bar{D})$,
$\bar {M}_{lr}=M_{lr}(\bar{D})$, and the effective exchange
constants are
\begin{eqnarray}
J^T_{a}&=&\frac{V_a^2}{\epsilon_F-\epsilon_a},\ \ \ \ \ \
J^{ST}_a=\alpha_aJ^T_a,\ \ \ \ \
J_{lr}=\frac{V_lV_r}{2}\left(\frac{1}{\epsilon_F-\epsilon_l}+\frac{1}
{\epsilon_F-\epsilon_r}\right).\label{J}
\end{eqnarray}
The vector operators ${\bf S}_a,{\bf R}_a,{\bf \tilde{R}}_a$ and
the permutation operator $\hat{P}$ manifest the dynamical symmetry
of TQD in a subspace $\mathbb{R}_8$. The permutation operator
\begin{equation}\label{perm}
{\hat P}=\sum_{a=l,r}\Big(X^{S_a S_{\bar a}}+\sum_{\mu=1,0,{\bar
1} } X^{\mu_a \mu_{\bar a}}\Big)
\end{equation}
commutes with ${\bf S}_l +{\bf S}_r$ and ${\bf R}_l +{\bf R}_r$.

The spherical components of these vectors are defined via Hubbard
operators connecting different states of the octet,
\begin{eqnarray}
S^{+}_a &=& \sqrt{2}( X^{1_a0_a}+X^{0_a\bar{1}_a}),\ \ \ \ \ \ \
S^{-}_a = (S^{+}_a)^\dagger, \ \ \
S^z_{a}=X^{1_a1_a}-X^{\bar{1}_a\bar{1}_a},\nonumber\\
R^{+}_a &=& \sqrt{2}( X^{1_aS_a}-X^{S_a\bar{1}_a}),\ \ \ \ \ \ \;
R^{-}_a = (R^{+}_a)^\dagger,\ \;
R^z_{a} =-(X^{0_aS_a}+X^{S_a0_a}),  \label{comm1} \\
\tilde{R}^{+}_{a} &=& \sqrt{2}(\alpha_{\bar a} X^{1_aS_{\bar
a}}-\alpha_{a} X^{S_a\bar{1}_{\bar a}}),\  \tilde{R}^{-}_{a} =
(\tilde{R}^{+}_{a})^\dagger,\   \tilde{R}^z_{a} = -(\alpha_{\bar
a} X^{0_aS_{\bar a}}+\alpha_{a} X^{S_a0_{\bar a}}).\nonumber
\end{eqnarray}
In addition to the spin operator (\ref{1.10}) for conduction
electrons, new spin operators are required,
\begin{equation}\label{1.100}
{\bf s}_{a \bar a}= \frac{1}{2}\sum_{kk'}\sum_{\sigma \sigma'}
c^\dag_{ak\sigma} \hat{\tau}_{\sigma \sigma'}c_{{\bar a}k'
\sigma'}.
\end{equation}
An extra symmetry element (l-r permutation) results in more
complicated algebra which involves new R-operator ${\bf
\tilde{R}}$ and the permutation operator $\hat{P}$ interchanging
$l$ and $r$ components of TQD.

One can derive from the generic Hamiltonian (\ref{generic}) more
symmetric effective Hamiltonians describing partly degenerate
configurations illustrated by the flow diagrams of Figs.
\ref{pdotso4}, \ref{SO5}, \ref{SO7}. These are the cases when the
level crossing occurs in the nearest vicinity of the SW line in
the flow diagram. It is important to distinguish between the cases
of generic and accidental symmetry. In the former case the device
possesses intrinsic $l-r$ and $s-d$ symmetry, i.e., the left and
right dots are identical, the corresponding tunnel parameters are
equal, and left and right leads also mirror each other, namely,
$\epsilon_{kl}=\epsilon_{kr}\equiv\epsilon_{k}$. In the latter
case the gate voltages violate $l-r$ symmetry, e.g., they make
$\epsilon_{l}\neq \epsilon_{r}$, $V_{l}\neq V_{r}$, etc. The level
degeneracy is achieved due to competition between the $l-r$
interdot tunneling and the lead-dot tunneling without changing the
symmetry of the Hamiltonian.

The basic spin Hamiltonian (\ref{generic}) acquires a more compact
form, when a TQD possesses generic or accidental degeneracy. In
these cases the operators (\ref{comm1}) form close algebras, which
predetermine the dynamical symmetry of Kondo tunneling.  We start
the discussion of the pertinent $SO(n)$ symmetries with the most
degenerate configuration (Fig. \ref{pdotso4}), where the TQD
possesses generic $l-r$ axial symmetry, i.e., the left and right
dots are completely equivalent. Then the energy spectrum of an
isolated TQD is given by Eqs.(\ref{degen}). The four-electron wave
functions are calculated in Appendix (\ref{diag}). Such TQD is a
straightforward generalization of the so called T-shaped DQD
introduced in Refs. \cite{KA01,KA02,Tshap,Tshap1,Tshap2}. It is
clear, that attachment of a third dot simply adds one more element
to the symmetry group $SO(4)$, namely the $l-r$ permutation $\hat
P$, which is parity sensitive.

To reduce the Hamiltonian (\ref{generic}) into a more symmetric
form, we rewrite the Hubbard operators in terms of new eigenstates
$\bar E_\Lambda$, recalculated with taking account of the generic
degeneracy (\ref{degen}) and $l-r$ mixing $\bar M_{lr}$. In
assuming that the latter coupling parameter is the smallest one,
it results in insignificant additional remormalization  $\sim
\mp|\bar M_{lr}|^2/(\varepsilon +Q-\varepsilon_c)$ of the states
$E_{S+}$ and $E_{Ex}$. Besides, it intermixes the triplet states
and changes their nomenclature from left/right to even/odd. The
corresponding energy levels are
\begin{equation}\label{lre}
E_{T\pm}(\bar D)=E_{T_a}\mp \bar M_{lr}.
\end{equation}

The flow trajectories for two pairs of states $(T_+, T_-)$ and
$(S_+, S_-)$ diverge slowly with decreasing $D$. If this
divergence is negligible in the scale of $T_K$, then three nearly
coincident trajectories $E_{T\pm},E_{S_-}$ cross the fourth
trajectory $E_{S+}$ at some point, since the inequality
$\Gamma_{S_+}<\Gamma_{T\pm}=\Gamma_{S-}$ with $\Gamma_{T\pm}=3\pi
\rho_o V^2,\ \ \Gamma_{S_+}=\alpha\Gamma_{T\pm}$ is valid
($\alpha=\sqrt{1-4\beta^2}<1$). If this level crossing happens
near the SW line, we arrive at a case of complete degeneracy of
the renormalized spectrum, and the whole octet $\mathbb{R}_8$ is
involved in the dynamical symmetry (Fig. \ref{pdotso4}). The fine
structure of the flow diagram in the region of avoided level
crossing is shown in the inset.

Since the tunneling occurs in even and odd channels independently,
the parity is conserved also in indirect SW exchange. As a result,
the effective spin Hamiltonian (\ref{generic}) acquires the form
\begin{eqnarray}
{H} &=&\sum_{\Lambda_{\eta}}{\bar
E}_{\Lambda_\eta}X^{\Lambda_\eta\Lambda_\eta}
+\sum_{k\sigma}\sum_{\eta=g,u}\epsilon_{k\eta} c^{\dagger}_{\eta
k\sigma }c_{\eta k \sigma} + \sum_{\eta=g,u}J^T_{1\eta} {\bf
S}_{\eta}\cdot {\bf s}_\eta+ \sum_{\eta=g,u}J^{ST}_{1\eta} {\bf
R}_{\eta}\cdot {\bf s}_\eta \nonumber\\
&+&J^T_{2} \sum_{\eta =g,u}{\bf S}_{\eta{\bar \eta}}\cdot {\bf
s}_{\eta{\bar \eta}} +\sum_{\eta=g,u}(J^{ST}_{2\eta}{\bf
R}^{(1)}_{\eta{\bar \eta}}+J^{ST}_{2{\bar\eta}}{\bf
R}^{(2)}_{\eta{\bar \eta}})\cdot {\bf s}_{\eta{\bar
\eta}}.\label{sym-h1-mlr}
\end{eqnarray}
Here $\epsilon_{kg}=\epsilon_{k}-t_{lr}$,
$\epsilon_{ku}=\epsilon_{k}+t_{lr}$ and the lead operators
$c_{\eta k\sigma}$ $(\eta=g,u)$ are defined in Appendix
\ref{GR-trans}. The operators ${\bf S}_\eta$, ${\bf R}_\eta$ are
defined analogously to ${\bf S}_a$, ${\bf R}_a$ in
Eq.(\ref{comm1}), and the vector operators ${\bf S}_{\eta{\bar
\eta}},$ ${\bf R}^{(1)}_{\eta{\bar \eta}}$, ${\bf
R}^{(2)}_{\eta{\bar \eta}}$ are defined as:
\begin{eqnarray}
{\bf S}_{\eta{\bar \eta}}&=&X^{\eta{\bar \eta}}{\bf S}_{{\bar
\eta}},\ \ \ \ \ {\bf R}^{(1)}_{\eta{\bar \eta}}+{\bf
R}^{(2)}_{\eta{\bar \eta}}=X^{\eta{\bar \eta}}{\bf R}_{{\bar
\eta}}.\label{eta2}
\end{eqnarray}
The spherical components of the operators ${\bf
R}^{(1)}_{\eta{\bar \eta}}$ and ${\bf R}^{(2)}_{\eta{\bar \eta}}$
are given by
\begin{eqnarray}
R^{(1)+}_{\eta{\bar \eta}}&=&-\sqrt{2}X^{S_{\eta}{\bar 1}_{\bar
\eta}}, \ \ \ \ {R}^{(1)-}_{\eta{\bar \eta}}=(R^{(1)+}_{\eta{\bar
\eta}})^{\dag},\ \ \ \ \ \ R^{(1)z}_{\eta{\bar\eta}}=-X^{S_{\eta}0_{\bar \eta}},\nonumber\\
R^{(2)+}_{\eta{\bar \eta}}&=&\sqrt{2}X^{1_{\eta}{S}_{\bar \eta}},
\ \ \ \ \ \; {R}^{(2)-}_{\eta{\bar \eta}}=(R^{(2)+}_{\eta{\bar
\eta}})^{\dag},\ \ \ \ \ \ R^{(2)z}_{\eta{\bar
\eta}}=-X^{0_{\eta}S_{\bar \eta}}.\label{R-eta12}
\end{eqnarray}

The spin operators for the electrons in the leads are introduced
by the obvious relations
\begin{eqnarray}
&&{\bf s}_{g}=\frac{1}{2}\sum_{kk'}\sum_{\sigma \sigma'}
c^\dag_{gk\sigma} \hat{\tau}_{\sigma \sigma'} c_{gk'\sigma'},\ \ \
\ \ {\bf s}_{u}=\frac{1}{2}\sum_{kk'}\sum_{\sigma \sigma'}
c^\dag_{uk\sigma} \hat{\tau}_{\sigma \sigma'} c_{uk'\sigma'},\nonumber\\
&&{\bf s}_{gu}=\frac{1}{2}\sum_{kk'}\sum_{\sigma \sigma'}
c^\dag_{gk\sigma} \hat{\tau}_{\sigma \sigma'} c_{uk'\sigma'},\ \ \
\ {\bf s}_{ug}=({\bf s}_{gu})^{\dag},\label{gu-operators}
\end{eqnarray}
instead of (\ref{1.10}). Now the operator algebra is given by the
closed system of commutation relations which is a generalization
of the $o_4$ algebra,
\begin{eqnarray}\label{algebra}
\lbrack S_{\eta j},S_{\eta' k}]=ie_{jkm}\delta_{\eta\eta'}S_{\eta
m},\   [R_{\eta j},R_{\eta' k}]=ie_{jkm}\delta_{\eta\eta'}¥S_{\eta
m},\   \lbrack R_{\eta j},S_{\eta'
k}]=ie_{jkm}\delta_{\eta\eta'}R_{\eta m}.
\end{eqnarray}
The operators ${\bf S}_\eta$ are orthogonal to ${\bf R}_\eta$, and
the Casimir operators in this case are ${\cal K}_\eta={\bf
S}_\eta^{2}+{\bf R}_\eta^{2}=3.$ This justifies the qualification
of such TQD as a {\it double spin rotator} which is obtained from
the spin rotator considered in Refs. \cite{KA01,KA02} by a mirror
reflection. The symmetry of such TQD is $P\times SO(4)\times
SO(4).$

Four additional vertices appear in the effective spin Hamiltonian
(\ref{sym-h1-mlr}) at the second stage of Haldane-Anderson scaling
procedure \cite{Anderson}. As a result, the exchange part of the
Hamiltonian (\ref{sym-h1-mlr}) takes the form
\begin{eqnarray}
{H}_{cot} &=&\sum_{\eta=g,u}J^T_{1\eta} {\bf S}_{\eta}\cdot {\bf
s}_\eta+ \sum_{\eta}J^{ST}_{1\eta} {\bf R}_{\eta}\cdot {\bf
s}_\eta
 +J^T_{2}
\sum_{\eta}{\bf S}_{\eta{\bar \eta}}\cdot {\bf s}_{\eta{\bar
\eta}} \nonumber\\
&+&\sum_{\eta}(J^{ST}_{2\eta}{\bf R}^{(1)}_{\eta{\bar
\eta}}+J^{ST}_{2{\bar\eta}}{\bf R}^{(2)}_{\eta{\bar \eta}})\cdot
{\bf s}_{\eta{\bar \eta}}
+ \sum_{\eta}J^T_{3\eta} {\bf S}_{\eta}\cdot {\bf s}_{\bar\eta}+
\sum_{\eta}J^{ST}_{3\eta} {\bf R}_{\eta}\cdot {\bf
s}_{\bar\eta}.\label{sym-h-mlr}
\end{eqnarray}

The coupling constants in the Hamiltonian (\ref{sym-h-mlr}) are
subject to renormalization. Their values at $D=\bar D$ are taken
as initial conditions
\begin{eqnarray}
J^T_{1\eta}({\bar D})&=&J^T_2({\bar D})=J^{ST}_{1u}({\bar
D})=J^{ST}_{2u}({\bar
D})=\frac{V^2}{\epsilon_{F}-\epsilon},\label{bcond}\\
J^T_{3\eta}({\bar D})&=&J^{ST}_{3u}({\bar D})=0, \
J^{ST}_{ig}({\bar D})=\alpha J^T_{ig}({\bar D}) \
(i=1,2,3)\nonumber
\end{eqnarray}
for solving the scaling equations. These can be written in the
following form:
\begin{eqnarray}
\frac{dj_{1\eta}^{T}}{d\ln d} &=& -\Big[(j^{T}_{1\eta})^2
+2(-1)^\eta m_{lr}j^{T}_{1\eta} 
+(j^{ST}_{1\eta})^2
+\frac{(j^{T}_2)^2}{2}+\frac{(j^{ST}_{2{\bar \eta}})^2}{2}\Big],\nonumber\\
\frac{dj_{2}^{T}}{d\ln d} &=&
-\frac{1}{2}\Big[\sum_{\eta=g,u}\{j^{T}_2(j^{T}_{1\eta}+
j^{T}_{3\eta}) + j^{ST}_{2\eta}(
j^{ST}_{1\eta}+j^{ST}_{3\eta})\}\Big],\nonumber\\
\frac{dj_{3\eta}^{T}}{d\ln d} &=& -\Big[(j^{T}_{3\eta})^2
+2(-1)^\eta m_{lr}j^{T}_{3\eta}
+(j^{ST}_{3\eta})^2+
\frac{(j^{T}_2)^2}{2}+\frac{(j^{ST}_{2{\bar \eta}})^2}{2}\Big],\nonumber\\
\frac{dj_{1\eta}^{ST}}{d\ln d} &=&-\Big[2j^{T}_{1\eta}
j^{ST}_{1\eta}+2(-1)^\eta m_{lr}j_{1\eta}^{ST}+ j^{T}_2
j^{ST}_{2\eta} \Big],\nonumber\\
\frac{dj_{2\eta}^{ST}}{d\ln d} &=&
-\frac{1}{2}\Big[\sum_{\eta}j^{T}_2(j^{ST}_{1\eta}+
j^{ST}_{3\eta})+ 2j^{ST}_{2\eta}
(j^{T}_{1{\bar \eta}}+j^{T}_{3{\bar \eta}}) \Big],\nonumber\\
\frac{dj_{3\eta}^{ST}}{d\ln d} &=&-\Big[2j^{T}_{3\eta}
j^{ST}_{3\eta}+2(-1)^\eta m_{lr}j_{3\eta}^{ST}+ j^{T}_2
j^{ST}_{2\eta} \Big],\label{sc-eq-so4}
\end{eqnarray}
where $j_{i\eta}=\rho_0 J_{i\eta}$ ($i=1,2,3$), $d=\rho_0 D$ and
$m_{lr}=\rho_0 M_{lr}.$ It should be noted that the terms
proportional to $m_{lr}$ arise in Eqs.(\ref{sc-eq-so4}) since the
dot Hamiltonian (the first term in Eq.(\ref{sym-h1-mlr})) is not
proportional to the unit matrix, and thus it does not commute with
the exchange terms (\ref{sym-h-mlr}). As can be seen from
Eq.(\ref{lre}), the deviation from the unit matrix is proportional
to $M_{lr}$.

Solution of Eqs. (\ref{sc-eq-so4}) yields the Kondo temperature
\begin{equation}
T_{K0}=\bar{D}\left(1-\frac{8m_{lr}}{(\sqrt{3}+1)(3j_{1g}^{T}+j_{1g}^{ST})}\right)
^{\displaystyle{\frac{1}{2m_{lr}}}}.\label{ Tsym}
\end{equation}
The limiting value of this relation for independent $l$, $r$
channels is
\begin{equation}\label{tko}
\lim_{m_{lr}\to 0}T_{K0} ={\bar D} \exp \left(
-\frac{4}{(\sqrt{3}+1)(3j_{1g}^T+j_{1g}^{ST})} \right).
\end{equation}
Here and below the coupling constants $j_i(D)$ in all equations
for $T_K$ are taken at $D=\bar D$. We see that avoided crossing
effect in the case of slightly violated $l-r$ symmetry of TQD
turns the Kondo temperature to be a function of the level
splitting (\ref{lre}). A similar situation has been noticed in
previous studies of DQD \cite{KA01,KA02} and planar QD with even
occupation \cite{Magn}, where $T_K$ turned out to be a
monotonically decreasing function of S/T splitting energy ${
\delta}={\bar E}_{S}-{\bar E}_{T}$ with a maximum at ${
\delta}=0$. Now the Kondo temperature is a function of two
parameters, $T_K(M_{lr},\delta).$
\begin{figure}[htb]
\centering
\includegraphics[width=70mm,height=55mm,angle=0,]{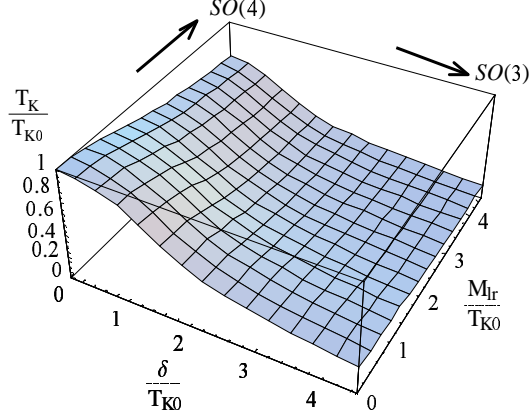}
\caption{Variation of $T_K$ with the parameters $\delta$ and
$M_{lr}$ (see text for further details).} \label{twopar}
\end{figure}
Looking at Fig.\ref{pdotso4} (which corresponds to $\delta=0$) we
notice that for large enough $M_{lr}$, when the inequality
(\ref{ineq1}) is violated, $M_{lr}\gg T_K$, the symmetry of TQD is
reduced to $SO(4)$ symmetry of S/T manifold with the Kondo
temperature
\begin{equation}\label{TK1}
T_{K1}=\bar{D}\exp\left\{-\frac{1}{j_{1}^T+j_{1}^{ST}}\right\}.
\end{equation}
Additional S/T splitting induced by the gate voltage
($\epsilon_l\neq \epsilon_r$) results in further decrease of $T_K$
as a function of $\delta$. The asymptotic form of the function
$T_K(\delta)$ is
\begin{equation} \label{Tdelta}
\frac{T_K}{T_{K1}}\approx
\left(\frac{T_{K1}}{\delta}\right)^{\alpha},
\end{equation}
where $\alpha=\sqrt{1-4\beta^2}<1$. In the limit of
$\delta\rightarrow \bar{D}$ the singlet state should be excluded
from the manifold, and the symmetry of the TQD with spin one in
this case is $SO(3).$ The general shape of $T_K(M_{lr},\delta)$
surface is presented in Fig. \ref{twopar}. Thus the Kondo effect
for the TQD with mirror symmetry is characterized by the stable
infinite fixed point characteristic for the {\it underscreened}
spin one dot, similar to that for DQD \cite{KA01,KA02}.

\subsection{$SO(5)$ Symmetry}\label{subsec.3.2.2}
Now we turn to asymmetric configurations
where $E_{lc}\neq E_{rc}$, ${\Gamma_{T_r}}\neq{\Gamma_{T_l}}$. In
this case the system loses the $l-r$ symmetry, and it is more
convenient to return to the initial variables used in the generic
Hamiltonian (\ref{generic}).
\begin{figure}[htb]
\centering
\includegraphics[width=75mm,height=60mm,angle=0,]{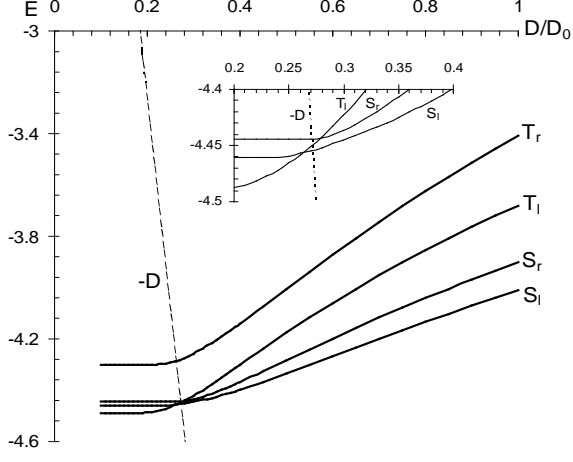}
\caption{Scaling trajectories resulting in an $SO(5)$ symmetry in
the SW regime. Inset: Zoomed in avoided level crossing pattern
near the SW line.} \label{SO5}
\end{figure}

When the Haldane renormalization results in an {\it accidental}
degeneracy of two singlets and one triplet, ${\bar E}_{S_l}\approx
{\bar E}_{T_l}\approx {\bar E}_{S_r}< {\bar E}_{T_r}$ (Fig.
\ref{SO5}), the TQD acquires an $SO(5)$ symmetry of a manifold
$\{T_l,S_l,S_r\}$. In this case the SW Hamiltonian (\ref{generic})
transforms into
\begin{eqnarray}
H&=&\sum_{\Lambda=T_l,S_l,S_r}{\bar{E}}_{\Lambda}X^{\Lambda\Lambda}
+M_{lr}(X^{S_lS_r}+X^{S_rS_l}) +\sum_{ k
\sigma}\sum_{a=l,r}(\epsilon_{ka}
c^{+}_{ak\sigma}c_{ak\sigma}+t_{lr} c^{+}_{ak\sigma }c_{{\bar
a}k\sigma })
\nonumber\\
&+&J_{1}{\bf S}_{l}\cdot {\bf s}_l+J_{2}{\bf R}_{l}\cdot {\bf
s}_l
+ J_3(\tilde{\bf R}_{1}\cdot {\bf s}_{rl}+\tilde{\bf R}_{2}\cdot
{\bf s}_{lr}),\label{sym5-tlr}
\end{eqnarray}
where $J_1=J^T_{l}$, $J_2=J^{ST}_{l}$ and $J_3=\alpha_r J_{lr}$.
The spherical components of the vector operators $\tilde{\bf R}_1$
and $\tilde{\bf R}_2$ are given by the following expressions,
\begin{eqnarray} \tilde{R}^{+}_1
&=&-\sqrt{2}X^{S_r\bar{1}_l},\ \ \ \ \tilde{R}^{-}_1=\sqrt{2}
X^{S_r1_l}, \ \ \ \ \tilde{R}_{1z} =-X^{S_r0_l},\nonumber\\
\tilde{R}^+_2&=&(\tilde{R}^{-}_1)^\dag,\ \ \ \ \ \ \ \ \ \;
\tilde{R}^-_2=(\tilde{R}^{+}_1)^\dag,\ \ \ \ \ \ \ \
\tilde{R}_{2z}=\tilde{R}_{1z}^\dag. \label{r-tild}
\end{eqnarray}
 The group generators of the $o_5$
algebra are the $l$-vectors ${\bf S}_l, {\bf R}_l$  from
(\ref{comm1}) and the operators intermixing $l$- and $r$-states,
namely the vector ${\bf \tilde{R}}={\bf \tilde{R}}_1+{\bf
\tilde{R}}_2$,
\begin{eqnarray}\label{tilt}
{\tilde R}^{+}=\sqrt{2}(X^{1_{l}S_{r}}-X^{S_{r}\bar{1}_{l}}),~~~~
{\tilde R}^{-}=({\tilde R}^{+})^{\dagger},~~~~
{\tilde R}^{z}=-(X^{0_{l}S_{r}}+X^{S_{r}0_{l}}),
\end{eqnarray}
and a scalar $A$ interchanging $l,r$ variables of degenerate
singlets
\begin{equation}\label{scal}
A=i(X^{S_rS_l}-X^{S_lS_r})~.
\end{equation}

The commutation relations (\ref{1.1}), (\ref{1.2}) in this
particular case acquire the form
\begin{eqnarray}
&&\lbrack S_{lj},S_{lk}]=ie_{jkm}S_{lm},\;\;\;\;~
[R_{lj},R_{lk}]=ie_{jkm}S_{lm},
\;\;\;\;\;~ \lbrack R_{lj},S_{lk}]=ie_{jkm}R_{lm},\nonumber\\
&&[\tilde{R}_{j},\tilde{R}_{k}] = ie_{jkm}S_{lm},\;\;\;\;\;\;
\lbrack \tilde{R}_{j},S_{lk}]=ie_{jkm}\tilde{R}_{m},
\;\;\;\;\;~~
[R_{lj},\tilde{R}_{k}]=i\delta_{jk}A, \nonumber\\
&&\lbrack \tilde{R}_{j},A] = iR_{lj}, \ \ \ \ \ \ \
[A,R_{lj}]=i\tilde{R}_{j},\ \ \ \ \ \ \ [A,S_{lj}]=0.\label{comm2}
\end{eqnarray}
The operators ${\bf R}_l$ and ${\bf \tilde{R}} $ are orthogonal to
${\bf S}_l$ in accordance with (\ref{orth}). Besides, ${\bf
R}_l\cdot {\bf\tilde{R}}=3X^{S_l S_r},$ and the Casimir operator
is ${\cal K}={\bf S}_l^{2}+{\bf R}_l^{2}+\tilde{\bf R}^{2}+A^2=4.$

Like in the case of double $SO(4)$ symmetry studied above, the
second step of RG procedure generates additional vertices in the
exchange part of the interaction Hamiltonian~(\ref{sym5-tlr}),
\begin{eqnarray}
H_{cot}&=&J_{1}{\bf S}_{l}\cdot {\bf s}_l+J_{2}{\bf R}_{l}\cdot
{\bf s}_l  + J_3(\tilde{\bf R}_{1}\cdot {\bf s}_{rl}+\tilde{\bf
R}_{2}\cdot {\bf s}_{lr}) +J_{4}{\bf S}_{l}\cdot {\bf s}_r
+J_5{\bf \tilde{R}}\cdot {\bf s}_l \nonumber\\
&+&J_6({\bf R}_{1l}\cdot {\bf s}_{rl}+ {\bf R}_{2l}\cdot {\bf
s}_{lr})
 +J_7{\bf S}_{l}\cdot({\bf s}_{lr}+{\bf s}_{rl})+
 J_8(\tilde{\bf R}_{1}\cdot {\bf s}_{lr}+\tilde{\bf
R}_{2}\cdot {\bf s}_{rl})\nonumber\\
&+&J_{9}{\bf R}_{l}\cdot {\bf s}_r+J_{10}{\bf \tilde{R}}\cdot {\bf
s}_r+J_{11}{\bf R}_{l}\cdot({\bf s}_{lr}+{\bf s}_{rl})
+J_{12}({\bf R}_{1l}\cdot {\bf s}_{lr}+{\bf R}_{2l}\cdot {\bf
s}_{rl}),\label{sym52-tlr}
\end{eqnarray}
where ${\bf R}_{1l}=X^{S_lS_r}\tilde{\bf R}_{1}$, ${\bf
R}_{2l}=\tilde{\bf R}_{2}X^{S_rS_l}$. The scaling properties of
the system are determined by a system of 12 scaling equations with
initial conditions
\begin{eqnarray}
J_1(\bar D)=J^T_{l},\ \ \ \ \ \ \ J_2(\bar D)=J^{ST}_{l},\ \ \ \ \
J_3(\bar D)=\alpha_r J_{lr}, \ \ \ \ \  J_{i}(\bar
D)=0~~~~(i=4-12)\label{bcond5}
 \end{eqnarray}
(see Eq. \ref{J} for definitions) specifically
\begin{eqnarray}
\frac{dj_{1}}{d\ln d}&=&-\left[j_1^2 + j_2^2
+j_5^2+j_7^2+j_{11}^2+j_{11}(j_6+j_{12})
+\frac{j_3^2+j_6^2+j_8^2+j_{12}^2}{2}\right],\nonumber\\
\frac{dj_{2}}{d\ln d} &=& -[2(j_1 j_2+j_7j_{11})+j_7(j_6+j_{12})-m_{lr}j_5],\nonumber\\
\frac{dj_{3}}{d\ln d} &=& -[j_3(j_1+j_4)+j_7(j_5+j_{10})-m_{lr}(j_6+j_{11})],\nonumber\\
\frac{dj_{4}}{d\ln d} &=& -\left[j_4^2+j_7^2+j_9^2+j_{10}^2+j_{11}^2+j_{11}(j_6+j_{12})
+\frac{j_3^2+j_6^2+j_8^2+j_{12}^2}{2}\right],\nonumber\\
\frac{dj_{5}}{d\ln d} &=& -[2j_1j_5+j_7(j_3+j_{8})-m_{lr}j_2],\nonumber\\
\frac{dj_{6}}{d\ln d} &=& -[j_6(j_1+j_{4})-m_{lr}j_3],\nonumber\\
\frac{dj_{7}}{d\ln d} &=& -\left[\frac{(j_3
+j_8)(j_5+j_{10})+(j_2+j_9)(j_6+j_{12})}{2}
+j_7(j_{1}+j_4)+j_{11}(j_2+j_{9})\right],\nonumber\\
\frac{dj_{8}}{d\ln d} &=& -[j_8(j_1+j_4)+j_7(j_5+j_{10})-m_{lr}(j_{11}+j_{12})],\nonumber\\
\frac{dj_{9}}{d\ln d} &=& -[2(j_4 j_9+j_7j_{11})+j_7(j_6+j_{12})-m_{lr}j_{10}],\nonumber\\
\frac{dj_{10}}{d\ln d} &=& -[2j_4j_{10}+j_7(j_3+j_{8})-m_{lr}j_9],\nonumber\\
\frac{dj_{11}}{d\ln d} &=& -[j_{11}(j_1+j_4)+j_7(j_2+j_{9})],\nonumber\\
\frac{dj_{12}}{d\ln d} &=& -[j_{12}(j_1+j_{4})-m_{lr}j_8].
 \label{lr-sc-eq5}
\end{eqnarray}
Here the terms proportional to $m_{lr}$ arise because the second
term in the Hamiltonian (\ref{sym5-tlr}) contains non-diagonal
terms.

From equations (\ref{lr-sc-eq5}), one deduces the Kondo
temperature,
\begin{eqnarray}
T_{K2}=\bar{D}\Big(1-\frac{2\sqrt{2}m_{lr}}{j_1+j_2+\sqrt{(j_1+j_2)^2+2j_3^2}}\Big)
^{\displaystyle{\frac{1}{\sqrt{2}m_{lr}}}}. \label{T5-lr}
\end{eqnarray}
Similarly to the previous case, this equation transforms into the
usual exponential form when the $l$ and $r$ channels are
independent,
\begin{eqnarray}\label{tko1}
\lim_{m_{lr}\to 0}T_{K2} ={\bar D} e^{\displaystyle{
-\frac{2}{j_1+ j_2+\sqrt{(j_1+j_2)^2+2j_3^2}}}}.
\end{eqnarray}
Upon increasing $m_{lr}$, the symmetry reduces from $SO(5)$ to
$SO(4)$. The same happens at small $m_{lr}$ but with increasing
 $\bar \delta_{l}={\bar E}_{S_l}-{\bar E}_{T_l}$. In the latter case the
energy $\bar E_{S_l}$ is quenched, and at $\bar \delta_l \gg
T_{K2}$ Eq. (\ref{T5-lr}) transforms into $ T_{K}=\bar \delta_l
\exp\{-[j_1(\bar\delta_l)+ j_3(\bar\delta_l)]^{-1}\} $ (cf.
\cite{KKA02}). On the other hand, upon decreasing $\bar
\delta_{r}={\bar E}_{T_r}-{\bar E}_{S_l}$ the symmetry ${P}\times
SO(4)\times SO(4)$ is restored at $\bar \delta_{r}<T_{K0}$. The
Kondo effect disappears when $\bar \delta_{l}$ changes sign (the
ground state becomes a singlet).

\subsection{$SO(7)$ Symmetry}\label{subsec.3.2.3}
\begin{figure}[htb]
\centering
\includegraphics[width=75mm,height=60mm,angle=0,]{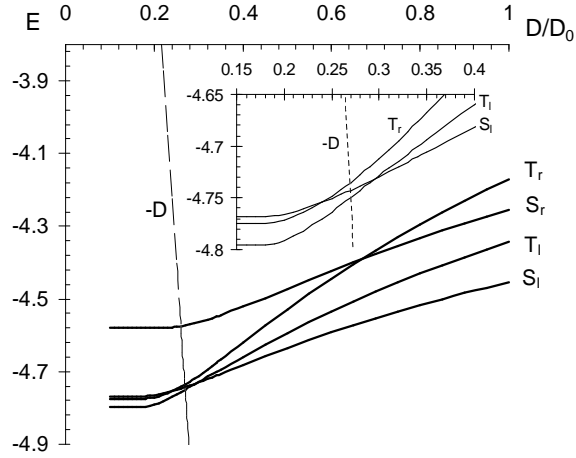}
\caption{Scaling trajectories for $ SO(7)$ symmetry in the SW
regime. Inset: Zoomed in avoided level crossing pattern near the
SW line.} \label{SO7}
\end{figure}

The next asymmetric configuration is illustrated by the flow
diagram of Fig.\ref{SO7}.

In this case, the manifold $\{T_l,S_l,T_r\}$ is involved in the
dynamical symmetry of TQD. The relevant symmetry group is $SO(7)$.
It is generated by six vectors and three scalars. These are spin
operators ${\bf S}_a$ $(a=l,r)$ and R-operator ${\bf R}_l$  (see
Eq. \ref{comm1}) plus three vector operators ${\bf \tilde{R}}_i$
and three scalar operators $A_i$ involving $l-r$ permutation. Here
are the expressions for the spherical components of these vectors
via Hubbard operators,
\begin{eqnarray}
&&\tilde{R}_1^{+} =\sqrt{2}( X^{1_r0_l}+X^{0_l\bar{1}_r}),\;\;\;
\tilde{R}_1^{z} =X^{1_l1_r}-X^{{\bar 1}_r{\bar 1}_l},\nonumber\\
&&\tilde{R}_2^{+}=\sqrt{2}( X^{1_l0_r}+X^{0_r\bar{1}_l}),\ \
\tilde{R}_2^{z} =X^{1_r1_l}-X^{{\bar 1}_l{\bar 1}_r}, \label{so7}\\
&&\tilde{R}^{+}_3 =\sqrt{2}( X^{1_rS_l}-X^{S_l\bar{1}_r}),\ \;
\tilde{R}_3^{z} =-(X^{0_rS_l}+X^{S_l0_r}).\nonumber
\end{eqnarray}
The scalar operators $A_1$, $A_2$, $A_3$ now involve the $l-r$
permutations for the triplet states. They are defined as
\begin{eqnarray}
A_1 &=&\frac{i\sqrt{2}}{2}\left(X^{1_r{\bar 1}_l}-X^{1_l{\bar
1}_r+
X^{{\bar 1}_r1_l}-X^{{\bar 1}_l1_r}}\right ),\nonumber\\
A_2 &=&\frac{\sqrt{2}}{2}\left(X^{1_l{\bar 1}_r}-X^{1_r{\bar
1}_l}+X^{{\bar 1}_r1_l}-
 X^{{\bar 1}_l1_r}\right),\nonumber\\
A_3&=&i\left(X^{0_l0_r}-X^{0_r0_l}\right).\label{A}
\end{eqnarray}
The (somewhat involved) commutation relations of $o_7$ algebra for
these operators and various kinematic constraints are presented in
Appendix \ref{algebra-o7}. The SW transformation results in the
effective Hamiltonian
\begin{eqnarray}
H&=&\sum_{\Lambda=T_l,S_l,T_r}{\bar{E}}_{\Lambda}X^{\Lambda\Lambda}
+M_{lr}(X^{T_lT_r}+X^{T_rT_l})
+\sum_{ k \sigma}\sum_{a=l,r}(\epsilon_{ka}
c^{+}_{ak\sigma}c_{ak\sigma}+t_{lr}
c^{+}_{ak\sigma}c_{{\bar a}k\sigma })
\nonumber\\
&+&\sum_{a=l,r} J_{1a}{\bf S}_{a}\cdot {{\bf s}}_a+
J_2\sum_{a=l,r}{\bf
{S}}_{a{\bar a}}\cdot {\bf s}_{a{\bar a}}
+J_{3}({\bf {\tilde R}}^{(1)}_{3}\cdot {\bf s}_{rl}+ {\bf {\tilde
R}}^{(2)}_{3}\cdot {\bf s}_{lr})+J_4{\bf {R}}_{l}\cdot {{\bf
s}_l}, \label{int7}
\end{eqnarray}
where $J_{1a}=J_a^T$, $J_{2}=J_{{lr}}$, $J_{3}=\alpha_l J_{lr}$,
$J_{4}=\alpha_l J_{l}^T$ and ${\bf {S}}_{a{\bar
a}}=\sum_{\mu}X^{\mu_a\mu_{\bar a}}{\bf {S}}_{{\bar a}}$. The
spherical components of the vector operators $\tilde{\bf R}_1$ and
$\tilde{\bf R}_2$ are
\begin{eqnarray}
\tilde{R}^{(1)+}_3 &=&\sqrt{2}X^{1_rS_l},\ \ \ \;
\tilde{R}^{(1)-}_3=-\sqrt{2} X^{{\bar 1}_rS_l},\ \ \ \
\tilde{R}^{(1)}_{3z} =-X^{0_rS_l},\nonumber\\
\tilde{R}^{(2)+}_3&=&(\tilde{R}^{(1)-}_3)^\dag,\ \ \ \
\tilde{R}^{(2)-}_3=(\tilde{R}^{(1)+}_3)^\dag, \ \ \ \ \ \ \ \;
\tilde{R}^{(2)}_{3z}=(\tilde{R}^{(1)}_{3z})^\dag.
\end{eqnarray}
It is easy to see that ${\bf S}_{lr}+{\bf S}_{rl}={\bf {\tilde
R}}_1+{\bf {\tilde R}}_2$ and ${\bf {\tilde R}}_3={\bf {\tilde
R}}^{(1)}_{3}+{\bf {\tilde R}}^{(2)}_{3}$.

Like in the case of $SO(5)$ symmetry, the tunneling terms
$M_{lr}X^{T_a T_{\bar a}}$ generate additional vertices in the
renormalized Hamiltonian $H_{cot}$. The number of these vertices
and the corresponding scaling equations is too wide to be
presented here. We leave the description of RG procedure for
$SO(7)$ group for the next chapter (as well as the case of TQD
with odd occupation), where the case of $M_{lr}=0$ is considered.
In that situation the scaling equations describing the Kondo
physics of TQD with $SO(n)$ symmetry are more compact.
\section{Conclusions}\label{sec.3.3}
The basic physics for all $SO(n)$ symmetries is the same, and we
summarize it here. The TQD in its ground state cannot be regarded
as a simple quantum top in the sense that beside its spin operator
other vector operators ${\bf R}_{n}$ are needed (in order to fully
determine its quantum states), which have non-zero matrix elements
between states of different spin multiplets $\langle S_{i}M_{i}
|{\bf R}_{n}|S_{j}M_{j} \rangle \ne 0$. These "Runge-Lenz"
operators do not appear in the isolated dot Hamiltonian (so in
some sense they are "hidden"). Yet, they are exposed when
tunneling between the TQD and leads is switched on. The effective
spin Hamiltonian which couples the metallic electron spin ${\bf
s}$ with the operators of the TQD then contains new exchange
terms, $J_{n} {\bf s} \cdot {\bf R}_{n}$ beside the ubiquitous
ones $J_{i} {\bf s}\cdot {\bf S}_{i}$. The operators ${\bf S}_{i}$
and ${\bf R}_{n}$ generate a dynamical group (usually $SO(n)$).

We have analyzed several examples of TQD with even occupation in
the parallel geometry (Fig. \ref{TQD-par}). Our analysis
demonstrates the principal features of Kondo effect in CQD in
comparison with the conventional QD composed of a single well.
These examples teach us that in Kondo tunneling through CQD, not
only the spin rotation but also the ''Runge-Lenz'' type operators
${\bf R}$ and $\tilde{\bf R}$ are involved. Physically, the
operators $\tilde{\bf R}$ describe left-right transitions, and
different Young schemes give different spin operators in the
effective co-tunneling Hamiltonians (see Appendix \ref{tab}).

\chapter{Trimer in Series} \label{IV}
{\it In this Chapter we expose the physics of Kondo tunneling
through sandwich-type molecules adsorbed on metallic substrate. In
Sec. \ref{sec.4.1} we introduce the system of our study and
demonstrate the correspondence between the low-energy tunnel
spectra of chemisorbed lanthanocene molecule and an artificial
trimer, i.e., TQD in series geometry. In Sec. \ref{sec.4.2} we
concentrate on the case of even occupation. The scaling equations
are derived and the Kondo temperatures are calculated for the
evenly occupied trimer in the cases of the $P\times SO(4)\times
SO(4)$, $SO(5)$, $SO(7)$ and $P\times SO(3)\times SO(3)$ dynamical
symmetries. The dynamical-symmetry phase diagram is displayed and
the possibility of its experimental realization is outlined. The
anisotropic Kondo effect induced by an external magnetic field is
discussed in Sec. \ref{sec.4.3}. It is shown that the symmetry
group for such magnetic field induced Kondo tunneling is $SU(3)$.
The case of odd occupation is considered in Sec. \ref{sec.4.4}.
When the ground state of the trimer is a doublet, the effective
spin Hamiltonian of the trimer manifests a two-channel Kondo
problem albeit only in the weak coupling regime. Analysis of the
Kondo effect in cases of higher spin degeneracy of the trimer
ground state is carried out in relation with dynamical symmetries.
In the Conclusions we underscore the main results obtained.}

\section{Introduction}\label{sec.4.1}

In this Chapter we extend the theory of single-electron tunneling
developed in Chapter \ref{III} for complex quantum dots to the
case of sandwich-type molecules adsorbed on metallic substrate.
The geometry of the nano-objects under investigation is: metallic
subsrate (MS) - molecule  - nanotip of scanning tunnel microscope
(STM). The specific objects of our studies are lanthanocene
molecules Ln(C$_8$H$_8)$$_2$ where the magnetic ion Ln=Ce, Yb
(central ion) is secluded in a cage of carbon--containing radicals
and only these radicals are in direct tunnel contact with MS and
STM. Ce is known as a mixed-valent ion in many molecular and
crystalline configurations. This means that the covalent bonding,
i.e., hybridization between strongly correlated 4f-electron and
weakly interacting molecular $\pi$-orbitals $\rm Mo=$~C$_8$H$_8$
in cerocene molecule Ce(C$_8$H$_8)$$_2$ characterized by the
parameter $V_h$, is noticeable. Besides, the difference between
the ionization energy $\varepsilon_f$ of 4f-electron and the
ionization energy $\varepsilon_{\pi}$ of an electron in molecular
$\pi$-orbital is large enough, so that $V_h\ll
\varepsilon_\pi-\varepsilon_f$. As is shown in
\cite{Dolg,Dolg1,Liu}, the direct consequence of this inequality
is unusual electronic and spin structure of the ground state and
low-energy excitations in Ce(C$_8$H$_8)$$_2$. Instead of purely
ionic bonding ${\rm Ce^{3+}(4f^1)(Mo)_2^{3-}}(e_{2u}^3)$ a mixed
valence state arises with admixture of configuration ${\rm
4f^0}e_{2u}^4$ to the ground state spin singlet. Due to the Pauli
principle, such admixture is forbidden in a triplet state. As a
result, the energy difference between the ground state singlet and
excited triplet is controlled by the small parameter
$V_h/(\varepsilon_\pi-\varepsilon_f)$.

\begin{figure}[htb]
\centering
\includegraphics[width=80mm,height=55mm,angle=0]{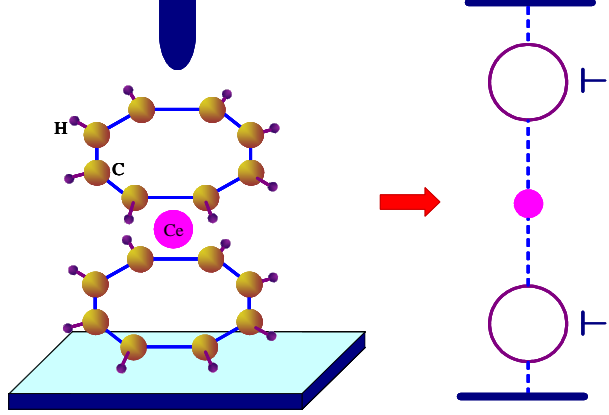}
\caption{A molecule with strong correlations is modelled by a TQD
in a series geometry.} \label{mol-TQD}
\end{figure}

This scenario reminds the mechanism of singlet-triplet splitting
in asymmetric DQD (Sec. \ref{sec.2.3}) and exactly coincides with
that for TQD with small central dot, whose properties are studied
in Chapter 3. The singly occupied central dot with strong Coulomb
blockade plays the same part as magnetic ion (Ce or Yb) in
lantanocene molecules. Two side dots in TQD play the same role as
two molecular rings C$_8$H$_8$ in formation of low energy spin
spectrum. Only a pair of the highest occupied molecular orbitals
(HOMO) is involved in formation of this spectrum. All other states
may be treated as molecular excitons, which are quenched within
the scale of $V_h^2/(\varepsilon_\pi-\varepsilon_f)$ above the
ground state. As a result, the orbital degrees of freedom are
irrelevant in the Kondo regime. The similarity between the
configurations "STM--Ce(C$_8$H$_8)$$_2$--metallic adsorbent" and
source -- TQD --drain" is illustrated by Figs. \ref{mol-TQD} and
\ref{TQD-s}. The only essential condition for modelling both
systems by the same Hamiltonian (\ref{Hser}) and treating them as
a trimer in contact with metallic leads is the inequality $W>V$,
i.e., the demand that the tunnel contact with the reservoir does
not destroy the coherent quantum mechanical state of a trimer.

It was mentioned already in Chapter \ref{III} that a TQD with
leads $l$ and $r$ representing independent tunneling channels can
be mapped onto a TQD in a series by means of geometrical conformal
transformation. Indeed, if the inter-channel tunneling amplitude
$t_{lr}$ in the Hamiltonian (\ref{tlead}) vanishes, one may apply
a rotation in source-drain space separately to each channel and
exclude the odd $s-d$ combination of lead states both in the $l$-
and $r$-channel \cite{Glaz}. Since now each lead is coupled to its
own reservoir, and one arrives at the series configuration shown
in Fig. \ref{TQD-s}.

It is virtually impossible to conceive an additional
transformation after which  the odd combination of lead states is
excluded from the tunneling Hamiltonian \cite{EPL}. As a result,
the challenging situation arises in case of odd occupation ${\cal
N}=3$, where the net spin of TQD is $S=1/2$, and the two leads
play part of two channels in Kondo tunneling Hamiltonian.
Unfortunately, despite the occurrence of two electron channels in
the spin Hamiltonian, the complete mapping on the two-channel
Kondo problem is not attained because there is an additional
cotunneling term $J_{lr} {\bf S}\cdot {\bf s}_{lr}+H.c.$ (${\bf
s}_{lr}$ is determined by (\ref{1.100})) which turns out to be
relevant, and the two-channel fixed point cannot be reached (see
Sec. \ref{sec.4.4}). And yet, from the point of view of dynamical
symmetry the series geometry offers a new perspective which we
analyze in the present chapter for the cases of even and odd
occupation.

\section{Even Occupation}\label{sec.4.2}
\begin{figure}[htb]
\centering
\includegraphics[width=75mm,height=70mm,angle=0]{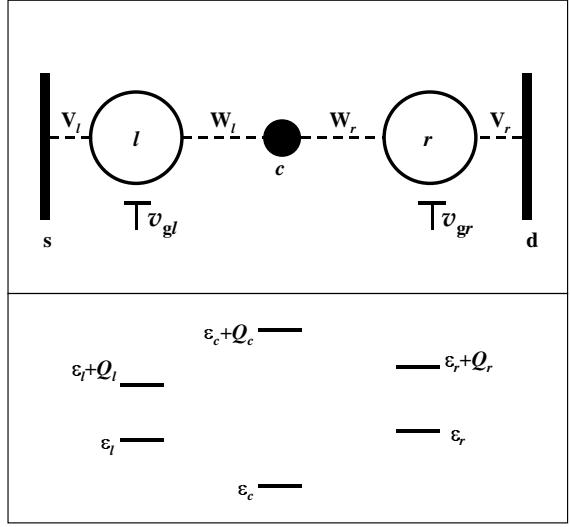}
\caption{Triple quantum dot in series. Left $(l)$ and right $(r)$
dots are coupled by tunneling $W_{l,r}$ to the central $(c)$ dot
and by tunneling $V_{l,r}$ to the source $(s)$ (left) and drain
$(d)$ (right) leads.}\label{TQD-s}
\end{figure}

Consider then a trimer in series (Fig. \ref{TQD-s}) with four
electron occupation ${\cal N}=4$. The Hamiltonian of the system
can be written in the form,
\begin{eqnarray}
H&=&\sum_{\Lambda_{a}}{E}_{\Lambda_a} X^{\Lambda_a\Lambda_a}
+\sum_{\lambda}{E}_\lambda X^{\lambda\lambda}
 +\sum_{
k \sigma}\sum_{b=s,d}\epsilon_{kb} c^{+}_{bk\sigma}c_{bk\sigma}
\nonumber\\
& +&
\sum_{\Lambda\lambda}\sum_{k\sigma}[(V^{\lambda\Lambda}_{l\sigma}
c^{+}_{sk\sigma }+V^{\lambda\Lambda}_{r\sigma}
c^{+}_{dk\sigma})X^{\lambda\Lambda}+ H.c.]. \label{Hser}
\end{eqnarray}
Here $|\Lambda\rangle$, $|\lambda\rangle$ are the four- and
three-electron eigenfunctions (\ref{eg-fun4}) and
(\ref{eg-func-trimer}), respectively; ${E}_\Lambda,$ ${E}_\lambda$
are the four- and three-electron energy levels, respectively (five
electron states cost much energy and are discarded);
$X^{\lambda\Lambda}=|\lambda\rangle\langle\Lambda|$ are number
changing dot Hubbard operators. The tunneling amplitudes
$V^{\lambda\Lambda}_{a\sigma}=V_a\langle\lambda|d_{a\sigma}|\Lambda\rangle$
($a=l,r$) depend explicitly on the respective $3-4$ particle
quantum numbers $\lambda$, $\Lambda$. Note that direct tunneling
through the TQD is suppressed due to electron level mismatch and
Coulomb blockade, so that only {\it cotunneling} mechanism
contributes to the current.

After a SW transformation the generic Hamiltonian (\ref{generic})
simplifies in this case to
\begin{eqnarray}
H&=&\sum_{\Lambda_a}{\bar E}_{\Lambda_a}X^{\Lambda_a\Lambda_a}
+\sum_{k\sigma}\sum_{b=l,r}\epsilon_{kb}
c^{\dagger}_{bk\sigma}c_{bk\sigma}+\sum_{a=l,r}J^T_{a} {\bf
S}_{a}\cdot {\bf s}_a
\nonumber\\
&+&
  J_{lr}{\hat P}
\sum_{a=l,r}{\bf S}_{a}\cdot {\bf s}_{{\bar a}a}
+\sum_{a=l,r}J^{ST}_a {\bf R}_{a}\cdot {\bf s}_a
+J_{lr}\sum_{a=l,r}\tilde{\bf R}_{a}\cdot {\bf s}_{a{\bar
a}},\label{ex-H}
\end{eqnarray}
(the notation $l,r$ is used for the electron states both in the
leads and in the TQD). The antiferromagnetic coupling constants
are defined by (\ref{J}). The vectors ${\bf S}_a$, ${\bf R}_a$ and
$\tilde{\bf R}_a$ are the dot operators (\ref{comm1}), $\hat P$ is
the permutation operator (\ref{perm}), and the components of the
vectors ${\bf s}_a$, ${\bf s}_{a\bar a}$ are determined in Eqs.
(\ref{1.10}) (with $\alpha=a=l,r$) and (\ref{1.100}). The vector
operators ${\bf S}_{a},$ ${\bf R}_{a}$, $\tilde{\bf R}_{a}$ and
the permutation operator ${\hat P}$ manifest the dynamical
symmetry of the TQD.

We now discuss possible realization of $P\times SO(4)\times
SO(4)$, $SO(5)$, $SO(7)$ and $P\times SO(3)\times SO(3)$
symmetries arising in the TQD with ${\cal N}=4$
\cite{KKA04,KKA05}. Due to the absence of interchannel mixing, the
avoided crossing effect does not arise in the series geometry. As
a result, the cases of $P\times SO(4)\times SO(4)$, $SO(5)$ and
$SO(7)$ symmetry are characterized by the same flow diagram of
Figs. \ref{pdotso4}, \ref{SO5} and \ref{SO7} but {\it without
avoided crossing effects shown in the insets.}

Let us commence the analysis of the Kondo effect in the series
geometry with the case ${P}\times SO(4)\times SO(4)$ where
$E_{lc}=E_{rc}$ and ${\Gamma_{T_r}}={\Gamma_{T_l}}$ (Fig.
\ref{pdotso4}). In this case the exchange part of the Hamiltonian
(\ref{ex-H}) is a simplified version of the Hamiltonian
(\ref{sym-h-mlr}) with the boundary conditions (\ref{bcond}). The
scaling equations are the same as (\ref{sc-eq-so4}) with
$m_{lr}=0$. Solving them one gets Eq.(\ref{tko}) for the Kondo
temperature.

When ${\bar E}_{S_l}\approx {\bar E}_{T_l}\approx {\bar E}_{S_r}<
{\bar E}_{T_r}$ (Fig. \ref{SO5}), the TQD possesses the $SO(5)$
symmetry. In this case the interaction Hamiltonian has the form
\begin{equation}
H_{cot}=J_{1}{\bf S}_{l}\cdot {\bf s}_l+J_{2}{\bf R}_{l}\cdot {\bf
s}_l+ J_3(\tilde{\bf R}_{1}\cdot {\bf s}_{rl}+\tilde{\bf
R}_{2}\cdot {\bf s}_{lr}), \label{sym5-ser}
\end{equation}
which is the same as in Eq.(\ref{sym5-tlr}) with $\tilde{\bf
R}_{1},\ \tilde{\bf R}_{2}$ determined by Eq.(\ref{r-tild}).
Respectively, the effective Hamiltonian for the Anderson scaling
is a reduced version of the Hamiltonian (\ref{sym52-tlr})
\begin{eqnarray}
H_{cot}&=&J_{1}{\bf S}_{l}\cdot {\bf s}_l+J_{2}{\bf R}_{l}\cdot
{\bf s}_l
+ J_3(\tilde{\bf R}_{1}\cdot {\bf s}_{rl}+\tilde{\bf R}_{2}\cdot
{\bf s}_{lr})+J_{4}{\bf S}_{l}\cdot {\bf s}_r, \label{sym52}
\end{eqnarray}
with the boundary conditions (\ref{bcond5}) for $J_i, ~i=1-4$.

The scaling equations have the form
\begin{eqnarray}
\frac{dj_{1}}{d\ln d} &=& -\left[j_1^2 + j_2^2
+\frac{j_3^2}{2}\right],\nonumber\\
\frac{dj_{2}}{d\ln d} &=& -2j_1 j_2,\nonumber\\
\frac{dj_{3}}{d\ln d} &=& -j_3(j_1+j_4),\nonumber\\
\frac{dj_{4}}{d\ln d} &=& -\left[j_4^2
+\frac{j_3^2}{2}\right].\label{ser-sc-eq55}
\end{eqnarray}
Of course, Eqs.(\ref{ser-sc-eq55}) for the Kondo temperature yield
the limiting value (\ref{tko1}).

 When ${\bar E}_{T_l}\approx {\bar E}_{T_r}\approx {\bar
E}_{S_l}< {\bar E}_{S_r}$ (Fig. \ref{SO7}), the TQD possesses the
$SO(7)$ symmetry. In this case the Anderson RG procedure adds
three additional vertices in the exchange part of the basic SW
Hamiltonian (\ref{int7}),
\begin{eqnarray}
H_{cot}&=&\sum_{a=l,r} J_{1a}{\bf S}_{a}\cdot {{\bf s}}_a+
J_2\sum_{a=l,r}{\bf {S}}_{a{\bar a}}\cdot {\bf s}_{a{\bar a}}
+J_{3}({\bf {\tilde R}}^{(1)}_{3}\cdot {\bf s}_{rl}+ {\bf {\tilde
R}}^{(2)}_{3}\cdot {\bf s}_{lr})\nonumber\\
&+&J_4{\bf {R}}_{l}\cdot
{{\bf s}_l}
+\sum_{a=l,r} J_{5a}{\bf S}_{a}\cdot {{\bf s}}_{\bar a}+J_6{\bf
{R}}_{l}\cdot {{\bf s}_r}.\label{int7-1}
\end{eqnarray}
The boundary conditions for solving the scaling equations are
\begin{eqnarray}
&&J_{1a}(\bar D)=J^T_a,~~~~~~~J_2(\bar D)=J_{lr}, ~~~~~~J_3(\bar
D)=\alpha_l J_{lr}, \nonumber\\
&&J_4(\bar D)=\alpha_l J_{l}^T, ~~~~~J_{5a}(\bar D)=J_6(\bar
D)=0~~ (a=l,r).\label{bound7}
\end{eqnarray}

The system of scaling equations
\begin{eqnarray}
\frac{dj_{1l}}{d\ln d}&=&-\left[j_{1l}^2 + \frac{j_2^2}{2}
+j_4^2\right],\nonumber\\
\frac{dj_{1r}}{d\ln d}&=&-\left[j_{1r}^2 + \frac{j_2^2}{2}
+\frac{j_3^2}{2}\right],\nonumber\\
\frac{dj_{2}}{d\ln d} &=& -\frac{j_2(j_{1l}+j_{1r}+j_{5l}+j_{5r})+ j_3(j_4+j_6)}{2},\nonumber\\
\frac{dj_{3}}{d\ln d}&=&-\left[j_2(j_4+j_6)+j_3(j_{1r}+j_{5r})\right],\nonumber\\
\frac{dj_{4}}{d\ln d} &=& -\left[2j_{1l}j_4+j_2j_3\right],\nonumber\\
\frac{dj_{5l}}{d\ln d}&=&-\left[j_{5l}^2 + \frac{j_2^2}{2}
+j_6^2\right],\nonumber\\
\frac{dj_{5r}}{d\ln d}&=&-\left[j_{5r}^2 + \frac{j_2^2}{2}
+\frac{j_3^2}{2}\right],\nonumber\\
\frac{dj_{6}}{d\ln d} &=& -\left[2j_{5l}j_6+j_2j_3\right]
\label{ser-sc-eq7}
\end{eqnarray}
is now solvable analytically, and the Kondo temperature is,
\begin{eqnarray}
T_{K}=\bar{D}\exp\left\{-\frac{4}{2j_{+}+\sqrt{4j_{-}^2+3(j_2+j_3)^2}}\right\}
, \label{T7-ser}
\end{eqnarray}
where $j_+=j_{1l}+j_4+j_{1r}$, $j_-=j_{1l}+j_4-j_{1r}$.

Like in the cases considered above, the Kondo temperature and the
dynamical symmetry itself depend on the level splitting. On
quenching the $S_l$ state (increasing $\bar\delta_{lr}={\bar
E}_{S_l}-{\bar E}_{T_r}$), the pattern is changed into a
${P}\times SO(3)\times SO(3)$ symmetry of two degenerate triplets
with a mirror reflection axis. Changing the sign of $\delta_{lr}$
one arrives at a singlet regime with $T_K=0$.
\begin{figure}[htb]
\centering
\includegraphics[width=72mm,height=62mm,angle=0,]{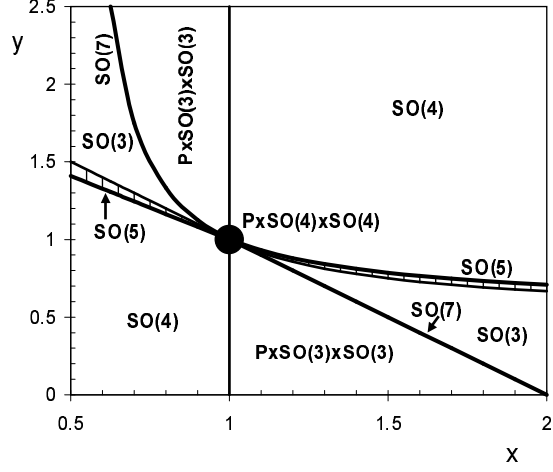}
\caption{Phase diagram of TQD. The numerous dynamical symmetries
of a TQD in the parallel geometry are presented in the plane of
experimentally tunable  parameters $x=\Gamma_l/\Gamma_r$ and
$y=E_{lc}/E_{rc}.$}\label{phasediag}
\end{figure}

When the lowest renormalized states in the SW limit are two
triplets $T_l$ and $T_r$, the TQD possesses the $P\times
SO(3)\times SO(3)$ symmetry with mirror reflection axis. The
corresponding co-tunneling spin Hamiltonian has the form,
\begin{eqnarray}
&&{H}_{cot} = \sum_{a=l,r}{\bf S}_{a}\cdot (J_{1a} {{\bf
s}_a}+J_{3a} {{\bf s}_{\bar a}}) + J_{2}{\hat P}\sum_{a=l,r}{\bf
S}_{a}\cdot {\bf s}_{a\bar a}.
 \label{intp33-2}
\end{eqnarray}
Here $J_{1a}({\bar D})=V_a^2/(\epsilon_F-\epsilon_a)$, $J_2({\bar
D})=\frac{V_lV_r}{2}\sum_{a}(\epsilon_F-\epsilon_a)^{-1}$ and
$J_{3a}({\bar D})=0$. The $P\times SO(3)\times SO(3)$ symmetry is
generated by the spin one operators ${\bf S}_a$ with projections
${\mu}_a =1_a,0_a,{\bar{1}_a}$, and the left-right permutation
operator $\hat P$ (\ref{perm}).

 The system of scaling equations for the Hamiltonian (\ref{intp33-2}) is,
\begin{eqnarray}
\frac{dj_{1a}}{d\ln d} &=&
-\left[j_{1a}^2+\frac{j_{2}^2}{2}\right],\ \ \ \frac{dj_{3a}}{d\ln
d} =
-\left[j_{3a}^2+\frac{j_{2}^2}{2}\right],\nonumber\\
\frac{dj_{2}}{d\ln d} &=&
-\frac{j_{2}}{2}\left(j_{1l}+j_{1r}+j_{3l}+j_{3r}\right),
 \label{scp33}
\end{eqnarray}
where $j=\rho_0 J,$ $a=l,r$. From Eqs.(\ref{scp33}) we obtain the
Kondo temperature, provided $|E_{T_l}-E_{T_r}|<T_K$,
\begin{equation}
T_{K}={\bar D}\exp\left[-\frac{2}{j_{1l}+j_{1r}+\sqrt{
(j_{1l}-j_{1r})^2+ 2j_{2}^2}}\right]. \label{Tp33}
\end{equation}

The results of calculations described in this section are
summarized in Fig. \ref{phasediag}. The central domain of size
$T_{K0}$ describes the fully symmetric state where there is
left-right symmetry. Other regimes of Kondo tunneling correspond
to lines or segments in the $\{x,y\}$ plane. These lines
correspond to cases of higher conductance (ZBA). On the other
hand, at some hatched regions, the TQD has a singlet ground state
and the Kondo effect is absent. These are marked by the vertically
hatched domain. Both the tunneling rates which enter the ratio $x$
and the relative level positions which determine the parameter $y$
depend on the applied potentials, so the phase diagram presented
in Fig. \ref{phasediag} can be scanned {\it experimentally} by
appropriate variations of $V_a$ and $v_{ga}.$ This is a rare
occasion where an abstract concept like dynamical symmetry can be
felt and tuned by experimentalists. The quantity that is measured
in tunneling experiments is the zero-bias anomaly (ZBA) in tunnel
conductance $g$ \cite{KKK,KKK1}. The ZBA peak is strongly
temperature dependent, and this dependence is scaled by $T_K$. In
particular, in a high temperature region $T>T_K$, where the
scaling approach is valid, the conductance behaves as
\begin{equation}\label{conductance}
g(T) \sim \ln^{-2}(T/T_K).
\end{equation}
As it has been demonstrated above, $T_K$ in CQD is a non-universal
quantity due to partial break-down of dynamical symmetry in these
quantum dots. It has a maximum value in the point of highest
symmetry ${P}\times SO(4)\times SO(4)$, and depends on the
parameters $\delta_a$ in the less symmetric phases (see, e.g.,
Eqs. \ref{tko}, \ref{Tdelta}, \ref{tko1}, \ref{T7-ser}.
\ref{Tp33}). Thus, scanning the phase diagram means changing
$T_K(\delta_a)$.
\begin{figure}[htb]
\centering
\includegraphics[width=70mm,height=50mm,angle=0,]{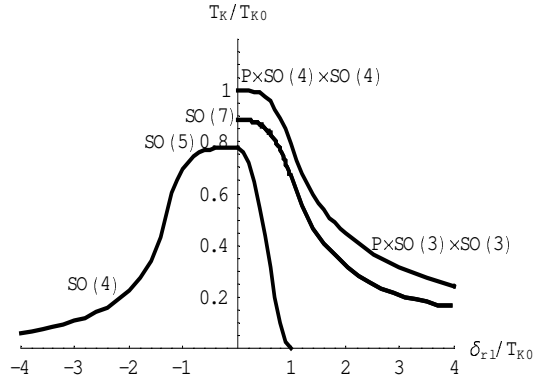}
\caption{Variation of Kondo temperature with $\delta_{rl}\equiv
v_{gr}-v_{gl}$. Increasing this parameter removes some of the
degeneracy and either "breaks" or reduces the corresponding
dynamical symmetry.} \label{vartek}
\end{figure}
These changes are shown in Fig. \ref{vartek} which illustrates the
evolution of $T_K$ with $\delta_{rl}$ for $x=0.96$, $0.8$ and 0.7
corresponding to a symmetry change from ${P}\times SO(4)\times
SO(4)$, $SO(7)$ to ${P}\times SO(3)\times SO(3)$ and from $SO(5)$
to $SO(4)$, respectively. It is clear that the conductance
measured at given $T$ should follow variation of $T_K$ in
accordance with (\ref{conductance}).
\section{Anisotropic Kondo Tunneling
through Trimer } \label{sec.4.3}
\subsection{Generalities}
In all examples of CQDs considered above the co-tunneling problem
is mapped on the specific spin Hamiltonian where both ${\bf S}$
and ${\bf R}$ vectors are involved in resonance cotunneling. There
are, however, more exotic situations where the effective spin
Hamiltonian is in fact a "Runge-Lenz" Hamiltonian in the sense
that the vectors ${\bf R}$ {\it alone} are responsible for Kondo
effect. Actually, just this aspect of dynamical symmetry in Kondo
tunneling was considered in the theoretical papers
\cite{Magn,Magn-v1,Magn-v2,Magn-v3} and observed experimentally in
Refs. \cite{Magn1,Magn-v}, in which the Kondo effect in planar and
vertical QDs induced by external magnetic field $B$ has been
studied. In this section we lay down the theoretical basis for
this somewhat unusual kind of Kondo effect.

Consider again the case of TQD in series geometry with ${\cal
N}=4$. In the previous sections the variation of spin symmetry was
due to the interplay of two contributions to indirect exchange
coupling between the spins ${\bf S}_a$. One source of such an
exchange is tunneling within the CQD (amplitudes $W_a$) and
another one is the tunneling between the dots and the leads
(amplitudes $V_a$). An appropriate tuning of these two
contributions results in accidental degeneracy of spin states
(elimination of exchange splitting), and various combinations of
these accidental degeneracies lead to the rich phase diagram
presented in Fig. \ref{phasediag}. A somewhat more crude approach,
yet more compatible with experimental observation of such
interplay is provided by the Zeeman effect. This mechanism is
effective for CQD which remains in a singlet ground state after
all exchange renormalizations have taken place. The negative
exchange energy $\delta_a$ may then be compensated by the Zeeman
splitting of the nearest triplet states, and Kondo effect arises
once this compensation is complete \cite{Magn}. From the point of
view of dynamical symmetry, the degeneracy induced by magnetic
field means realization of one possible subgroup of the
non-compact group $SO(n)$ (see Eq. \ref{1.14} and corresponding
discussion in Section \ref{sec.2.4}). The transformation $SO(4)\to
SU(2)$ for DQD in magnetic field was discussed in Ref.
\cite{KA02}.

\subsection{Trimer with SU(3) Dynamical Symmetry}
In similarity with DQD, the Kondo tunneling may be induced by
external field $B$ in the non-magnetic sector of the phase diagram
of Fig. \ref{phasediag}. A very peculiar Kondo tunneling is
induced by an external magnetic field $B$ in the non-magnetic
sector of the phase diagram of Fig. \ref{phasediag} close to the
$SO(5)$ line. In this case, a remarkable symmetry reduction occurs
when the Zeeman splitting compensates negative
$\delta_{l,r}=E_{S_{l,r}}-E_{T_l}$.
 Then we are left with the subspace of states
$\{T{1}_l,S_l, S_r\}$, and the interaction Hamiltonian has the
form,
\begin{eqnarray}
\widetilde{H}_{cot}
&=&(J_{1}R^z_{1}+J_{2}R^z_{2})s^z_l+\frac{\sqrt{2}}{2}J_{3l}
\left(R^{+}_{1}s^{-}_l+R^{-}_{1}s^{+}_l\right)
+\frac{\sqrt{2}}{2}J_{3r}(R^{+}_{2}s^{-}_{lr}+R^{-}_{2}s^{+}_{rl})\nonumber\\
&+&
J_{4}\left(R_{3}s^{z}_{lr}+R_{4}s^{z}_{rl}\right)
+(J_5R^z_1+J_6R^z_2)s^z_r
+J_7(R^{+}_{1}s^{-}_r+R^{-}_{1}s^{+}_r).\label{int7m-s}
\end{eqnarray}
Here
\begin{eqnarray}
 J_{1}(\bar D)&=&J_{2}(\bar D) =\frac{2J^{T}_{l}}{3},\ \ \ \ \ J_{3l}(\bar
D)=J^{ST}_{l},\nonumber\\
J_{3r}(\bar D)&=&\alpha_r J_{lr},\ \ \ \ \ \ J_i(\bar D)=0 \ \ \
(i=4-7).
\end{eqnarray}
The operators ${\bf R}_{1}$, ${\bf R}_{2}$, $R_{3}$ and $R_{4}$
are defined as,
\begin{eqnarray}
R^z_{1}&=&\frac{1}{2}(X^{1_l1_l}-X^{S_lS_l}),\ \;
R^{+}_{1}=X^{1_lS_l}, \ \;
R^{-}_{1}=(R^{+}_{1})^{\dag},\nonumber\\
R^z_{2}&=&\frac{1}{2}(X^{1_l1_l}-X^{S_rS_r}),\
R^{+}_{2}=X^{1_lS_r}, \ \;
R^{-}_{2}=(R^{+}_{2})^{\dag},\nonumber\\
R_{3}&=&\frac{\sqrt 3}{2}X^{S_lS_r},\ \ \ \ \ \ \ \ \ \ \ \ \ \ \
\ R_{4}=\frac{\sqrt 3}{2}X^{S_rS_l}. \label{RU3}
\end{eqnarray}
We see that the anisotropic Kondo Hamiltonian (\ref{int7m-s}) is
quite unconventional. There are several different terms
responsible for transverse and longitudinal exchange involving the
R-operators which generate both S$_a$/T and S$_a$/S$_{\bar a}$
transitions.

The operators (\ref{RU3}) obey the following commutation
relations,
\begin{eqnarray}
&&\lbrack R_{1j},R_{1k}]=ie_{jkm}R_{1m},\ \ \ \ \ \ \;\;
[R_{2j},R_{2k}]=i e_{jkm}R_{2m}, \nonumber\\
&&\lbrack R_{1j},R_{2k}]=
\frac{\sqrt{3}}{6}(R_3-R_4)\delta_{jk}(1-\delta_{jz})\nonumber\\
&&\ \ \ \ \ \ \ \ \ \ \ \ \
+\frac{i}{2}e_{jkm}\Big(R_{1m}\delta_{kz}+
R_{2m}\delta_{jz}-\frac{\sqrt{3}}{3}\delta_{mz}(R_3+R_4)\Big), \nonumber\\
&&\lbrack R_{1j},R_{3}]=-\frac{1}{2}R_3\delta_{jz}+
\frac{\sqrt{3}}{4}(R_{2x}+i R_{2y})(\delta_{jx}-i\delta_{jy}),\nonumber\\
&&\lbrack R_{1j},R_{4}]=\frac{1}{2}R_4\delta_{jz}-
\frac{\sqrt{3}}{4}(R_{2x}-i R_{2y})(\delta_{jx}+i\delta_{jy}),\nonumber\\
&&\lbrack R_{2j},R_{3}]=\frac{1}{2}R_3\delta_{jz}-
\frac{\sqrt{3}}{4}(R_{1x}+i R_{1y})(\delta_{jx}+i\delta_{jy}),\nonumber\\
&&\lbrack R_{2j},R_{4}]=-\frac{1}{2}R_4\delta_{jz}+
\frac{\sqrt{3}}{4}(R_{1x}+i R_{1y})(\delta_{jx}-i\delta_{jy}),\nonumber\\
 &&\lbrack R_{3},R_{4}]=\frac{3}{2}(R_{2}^z-R_{1}^z).\label{comu3-s}
\end{eqnarray}
 These operators generate
the algebra $u_3$ in the reduced spin space $\{T1_l,S_l,S_r\}$
specified by the Casimir operator
$${\bf R}^2_1+{\bf R}^2_2+R^2_{3}+R^2_{4}=\frac{3}{2}.$$
Therefore, in this case the TQD possesses $SU(3)$ symmetry. These
$R$ operators may be represented via the familiar Gell-Mann
matrices $\lambda_i$ ($i=1,...,8$) for the $SU(3)$ group,
\begin{eqnarray*}
R^{+}_{1}&=&\frac{1}{2}\left(\lambda_1+i\lambda_2\right), \ \ \ \
\ \
R^{-}_{1}=\frac{1}{2}\left(\lambda_1-i\lambda_2\right), \\
R^z_1&=&\frac{\lambda_3}{2}, \ \ \ \ \ \ \ \ \ \ \ \ \ \ \ \ \ \
R^z_2=\frac{1}{4}(\lambda_3+\sqrt{3}\lambda_8),\\
R^{+}_{2}&=&\frac{1}{2}\left(\lambda_4+i\lambda_5\right), \ \ \ \
\ \
R^{-}_{2}=\frac{1}{2}\left(\lambda_4-i\lambda_5\right), \\
R_{3}&=&\frac{\sqrt{3}}{4}\left(\lambda_6+i\lambda_7\right), \ \ \
 R_{4}=\frac{\sqrt{3}}{4}\left(\lambda_6-i\lambda_7\right).
\end{eqnarray*}

As far as the RG procedure for the "Runge-Lenz" exchange
Hamiltonian (\ref{int7m-s}) is concerned, the poor-man scaling
procedure is applicable also for the $R$ operators. The scaling
equations have the form,
\begin{eqnarray}
&&\frac{dj_{1}}{d\ln d} = -2j_{3l}^2,\ \ \ \ \ \ \
 \frac{dj_{2}}{d\ln d} =-j_{3r}^2,  \nonumber\\
&&\frac{dj_{3l}}{d\ln d} =-\Big[j_{3l}\Big(j_1+\frac{j_2}{2}\Big)
 -\frac{\sqrt{3}}{4}j_{3r}j_4\Big],  \nonumber\\
&&\frac{dj_{3r}}{d\ln d}=-\frac{j_{3r}(j_1+2j_2
+j_5+2j_6)-\sqrt{3}j_{4}(j_{3l}+\sqrt{2}j_{7})}{4},\nonumber\\
&&\frac{dj_{4}}{d\ln
d}=j_{3r}\Big(\frac{\sqrt{3}}{3}j_{3l}+\frac{\sqrt{2}}{2}j_{7}\Big),
\nonumber\\
&&\frac{dj_{5}}{d\ln d} = -4j_{7}^2,\ \ \ \ \ \ \ \
 \frac{dj_{6}}{d\ln d} =-j_{3r}^2, \nonumber\\
&&\frac{dj_{7}}{d\ln d}
=-\Big[j_{5}j_7+\frac{j_6j_7}{2}-\frac{\sqrt{6}}{8}j_{3r}j_4\Big],
\label{sc7m}
\end{eqnarray}
where $j=\rho_0 J$. We cannot demonstrate analytical solution of
this system, but the numerical solution shows that stable infinite
fixed point exists in this case like in all previous
configurations.

Another type of field induced Kondo effect is realized in the
symmetric case of ${\delta}=E_{S_{g}}-E_{T_{g,u}}<0$. Now the
Zeeman splitting compensates negative $\delta$. Then the two
components of the triplets, namely $E_{T1_{g,u}}$ cross with the
singlet state energy $E_{S_{g}}$, and the symmetry group of the
TQD in magnetic field is $SU(3)$ as in the case considered above.

\subsection{Summary}
It has been demonstrated that the loss of rotational invariance in
external magnetic field radically changes the dynamical symmetry
of TQD. We considered here two examples of symmetry reduction,
namely $SO(5)\to SU(3)$ and $P\times SO(4)\times SO(4)\to SU(3)$.
In all cases the Kondo exchange is anisotropic, which, of course,
reflects the axial anisotropy induced by the external field. These
examples as well as the $SO(4)\to SU(2)$ reduction considered
earlier \cite{KA01,KA02} describe the magnetic field induced Kondo
effect owing to the dynamical symmetry of complex quantum dots.
Similar reduction $SO(n)\to SU(n')$ induced by magnetic field may
arise also in more complicated configurations, and in particular
in the parallel geometry. The immense complexity of  scaling
procedure adds nothing new to the general pattern of the field
induced anisotropy of Kondo tunneling, so we confine ourselves
with these two examples.

Although the anisotropic Kondo Hamiltonian was introduced formally
at the early stage of Kondo physics \cite{AYH,AYH1}, it was rather
difficult to perceive how such Hamiltonian is derivable from the
generic Anderson-type Hamiltonian. It was found that the effective
anisotropy arises in cases where the pseudo-spin degrees of
freedom (like a two-level system) are responsible for anomalous
scattering. Another possibility is the introduction of magnetic
anisotropy in the generic spin Hamiltonian due to spin-orbit
interaction (see Ref. \cite{AZ} for a review of such models). One
should also mention the remarkable possibility of magnetic field
induced anisotropic Kondo effect on a magnetic impurity in
ferromagnetic rare-earth metals with easy plane magnetic
anisotropy \cite{Sandal}. This model is close to our model from
the point of view of effective spin Hamiltonian, but the sources
of anisotropy are different in the two systems. In our case the
interplay between singlet and triplet components of spin multiplet
is an eventual source both of the Kondo effect itself and of its
anisotropy in external magnetic field. Previously, the
manifestation of $SU(3)$ symmetry in anisotropic magnetic systems
were established in Refs. \cite{Onufr,Onufr1}. It was shown, in
particular, that this dynamical symmetry predetermines the
properties of collective excitations in anisotropic Heisenberg
ferromagnet. In the presence of single-ion anisotropy the relation
between  the Hubbard operators for $S=1$ and Gell-Mann matrices
$\lambda$ were established. It worth also mentioning in this
context the $SU(4)\supset SO(5)$ algebraic structure of
superconducting and antiferromagnetic coherent states in cuprate
High-T$_c$ materials \cite{foot2}.

\section{Odd Occupation}\label{sec.4.4}
We now turn our attention to investigation of the dynamical
symmetries of TQD in series with odd occupation ${\cal N}=3$,
whose low-energy spin multiplet contains two spin 1/2 doublets
$|B_{1,2}\rangle$ and a spin quartet $|Q\rangle$ with
corresponding energies
\begin{eqnarray}
E_{B_1}&=&{\varepsilon}_c +{\varepsilon}_l +{\varepsilon}_r- \frac
{3}{2}\left[W_l\beta_l+W_r\beta_r \right], \nonumber \\
E_{B_2}&=&{\varepsilon}_c +{\varepsilon}_l +{\varepsilon}_r-
\frac{1}{2}\left[W_l\beta_l+W_r\beta_r\right], \nonumber \\
E_{Q}&=&{\varepsilon}_c +{\varepsilon}_l +\varepsilon_r.
\label{Edot}
\end{eqnarray}
There are also four charge-transfer excitonic counterparts of the
spin doublets separated by the charge transfer gaps $\sim
\varepsilon_l-\varepsilon_c +Q_l$ and $\varepsilon_r-\varepsilon_c
+Q_r$ from the above states (see Appendix \ref{diag}).

Like in the four-electron case, the scaling equations (\ref{DE-2})
may be derived with different tunneling rates for different spin
states ($\Gamma_{Q}$ for the quartet and $\Gamma_{B_{i}} (i=1,2)$
for the doublets).
\begin{eqnarray}
&&\Gamma_{Q} = \pi\rho_0 \left(V_{l}^2+ V_{r}^2\right), \ \ \ \ \
\Gamma_{B_1} = \gamma_1^2\Gamma_{Q}, \ \ \ \ \ \Gamma_{B_2} =
\gamma_2^2\Gamma_{Q},\label{Gamma}
\end{eqnarray}
with
\begin{equation}
\gamma_1=\sqrt{1-\frac{3}{2}\left(\beta_l^2+\beta_r^2\right)},~~
\gamma_2=\sqrt{1-\frac{1}{2}\left(\beta_l^2+\beta_r^2\right)}.\label{gam}
\end{equation}
\begin{figure}[htb]
\centering
\includegraphics[width=70mm,height=45mm,angle=0,]{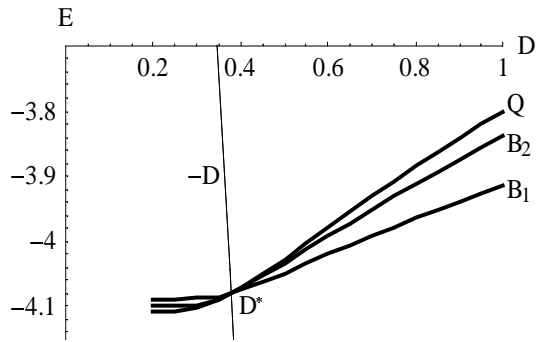}
\caption{Scaling trajectories resulting in $SO(4)\times SU(2)$
symmetry of TQD with ${\cal N}=3$.}\label{crossym}
\end{figure}
Since $\Gamma_{Q}>\Gamma_{B_1},\Gamma_{B_2 }$, the scaling
trajectories cross in a unique manner: This is the complete
degenerate configuration where all three phase trajectories
$E_{\Lambda}$ intersect
[$E_{Q}(D^{\star})=E_{B1}(D^{\star})=E_{B2}(D^{\star})$] at the
same point $D^{\star}.$ This happens at bandwidth $D=D^{\star}$
(Fig.\ref{crossym}) whose value is estimated as
\begin{eqnarray}
D^{\star} &=& D_0\exp \left(-\frac{\pi
r}{\Gamma_{Q}}\right),\label{Dc}
\end{eqnarray}
where
$$r=\frac{W_l^2E_{rc}+W_r^2E_{lc}}{W_l^2E_{rc}^2+W_r^2E_{lc}^2}E_{lc}E_{rc}.$$
This level crossing may occur either before or after reaching the
SW limit $\bar D$ where scaling terminates \cite{SW}. Below we
discuss the Kondo physics arising in the cases: $D^{\star}<{\bar
D}$, $D^{\star}={\bar D}$ and $D^{\star}>{\bar D}$.
\subsection{Towards Two-Channel Kondo Effect}
When $D^{\star}<{\bar D}$, the lowest renormalized state in the SW
limit is a doublet $B_1$. Following an RG procedure and a SW
transformation, the spin Hamiltonian in this case reads
\begin{eqnarray}
H_{cot} &=& J_l{\bf S}\cdot {\bf s}_l+J_r{\bf S}\cdot {{\bf
s}_r}+J_{lr} {\bf S}\cdot({\bf s}_{lr}+{\bf s}_{rl}). \label{Hs}
\end{eqnarray}
The spin 1/2 operator ${\bf S}$ acts on
$|B_{1\sigma=\uparrow,\downarrow} \rangle$
(Eq.(\ref{eg-func-trimer})), whereas the lead electrons spin
operators ${\bf s}_{a}$ and ${\bf s}_{a{\bar a}}$ are determined
in Eqs. (\ref{1.10}) (with $\alpha=a=l,r$) and (\ref{1.100}). The
exchange coupling constants are
\begin{eqnarray}
J_{a}= \frac{8 \gamma_1
V_{a}^{2}}{3(\epsilon_{F}-\varepsilon_{a})},\ \ \ J_{lr}=-\frac{4
\beta_l \beta_r V_{l}V_r}{3(\epsilon_{F}-\varepsilon_{a})}.
\end{eqnarray}
The Hamiltonian (\ref{Hs}) then encodes a two-channel Kondo
physics, where the leads serve as two independent channels and
$T_{K}=max\{T_{Kl},T_{Kr}\}$ with $T_{Ka}={\bar D}e^{-1/j_{a}}$
and $j_a=\rho_0 J_a$.

\begin{figure}[htb]
\centering
\includegraphics[width=80mm,height=45mm,angle=0,]{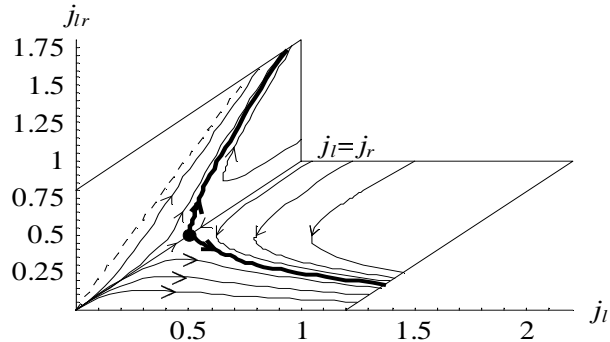}
\caption{Scaling trajectories for two-channel Kondo effect in
TQD.} \label{flow}
\end{figure}

A poor-man scaling technique is used to renormalize the exchange
constants by reducing the band-width ${\bar D} \to D$. The
pertinent fixed points are then identified as $D \to T_{K}$
\cite{Anderson}. Unlike the situation encountered in the
single-channel Kondo effect, third order diagrams in addition to
the usual single-loop ones should be included (see Fig. 5 in Ref.
\cite{NB} and Fig. 9 in Ref. \cite{Coleman}). With $a=l,r$ and
${\bar a}=r,l$ the three RG equations for $j_l, j_r, j_{lr}$ are
\begin{eqnarray}
&& \frac {d j_{a}} {d \ln d}=-(j_{a}^{2}+j_{lr}^2)+
j_{a}(j_{a}^{2}+j_{\bar{a}}^{2}+ 2 j_{lr}^{2}), \nonumber \\
&&\frac{d j_{lr}}{d\ln d}= -j_{lr}\left(j_l+j_r
\right)+j_{lr}(j_l^2+j_r^2+ 2j_{lr}^2). \label{RG3}
\end{eqnarray}
On the symmetry plane $j_l=j_r \equiv j$, Eqs. (\ref{RG3}) reduce
to a couple of RG equations for $j_{1,2}=j \pm j_{lr}$
\begin{eqnarray}
&&\frac{d j_i}{d\ln d}= -j_i^2+ j_i(j_1^2 + j_2^2) \ \ \ (i=1,2),
\label{RG4}
\end{eqnarray}
subject to $j_{i}(D={\bar D})\equiv j_{i0}=\rho_0 J_i$. These are
the well-known equations for the anisotropic two-channel Kondo
effect \cite{NB}. With $\phi_{i} \equiv (j_{1}+j_{2}-1)/j_{i}$,
$C_i \equiv \phi_{{\bar i}0}-\phi_{i0}$ ($\phi_{i0}=\phi_i({\bar
D})$) and $L_i(x) \equiv x-\ln(1+C_i/x)-2 \ln x$, the solution of
the system (\ref{RG4}) is
\begin{eqnarray}
&& L_i(\phi_{i})-L_i(\phi_{i0})=\ln\left(\frac{\bar D}{D}\right) \
\ \ (i=1,2). \label{RGS}
\end{eqnarray}
The scaling trajectories in the sector $(j_l\geq j_r \geq
0,j_{lr}=0)$ and in the symmetry plane with $0<j_{lr}<j$ are shown
in Fig.~\ref{flow}.
Although the fixed point (1/2, 1/2, 0) remains inaccessible if
$j_{lr} \ne 0$, one may approach it close enough starting from an
initial condition $j_{lr0}\ll j_{l0},j_{r0}$. Realization of this
inequality is a generic property of TQD in series shown in Fig.
\ref{TQD-s}. A similar scenario was offered in \cite{Kim,KSK} for
a QD between two interacting wires.

According to general perturbative expression for the dot
conductance \cite{kng}, its zero-bias anomaly is encoded in the
third order term,
\begin{eqnarray}
&& G^{(3)}=G_0 j_{lr}^{2}\left[j_{l}(T) + j_{r}(T) \right], \ \ \
(G_{0}=\frac{2 e^2} {h}). \label{Gpeak}
\end{eqnarray}
Here the temperature $T$ replaces the bandwidth $D$ in the
solution (\ref{RGS}). Let us present a qualitative discussion of
the conductance $G[j_{a}(T)]$ (or in an experimentalist friendly
form, $G(v_{ga},T)$) based on the flow diagram displayed in Fig.
\ref{flow}. (Strictly speaking, the RG method and hence the
discussion below, is mostly reliable in the weak-coupling regime
$T> T_{K}$). Varying $T$ implies moving on a curve
$[j_{l}(T),j_{r}(T),j_{lr}(T)]$ in three-dimensional parameter
space (Fig. \ref{flow}), and the corresponding values of the
exchange parameters determine the conductance according to
equation (\ref{Gpeak}). Note that if, initially,
$j_{l0}=j_{r0}\equiv j_{0}$ the point will remain on a curve
$[j(T),j(T),j_{lr}(T)]$ located on the symmetry plane. By varying
$v_{ga}$ it is possible to tune the initial condition
$(j_{l0},j_{r0})$ from the highly asymmetric case $j_{l0} \gg
j_{r0}$ to the fully symmetric case $j_{l0}=j_{r0}$. For a fixed
value of $j_{lr0}$ the conductance shoots up (logarithmically) at
a certain temperature $T^{*}$ which decreases toward $T_{K}$ with
$|j_{l0}-j_{r0}|$ and $j_{lr0}$. The closer is $T^{*}$ to $T_{K}$,
the closer is the behavior of the conductance to that expected in
a generic two-channel situation. Thus, although the isotropic
two-channel Kondo physics is unachievable in the strong-coupling
limit, its precursor might show up in the intermediate-coupling
regime.
\begin{figure}[htb]
\centering
\includegraphics[width=60mm,height=40mm,angle=0,]{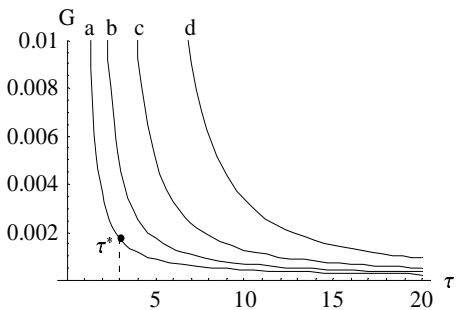}
\caption{Conductance $G$ in units of $G_{0}$ as a function of
temperature ($\tau=T/T_{K}$), at various gate voltages. The lines
correspond to:
 (a) the symmetric case $j_l=j_r$ ($v_{gl}=v_{gr}$),
(b-d) $j_l \gg j_r$, with $v_{gl}-v_{gr}=0.03,\ 0.06$ and $0.09$.
At $\tau \to \infty$ all lines converge to the bare conductance.}
\label{GT}
\end{figure}

\begin{figure}[htb]
\centering
\includegraphics[width=60mm,height=40mm,angle=0,]{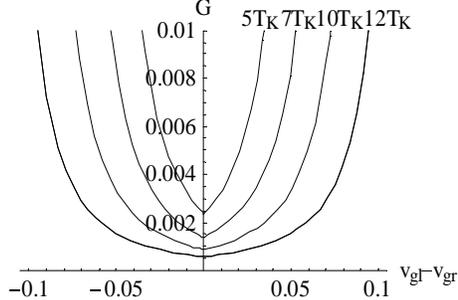}
\caption{Conductance $G$ in units of $G_{0}$ as a function of gate
voltage at various temperatures (at the origin $j_{l}=j_{r}$).}
\label{GV}
\end{figure}

The conductance $G(v_{ga},T)$ as function of $T$ for several
values of $v_{ga}$ and the same value of $j_{lr0}$ is displayed by
the family of curves in Fig. \ref{GT}. For $G$ displayed in curve
$a$, $T^{*}/T_{K}\approx 3$ and for $T > T^{*}$ it is very similar
to what is expected in an isotropic two-channel system.
Alternatively, holding $T$ and changing gate voltages $v_{ga}$
enables an experimentalist to virtually cross the symmetry plane.
This is equivalent to moving vertically downward on Fig. \ref{GT}.
At high temperature the curves almost coalesce and the conductance
is virtually flat. At low temperature (still above $T_{K}$) the
conductance exhibits a sharp minimum. This is summarized in Fig.
\ref{GV}.

\subsection{Higher Degeneracy and Dynamical Symmetries}
If the degenerate point (\ref{Dc}) occurs in the SW crossover
region, i.e., if $D^{\star}\approx \bar{D}$, the SW procedure
involves all three spin states (Fig. \ref{crossym}), and it
results in the following cotunneling Hamiltonian
\begin{eqnarray}
&&H_{cot}=\sum_{a=l,r}(J_{a}^{T}{\bf S} +J_{a}^{ST} {\bf R}
 )\cdot {\bf s}_a , \label{HSO4}
\end{eqnarray}
where ${\bf S}$ is the spin 1 operator and ${\bf R}$ is the
R-operator describing S/T transition similar to that for spin
rotator \cite{KA02}. The coupling constants are
\begin{eqnarray}
J_{a}^{T}= \frac{4 \gamma_1
V_{a}^{2}}{3(\epsilon_{F}-\varepsilon_{a})},\ \ \
J_{a}^{ST}=\gamma_2 J_{lr}^{T}.
\end{eqnarray}

This is a somewhat unexpected situation where Kondo tunneling in a
quantum dot with {\it odd} occupation demonstrates the exchange
Hamiltonian of a quantum dot with {\it even} occupation. The
reason for this scenario is the specific structure of the wave
function of TQD with ${\cal N}=3$. The corresponding wave
functions $|\Lambda\rangle$ (see Appendix \ref{diag}) are vector
sums of states composed of a "passive" electron sitting in the
central dot and singlet/triplet (S/T) two-electron states in the
$l,r$ dots. Constructing the eigenstates $|\Lambda \rangle$ using
certain Young tableaux (see Appendix D), one concludes that the
spin dynamics of such TQD is represented by the spin 1 operator
${\bf S}$ corresponding to the $l-r$ triplet, the corresponding
R-operator ${\bf R}$ and the spin 1/2 operator ${\bf s}_c$ of a
passive electron in the central well. The latter does not enter
the effective Hamiltonian $H_{cot}$ (\ref{HSO4}) but influences
the kinematic constraint via Casimir operator ${\cal K}={\bf
S}^{2}+{\bf M}^{2}+{\bf s}^{2}_c=\frac {15} {4}$. The dynamical
symmetry is therefore $SO(4) \times SU(2)$, and only the $SO(4)$
subgroup is involved in Kondo tunneling.

The scaling equations have the form,
\begin{eqnarray}
\frac{dj_{1a}}{d\ln d} &=&
-[j_{1a}^2+j_{2a}^2],\nonumber\\
\frac{dj_{2a}}{d\ln d} &=& -2j_{1a}j_{2a},
 \label{sc-42}
\end{eqnarray}
where $j_{1a}=\rho_0J_{a}^T$, $j_{2a}=\rho_0J_{a}^{ST}$ $(a=l,r)$.
From Eqs. (\ref{sc-42}) we obtain the Kondo temperature,
\begin{equation}
T_{K}=\rm{max}\{T_{Kl},T_{Kr}\}, \label{T3s}
\end{equation}
with $T_{Ka}={\bar D}\exp\left[-1/(j_{1a}+j_{2a})\right]$.

 An additional dynamical symmetry arises in the case when $D^\star>
\bar{D}$. In this case the ground state of TQD is a quartet S=3/2,
and we arrive at a standard underscreened Kondo effect for $SU(2)$
quantum dot as an ultimate limit of the above highly degenerate
state.

\subsection{Summary}
To conclude this section, it might be useful here to underscore
the following points: (1) In a TQD (Fig. \ref{TQD-s}), the
two-channel (left-right leads) Kondo Hamiltonian (\ref{Hs})
emerges in which the impurity is a {\it real} spin and the current
is due solely to co-tunneling. The corresponding exchange constant
$J_{lr}$ is a relevant parameter: by taking even and odd
combinations, the system is mapped on an anisotropic two-channel
Kondo problem where $J_{lr}$ determines the degree of anisotropy.
(2) Although the generic two-channel Kondo fixed-point is not
achievable in the strong coupling limit, inspecting the
conductance $G(v_{ga},T)$ as function of temperature (Fig.
\ref{GT}) and gate voltage (Fig. \ref{GV}) suggests an
experimentally controllable detection of its precursor in the weak
and intermediate coupling regimes. Apparently, genuine
multichannel Kondo regime with finite fixed point may be achieved
for configurations with more than two terminals. (3) There exists
a scenario of level degeneracy in which TQD with half-integer spin
behaves as a dot with integer spin in Kondo tunneling regime.

\section{Conclusions}\label{sec.4.5}

We have analyzed the occurrence of dynamical symmetries in
collective phenomena, which accompany quantum tunneling through
chemisorbed sandwich-type molecules and complex quantum dots in
configurations having a form of linear trimer. These symmetries
emerge when the trimer is coupled with metallic electrodes under
the conditions of strong Coulomb blockade in one of its three
constituents and nearly degenerate low energy spin spectrum.

As a prototype of complex molecules, where the magnetic ion is
sandwiched between two molecular radicals, we have chosen
lanthanocene molecule Ln(C$_8$H$_8)$$_2$ (Ln=Ce, Yb), following P.
Fulde's proposal \cite{Fulde}. This molecule is characterized by
anomalously soft spectrum of singlet-triplet excitations.
Artificial "Fulde molecule" in a form of linear triple dot (Fig.
\ref{TQD-s}) with even electron occupation mimics the low-energy
spectrum of cerocene. Fulde {\it et. al.} \cite{Dolg,Dolg1}
considered the interplay between two spin states (singlet and
triplet) and singlet charge exciton as a predecessor of genuine
Kondo singlet/triplet pairs, which arise in classical Kondo effect
for a local spin immersed into a Fermi sea. We have shown that the
Fulde trimer {\it as a whole} may be attached to a Fermi basin
(metallic layer or electrodes). As a result, the pseudo Kondo S/T
pairs becomes a new source of Kondo scattering/tunneling with
quite sophisticated structure of a ''scatterer''. This structure
is elegantly described in terms of non-compact dynamical symmetry
groups with complicated $o_n$ algebras. Application of an external
magnetic field results in additional accidental degeneracies. Due
to the loss of spin-rotational symmetry the effective cotunneling
Hamiltonian acquires spin anisotropy, and the dynamical symmetry
of a trimer is radically changed. Although the main focus in this
Chapter is related to the study of triple quantum dots, the
generalization to other quantum dot structures is indeed
straightforward.

The difference between series and parallel geometries of TQD
coupled to the leads by two channels exists only at non-zero
interchannel mixing in the leads, $t_{lr}\neq 0$. One may control
the dynamical symmetry of Kondo tunneling through TQD by varying
the gate voltage and/or lead-dot tunneling rate. In the case of
odd electron occupation (${\cal N}=3$) when the ground-state of
the isolated TQD is a doublet and higher spin excitations can be
neglected, the effective low-energy Hamiltonian of a TQD in series
manifests a two-channel Kondo problem albeit {\it only in the weak
coupling regime} \cite{EPL}. To describe the flow diagram in this
case, one should go beyond the one-loop approximation in RG flow
equations \cite{NB}. The nominal spin of CQD does not necessarily
coincide with that involved in Kondo tunneling. A simple albeit
striking realization of this scenario in this context is the case
of TQD with ${\cal N}=3$, which manifests itself as a dot with
integer or half-integer spin (depending on gate voltages).

Since we were interested in the symmetry aspect of Kondo tunneling
Hamiltonian, we restricted ourselves by derivation of RG flow
equations and solving them for obtaining the Kondo temperature. In
all cases the TQDs possess strong coupling fixed point
characteristic for spin 1/2 and/or spin 1 case. We did not
calculate the tunnel conductance in details, because it reproduces
the main features of Kondo-type zero bias anomalies studied
extensively by many authors (see, e.g.,
\cite{Magn,Magn-v1,Magn-v2,Magn-v3,Pust,Eto02}). The novel feature
is the possibility of changing $T_K$ by scanning the phase diagram
of Fig. \ref{phasediag}. Then the zero bias anomaly follows all
symmetry crossovers induced by experimentally tunable gate
voltages and tunneling rates.

The main message of this Chapter is that symmetry enters the realm
of mesoscopic physics in a rather non-trivial manner. Dynamical
symmetry in this context is not just a geometrical concept but,
rather, intimately related with the physics of strong correlations
and exchange interactions. The relation with other branches of
physics makes it even more attractive. The groups $SO(n)$ play an
important role in Particle Physics as well as in model building
for high temperature superconductivity (especially $SO(5)$). The
role of the group $SU(3)$ in Particle Physics cannot be
overestimated and its role in Nuclear Physics in relation with the
interacting Boson model is well recognized. This work extends the
role of these Lie groups in Condensed Matter Physics.

\chapter{Kondo Tunneling through Triangular Trimer in Ring Geometry}
{\it In this Chapter we consider a ring-like triangular trimer,
i.e., triangular triple quantum dot (TTQD), and focus on its
symmetry properties, which influence the Kondo tunneling. The
basic concepts are introduced in Sec. \ref{sec.5.1}. In Sec.
\ref{sec.5.2} we construct the Hamiltonian of the TTQD both in
three- and two-terminal geometry and expose its energy spectrum.
In Sec. \ref{sec.5.3} we focus on the modification of the symmetry
of TTQD in an external magnetic field. It is shown that TTQD in a
magnetic field demonstrates unique combination of Kondo and
Aharonov-Bohm features. The poor-man scaling equations are solved
and the Kondo temperatures are calculated for the cases of $SU(2)$
and $SU(4)$ symmetries. The conductance of a TTQD strongly depends
on the underlying dynamical symmetry group. We show that the
interplay between continuous spin-rotational symmetry $SU(2)$,
gauge symmetry $U(1)$ and discrete symmetry $C_{3v}$ of triangle
may result in sharp enhancement or complete suppression of tunnel
conductance as a function of magnetic flux through the TTQD. In
Conclusions we summarize the results obtained.}

\section{Introduction}\label{sec.5.1}

In the previous chapters we studied tunneling through linear
trimers in parallel and serial configurations, where the dots are
ordered linearly either parallel or perpendicular to the metallic
leads. Meanwhile, modern experimental methods allow also
fabrication of quantum dots in a ring geometry. This ring may have
a form of closed gutter \cite{gutter,gutter1} or be composed by
several separate dots coupled by tunnel channels. In the latter
case the simplest configuration is a triangle. Triangular triple
quantum dot (TTQD) was considered theoretically \cite{stop1} and
realized experimentally very recently \cite{stop2}, in order to
demonstrate the ratchet effect in single electron tunneling. To
achieve this effect the authors proposed a configuration, where
two of the three puddles are coupled in series with the leads
(source and drain), while the third one has a tunnel contact with
one of its counterparts and only a capacitive coupling with the
other. From the point of view of Kondo effect such configuration
can be treated as an extension of a T-shape quantum dot
\cite{KA01,KA02,Tshap,Tshap1,Tshap2}. Triangular trimers of Cr
ions on a gold surface were also studied
\cite{Cr-ex,Cr-th,Cr-th1,Cr-th2}. The electronic and magnetic
structure of these trimers is described in terms of a three-site
Kondo effect. The orbital symmetry of triangle is discrete. It
results in additional degeneracies of the spectrum of trimer,
which may be the source of non-Fermi-liquid fixed point
\cite{Cr-th2,Cr-th3}.

In this Chapter we concentrate on the {\it point symmetry} of
ring-like TTQD and its interplay with the {\it spin rotation
symmetry} in a context of Kondo tunneling through this artificial
molecule. Indeed, the generic feature of Kondo effect is the
involvement of internal degrees of freedom of localized
"scatterer" in the interaction with continuum of electron-hole
pair excitations in the Fermi sea of conduction electrons. These
are spin degrees of freedom in conventional Kondo effect, although
in some cases the role of pseudospin may be played by
configuration quantum numbers, like in two-level systems and
related objects \cite{AZ}. TTQD may be considered as a specific
Kondo object, where both spin and configuration (orbital)
excitations are involved in cotunneling on an equal footing
\cite{TTQD,TTQD-AB}.

\begin{figure}[htb]
\centering
\includegraphics[width=110mm,height=34mm,angle=0]{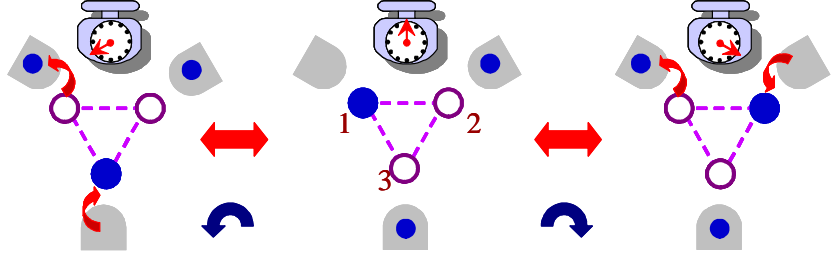}
\caption{Clockwise (c) and anti-clockwise (a) "rotation" of TTQD
due to cotunneling through the channels $2$ and $3$,
respectively.}\label{tik-tak}
\end{figure}

To demonstrate this interplay, we consider a fully symmetric TTQD
consisting of three identical puddles with the same individual
properties (energy levels and Coulomb blockade parameters) and
inter-dot coupling (tunnel amplitudes and electrostatic
interaction). Like in the above mentioned triangular ratchet
\cite{stop1,stop2}, we assume that the TTQD in the ground state is
occupied by one electron and Coulomb blockade is strong enough to
completely suppress double occupancy of any valley $j=1,2,3$. This
means that the only mechanism of electron transfer through TTQD is
cotunneling, where one electron leaves the valley $j$ into the
metallic leads, whereas another electron tunnels from the
reservoir to the same valley $j$ or to another valley $l$. In the
former case only the spin reversal is possible, whereas in the
latter case not only the spin is affected but also the TTQD is
effectively "rotated" either clockwise or anti-clockwise (see Fig.
\ref{tik-tak}).

\begin{figure}[htb]
\centering
\includegraphics[width=80mm,height=60mm,angle=0]{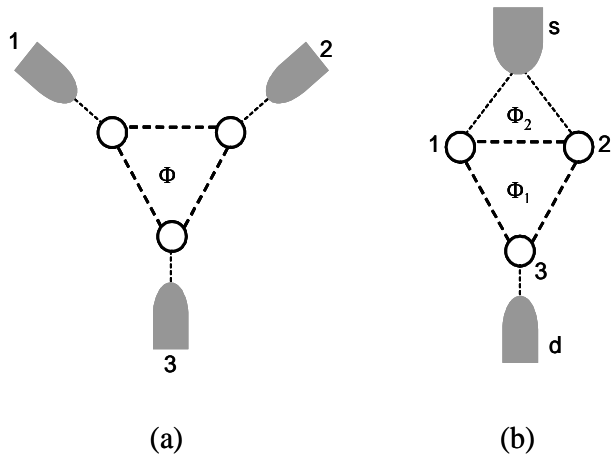}
\caption{Triangular triple quantum dot (TTQD) in three-terminal
(a) and two-terminal (b) configurations.}\label{figTQD}
\end{figure}

Discrete rotation in real space and continuous rotation in spin
space may be encoded in terms of group theory. The group $C_{3v}$
characterizes the symmetry of a triangle, and the group $SU(2)$
describes the spin symmetry. As a result, the total symmetry of
TTQD is determined by the direct product of these two groups. One
may use an equivalent language of permutation group $P_3$ for
description of the configurations of TTQD with an electron
occupying one of three possible positions in its wells. The
discrete group $P_3$ is characterized by three representations
$A,B,E$ or three Young tableaux $[3],[1^3],[21]$. Here $A$ and $B$
are one-dimensional representations with symmetric and
antisymmetric basis functions, respectively, and $E$ is the
two-dimensional representation with basis functions, which are
symmetric or antisymmetric with respect to mirror reflections. In
the configuration shown in Fig. \ref{figTQD}a, the perfect
triangular symmetry characterizes both the triple dot and three
leads, whereas in the configuration illustrated by Fig.
\ref{figTQD}b, the permutation symmetry between the sites (2,3)
and (1,3) is broken, while the system remains invariant relative
to (1,2) permutation.

Since the TTQD has a ring geometry, it is sensitive to an external
magnetic field directed perpendicular to its plane, because the
electron acquires additional gauge phase as it tunnels between the
wells. The phase is determined by the magnetic flux through the
ring. In a two-terminal geometry (Fig. \ref{figTQD}b), the
Aharonov-Bohm (AB) effect is possible due to the interference
between the channels (s13d) and (s23d). Since the magnetic flux in
this case is separated between two rings, the AB oscillations
should have more complicated character than in a text-book AB
interferometer. The $U(1)$ symmetry of an electron in an external
magnetic field also influences the Kondo resonance because it
changes the total symmetry of TTQD (breaks the chiral symmetry).
In this Chapter we show that the interplay between continuous
$SU(2)$ symmetry, discrete $C_{3v}$ symmetry and gauge $U(1)$
symmetry can be described in terms of {\it dynamical symmetry
group} approach \cite{KA01,KA02,KKAv,KKA02,KKA04} discussed in
Chapters 3 and 4.

It should be stressed that the interplay between the Kondo
tunneling and AB interference differs from the effects considered
recently in the context of mesoscopic quantum interferometer
containing a Kondo impurity on one of its arms
\cite{Fano,Fano1,Fano2,Fano3,Fano4}. Unlike the Fano-like effects
which take place in the latter case, the coherent TTQD {\it as a
whole} plays part in the physics of an AB interferometer and a
resonance Kondo-scatterer.

\section{Hamiltonian}\label{sec.5.2}

A symmetric TTQD in a contact with source and drain leads (Fig.
\ref{figTQD}) is described by the Anderson Hamiltonian for lead
electrons $c_{kj\sigma}$ and dot electrons $d_{j\sigma}$,
\begin{equation}
H=H_{d}+H_{lead}+H_t. \label{H-sys}
\end{equation}
The first term, $H_d$, is the Hamiltonian of the isolated TTQD,
\begin{eqnarray}
H_{d}=\epsilon\sum_{j=1}^3\sum_{\sigma}d^\dagger_{j
\sigma}d_{j\sigma }+Q\sum_{j}n_{j\uparrow}n_{j\downarrow}
+Q'\sum_{\langle jl\rangle}\sum_\sigma n_{j\sigma}n_{l\sigma'}
+W\sum_{\langle jl\rangle}\sum_{\sigma}(d^\dagger_{j
\sigma}d_{l\sigma }+H.c.), \label{H-dot}
\end{eqnarray}
 where ${\sigma}=\uparrow,\downarrow$ is the spin index, $\langle jl \rangle=
\langle12\rangle,\langle23\rangle,\langle31\rangle$. Here $Q$ and
$Q'$ are intra-dot and inter-dot Coulomb blockade parameters
$(Q\gg Q')$, and $W$ is the inter-dot tunneling parameter. The
second term, $H_{lead}$, describes the electrons in the leads
labelled by the same indices $j=1,2,3$ as the dots in the case of
Fig. \ref{figTQD}a. In the case of Fig. \ref{figTQD}b $j=s,d$ for
source and drain electrodes respectively,
\begin{eqnarray}
H_{lead}&=&\sum_{k\sigma}\sum_{j}\epsilon_{kj}c^\dagger_{kj
\sigma}c_{kj\sigma}. \label{H-l}
\end{eqnarray}
The last term, $H_t$, is the tunneling Hamiltonian between the dot
and the leads. It has the form
\begin{eqnarray}
H_t&=&\sum_{k\sigma}\sum_{j=1,2,3}\left(V_{j}c^\dagger_{kj\sigma}
d_{j\sigma}+H.c.\right),\label{H-tun-3t}
\end{eqnarray}
in the 3-terminal geometry (Fig. \ref{figTQD}a) and
\begin{eqnarray}
H_t&=&\sum_{k\sigma}\sum_{j=1,2}\left(V_{sj}c^\dagger_{ks\sigma}d_{j\sigma}+
V_dc^\dagger_{kd\sigma}d_{3\sigma} +H.c.\right)\label{H-tun-2t}
\end{eqnarray}
in the 2-terminal geometry (Fig. \ref{figTQD}b). We assume that in
the latter case the mean-free path for electrons near the "tip" of
the source electrode exceeds the size of this tip. The 2-terminal
device has the symmetry of isosceles triangle with one mirror
reflection axis $(1\leftrightarrow 2)$.

First we consider TTQD with three leads (Fig. \ref{figTQD}a) whose
ground state corresponds to a single electron occupation ${\cal
N}=1,$ and assume that all three channels are equivalent with
$V\ll W$ so that the tunnel contact preserves the rotational
symmetry of TTQD, which is thereby imposed on the itinerant
electrons in the leads. It is useful to re-write the Hamiltonian
in the special basis which respects the $C_{3v}$ symmetry,
employing an approach widely used in the theory of Kondo effect in
bulk metals \cite{CqS,CoqC}.
 The Hamiltonian $H_d+H_{lead}$
is diagonal in the basis
\begin{eqnarray}
    &d^\dag_{A,\sigma}=\displaystyle{\frac{1}{\sqrt{3}}}
\left(d^\dag_{1\sigma}+d^\dag_{2\sigma}+d^\dag_{3\sigma}\right),\
\ \ d^\dag_{E_\pm,\sigma}=
\frac{1}{\sqrt{3}}\left(d^\dag_{1\sigma}+e^{\pm
2i\varphi}d^\dag_{2\sigma}
    +e^{\pm i\varphi}d^\dag_{3\sigma}\right); \label{A1}\\
&c^\dag_{A,{k}\sigma}
 =\displaystyle{\frac{1}{\sqrt{3}}}
  \left(
      c^\dag_{1{k}\sigma}+
      c^\dag_{2{k}\sigma}+
      c^\dag_{3{k}\sigma}
 \right),\ \ \  c^\dag_{E_{\pm},{k}\sigma}
 =\frac{1}{\sqrt{3}}
  \left(
      c^\dag_{1{k}\sigma}+
      e^{\pm 2i\varphi}c^\dag_{2{k}\sigma}+
      e^{\pm i\varphi}c^\dag_{3{k}\sigma}\right).
\label{partw}
\end{eqnarray}
Here $\varphi=2\pi/3$, while $A$ and $E$ form bases for two
irreducible representations of the group $C_{3v}$. Only a
symmetric representation $A$ of the $P_3$ group arises in this
case. The antisymmetric state $B$ cannot be constructed due to the
well known frustration property of triangular cells. The spin
states with ${\cal N}=1$ are spin doublets ($D$), so the
Hamiltonian of the isolated TTQD in this charge sector has six
eigenstates. They correspond to a spin doublet $|A\rangle$ with
fully symmetric "orbital" wave function $(A)$ and two degenerate
doublets $|E_\pm\rangle$. The corresponding single electron
energies are,
\begin{equation}\label{energy-1}
E_{DA}=\epsilon+2W, ~~~~~~~~~~~E_{DE}=\epsilon-W.
\end{equation}

The Anderson Hamiltonian (\ref{H-sys}) rewritten in the variables
(\ref{A1}) and (\ref{partw}) may be expressed by means of Hubbard
operators $X^{\lambda\lambda'}=|\lambda\rangle\langle \lambda'|$,
with $\lambda=0,\Gamma,\Lambda$:
\begin{eqnarray}
H= \sum_\lambda E_\lambda X^{\lambda\lambda}+
\sum_{k\sigma}\sum_{\Gamma,k} \varepsilon_{k}n_{\Gamma,k}
+\sum_{\Gamma,k\sigma} \Big[V^{\Gamma 0} X^{\Gamma 0}
c_{\Gamma,k\sigma}+\sum_{\Lambda\Gamma'}
V^{\Gamma\Lambda}c^\dagger_{\Gamma',k\sigma}X^{\Gamma\Lambda}+H.c.\Big].\label{hah}
\end{eqnarray}
Here $|0\rangle$ stands for an empty TTQD,
$|\Gamma\rangle=|DA\rangle,~|DE\rangle$ belong to the single
electron charge sector, and $|\Lambda\rangle$ are the eigenvectors
of two-electron states. The eigenstates $E_\Lambda$ for ${\cal
N}=2$ are
\begin{eqnarray}
&&E_{SA} ={\epsilon_2}+2W- \frac{8W^2}{Q}, \ \ \ \ \ \ \
E_{TE} = {\epsilon_2} + W, \nonumber \\
&&E_{SE} = {\epsilon_2}- W - \frac{2W^2}{Q}, \ \ \ \ \ \ \
E_{TB} = {\epsilon_2} -2W  .\label{energy-n}
\end{eqnarray}
Here $\epsilon_2=2\epsilon+Q'$, the indices $S,T$ denote spin
singlet and spin triplet configurations of two electrons in TTQD,
and the inequality $W\ll Q$ is used explicitly. The irreducible
representation $B$ contains two-electron eigenfunction, which is
odd with respect to permutations $j \leftrightarrow l$. The tunnel
matrix elements are redefined accordingly,
$$V^{0\Gamma}=V\langle
0|d_{\Gamma,\sigma}|\Gamma\rangle, \ \ \
V^{\Gamma\Lambda}=V\langle
\Gamma|d_{\Gamma',\sigma}|\Lambda\rangle.$$
A peculiar feature of ring configuration is an {\it explicit
dependence of the order of levels} within the multiplets
(\ref{energy-1}) and (\ref{energy-n}) on the sign of the tunnel
integral $W$. For ${\cal N}=1$ the doublet $|DA\rangle$ is a
ground state, provided $W<0$. In the case of $W>0$ the lowest
levels are the doublets $|DE_\pm\rangle$. The orbital degeneracy
of the states $E_\pm$ is a manifestation of rotation/permutation
degrees of freedom of the TTQD. In the next section we will show
that these discrete rotations are explicitly involved in Kondo
tunneling.

\section{Magnetically Tunable Spin and Orbital Kondo
Effect}\label{sec.5.3}

\begin{figure}[htb]
\centering
\includegraphics[width=80mm,height=100mm,angle=0]{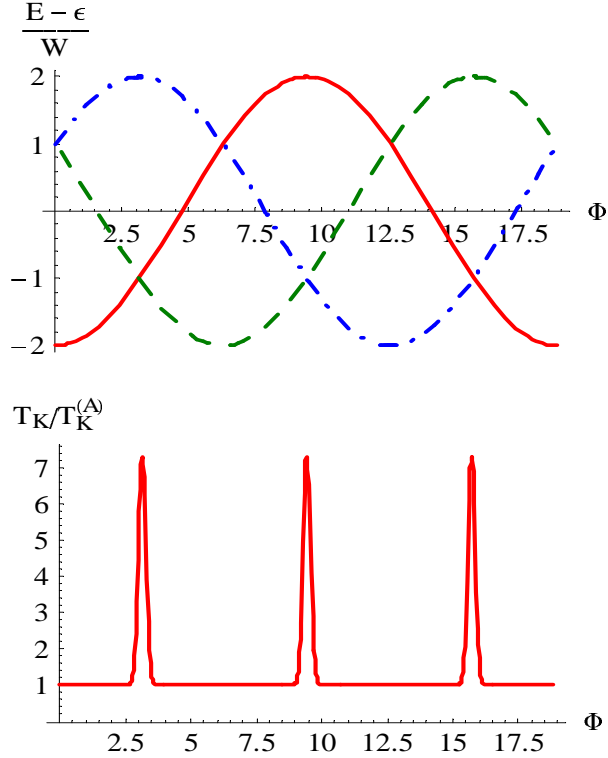}
\caption{Upper panel: Evolution of the energy levels $E_A$ (solid
line) and $E_\pm$ (dashed and dash-dotted line, respectively.)
Lower panel: corresponding evolution of Kondo
temperature.}\label{T-phi}
\end{figure}

To describe the influence of an external magnetic field $B$
(perpendicular to the TTQD plane) on the dot spectrum, one may
treat the TTQD as a three-site cyclic chain with nearest-neighbor
hopping integrals $-|W|$ connecting these sites. The spectrum of
this "chain" is
\begin{eqnarray}\label{Ep}
E_{D\Gamma}(p)&=&\epsilon-2W\cos p, \ \ \ p=0,~2\pi/3,~4\pi/3.
\end{eqnarray}
This equation is the same as (\ref{energy-1}) with negative $W$,
where $p=0,~2\pi/3,~4\pi/3$ correspond respectively to $\Gamma
=A,E_+,E_-$. A perpendicular magnetic field modifies this
spectrum. If the Zeeman splitting is weak, the only effect of the
magnetic field is reflected by an additional phase acquired by the
tunneling integral $W$. This phase is determined by the magnetic
flux $\Phi$ through the triangle, so that the spectrum of the TTQD
becomes,
\begin{equation}\label{emag}
E_{D\Gamma}(p,\Phi)=\epsilon-2W\cos \left(p-\frac{\Phi}{3}\right).
\end{equation}
Figure \ref{T-phi} illustrates the evolution of
$E_{D\Gamma}(p,\Phi)$ induced by $B$. Variation of $B$ between
zero and $B_0$ (the value of $B_0$ corresponding to the quantum of
magnetic flux $\Phi_0$ through the triangle) results in multiple
crossing of the levels $E_{D\Gamma}$. The periodicity in $\Phi$ of
the energy spectrum is $2\pi$ since each crossing point (Fig.
\ref{T-phi}) may be considered as an orbital doublet $E_{\pm}$ by
regauging phases $\varphi$ in Eq. (\ref{A1}).

The accidental degeneracy of spin states induced by the magnetic
phase $\Phi$ introduces new features into the Kondo effect.
 In the conventional Kondo problem, the effective low-energy exchange
Hamiltonian has the form $J{\bf S} \cdot {\bf s}$, where ${\bf S}$
and ${\bf s}$ are the spin operators for the dot and lead
electrons, respectively \cite{Anderson}. Here, however, the
low-energy states of TTQD form a multiplet characterized by both
spin and orbital quantum numbers. The effective exchange
interaction reflects the {\it dynamical symmetry} of the TTQD
\cite{KA01,KA02,KKAv,KKA02,KKA04}. As was mentioned in Sec.
\ref{sec.2.4}, the corresponding dynamical symmetry group is
identified not only by the operators which commute with the
Hamiltonian but also by operators inducing transitions between
different states of its multiplets. Hence, it is determined by the
set of dot energy levels which reside within a given energy
interval (its width is related to the Kondo temperature $T_{K}$).
Since the position of these levels is controlled in this case by
the magnetic field, we arrive at a remarkable scenario: Variation
of a magnetic field determines the dynamical symmetry of the
tunneling device. Generically, the dynamical symmetry group which
describes all possible transitions within the set $\{DA,DE_\pm\}$
is $SU(6)$. However, this symmetry is exposed at too high energy
scale $\sim W$, while only the low-energy excitations at energy
scale $T_K\ll W$ are involved in Kondo tunneling. It is seen from
Fig. \ref{T-phi} that the orbital degrees of freedom are mostly
quenched, but the ground state becomes doubly degenerate both in
spin and orbital channels around $\Phi=(2n+1) \pi$, $(n=0,\pm
1,\ldots)$.

Let us compare two cases:  $\Phi=0$, (the ground state is a spin
doublet ${DA}$), and $\Phi=\pi$, (the ground state is both orbital
and spin doublet $DE$). It is useful at this point to generalize
the notion of localized spin operator ${ S}^i= |\sigma\rangle \hat
\tau_i \langle\sigma'|$ (employing Pauli matrices $\hat\tau_i~
(i=x,y,z)$) to ${S}^i_{\Gamma\Gamma'}= |\Gamma\sigma\rangle \hat
\tau_i \langle\sigma'\Gamma'|$ ($\Gamma,\Gamma'=A,E_\pm$), in
terms of the eigenvectors (\ref{A1}). Similar generalization
applies for the spin operators of the lead electrons:
${s}^i_{\Gamma\Gamma'}=\sum_{kk'}c^\dag_{\Gamma,k\sigma}
\hat\tau_i c_{\Gamma',k'\sigma'}$ with $c_{\Gamma,k\sigma}$
defined in (\ref{partw}). First, when $\Phi=0$, the rotation
degrees of freedom are quenched at the low-energy scale. The only
vector, which is involved in Kondo co-tunneling through TTQD is
the spin ${\bf S}_{AA}$. Applying SW procedure \cite{SW}, the
effective exchange Hamiltonian reads,
\begin{eqnarray}\label{SWA}
 H_{SW} =
 J_E\left({\bf S}_{AA}\cdot{\bf s}_{E_+E_+}+
 {\bf S}_{AA}\cdot{\bf s}_{E_-E_-}\right)+
 J_{A}{\bf S}_{AA}\cdot{\bf s}_{AA}.
\end{eqnarray}
The exchange vertices $J_\Gamma$ are
\begin{eqnarray}\label{coupl-Ja}
 J_E &=&-\frac{2V^2}{3}
   \left(\frac{1}{\epsilon + Q'-\epsilon_F}
   -\frac{1}{\epsilon+Q- \epsilon_F}\right), \label{new-J}\\
  J_{A}&=& \frac{2V^2}{3}
   \left(\frac{3}{\epsilon_F
   -\epsilon}+
   \frac{1}{\epsilon+Q-\epsilon_F}
   +\frac{2}{\epsilon+Q'-\epsilon_F}\right).\nonumber
\end{eqnarray}
Note that $J_A>0$ as in the conventional SW transformation of the
Anderson Hamiltonian. On the other hand, $J_E<0$ due to the
inequality $Q\gg Q'$. Thus, two out of three available exchange
channels in the Hamiltonian (\ref{SWA}) are irrelevant. As a
result, the conventional Kondo regime emerges with the doublet
$DA$ channel and a Kondo temperature,
\begin{eqnarray}
T_{K}^{(A)}={\bar
D}\exp{\left\{-\frac{1}{j_A}\right\}}.\label{TKA}
\end{eqnarray}
where $j_A=\rho_0 J_A$.

\noindent Second, when $\Phi=\pi$, the doublet ${DA}$ is quenched
at low energy, and the Kondo effect is governed by tunneling
through the TTQD in the state $|DE\rangle$ whose symmetry is
$SU(4)$. This scenario of {\em orbital} degeneracy is different
from that of {\em occupation} degeneracy studied in double quantum
dot systems \cite{Bord03}. The 15 generators of $SU(4)$ include
four spin vector operators ${\bf S}_{E_aE_b}$ with $a,b=\pm$ and
one pseudospin vector ${\boldsymbol {\cal T}}$ defined as
\begin{eqnarray}
{\cal T}^{+}=\sum_\sigma |E_+,\sigma\rangle\langle \sigma,E_-|,\ \
\ {\cal T}^z=\frac{1}{2}\sum_\sigma\left(
|E_+,\sigma\rangle\langle \sigma,E_+|-|E_-,\sigma\rangle\langle
\sigma,E_-| \right).
\end{eqnarray}
Its counterpart for the lead electrons is
\begin{eqnarray}
 &&\tau^{+} =\sum_\sigma
         c^{\dag}_{E_+,k\sigma}c_{E_-,k\sigma},\ \ \
         \tau_z =\frac{1}{2}
  \sum_\sigma
  (c^{\dag}_{E_+,k\sigma}
       c_{E_+,k\sigma}
       -
       c^{\dag}_{E_-,k\sigma}
       c_{E_-,k\sigma}).
\end{eqnarray}

 Due to $SU(4)$ symmetry of the ground state, the
SW Hamiltonian acquires a rather rich structure,
\begin{eqnarray}
&&H_{SW}=J_1({\bf S}_{E_+E_+}\cdot {\bf s}_{E_+E_+}+{\bf
S}_{E_-E_-}\cdot {\bf s}_{E_-E_-})\nonumber\\
&&+J_2({\bf S}_{E_+E_+}\cdot {\bf s}_{E_-E_-}+{\bf
S}_{E_-E_-}\cdot {\bf
s}_{E_+E_+})
+J_3({\bf S}_{E_+E_+}+{\bf S}_{E_-E_-})\cdot {\bf
s}_{AA}\nonumber\\
&&+J_4({\bf S}_{E_+E_-}\cdot {\bf s}_{E_-E_+}+{\bf
S}_{E_-E_+}\cdot {\bf
s}_{E_+E_-})\label{SWE}\\
&&+J_5({\bf S}_{E_+E_-}\cdot ({\bf s}_{AE_-}+{\bf s}_{E_+A})+{\bf
S}_{E_-E_+}\cdot ({\bf s}_{AE_+}+{\bf s}_{E_-A}))
+ J_6 \boldsymbol {\cal T}\cdot{\boldsymbol \tau}
,\nonumber
\end{eqnarray}
where the coupling constants are
\begin{eqnarray}
 &&J_1=J_4=J_A,\ \ \ \
J_2=J_3=J_5
 =J_E,\ \ \ \ J_6=
 \frac{V^2}{{\epsilon_F
   -\epsilon}}+
 \frac{V^2}{\epsilon+Q'-\epsilon_F}.\label{J-E}
\end{eqnarray}
Thus, spin and orbital degrees of freedom of TTQD interlace in the
exchange terms. The indirect exchange coupling constants arise due
to co-tunneling processes with virtual excitations of states with
zero and two electrons. These constants include both diagonal
($jj$) and non-diagonal ($jl$) terms describing reflection and
transmission co-tunneling amplitudes. (Our representation of spin
operators and therefore the form of spin Hamiltonians
(\ref{SWA}),(\ref{SWE}) differs from that used in
\cite{Cr-th2,Cr-th3} and in \cite{Zarand1,Zarand}.)

\begin{figure}[htb]
\centering
\includegraphics[width=80mm,height=50mm,angle=0]{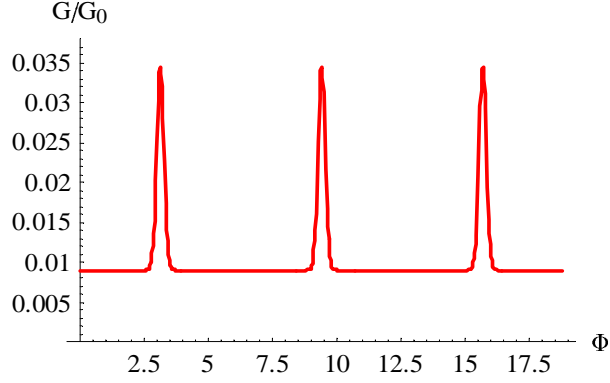}
\caption{Evolution of conductance ($G_0=\pi e^2/\hbar
$).}\label{C-phi}
\end{figure}

The interplay between spin and pseudospin channels naturally
affects the scaling equations obtained within the framework of
Anderson's "poor man scaling" procedure \cite{Anderson}. The
system of scaling equations has the following form:
\begin{eqnarray}
&& \frac{dj_1}{dt} =
 -\left[j_1^2+\frac{j_4^2}{2}+j_4j_6+\frac{j_5^2}{2}\right],\ \ \
 \
 \frac{dj_2}{dt} =
 -\left[j_2^2+\frac{j_4^2}{2}-j_4j_6+\frac{j_5^2}{2}\right],
 \nonumber \\
&& \frac{dj_3}{dt} =
 -\left[j_3^2+j_5^2\right],
  \ \ \ \ \ \ \
 \frac{dj_4}{dt} =
 -\left[j_4(j_1+j_2+j_6)+j_6(j_1-j_2)\right],
 \nonumber\\
&& \frac{dj_5}{dt} =
 -\frac{j_5}{2}\left[j_1+j_2+j_3-j_6\right],\ \ \ \
 \frac{dj_6}{dt} =
 -j_6^2. \label{scaling} 
 \end{eqnarray}
Here $j_i=\rho_0 J_i$, and the scaling variable is $t=\ln (\rho_0
D)$. Analysis of solutions of the scaling equations
(\ref{scaling}) with initial values of coupling parameters listed
in Eq. (\ref{J-E}), shows that the symmetry-breaking vertices
$j_{3}$ and $j_{5}$ are irrelevant, but the vertex $j_2$, which is
negative at the boundary $D =\bar D$ evolves into positive domain
and eventually enters the Kondo temperature. The latter is given
by the following equation
\begin{equation}\label{kondoe}
T_K^{(E)}={\bar
D}\exp\left\{{-\frac{2}{j_1+j_2+\sqrt{2}j_4+2j_6}}\right\}.
\end{equation}
We see from (\ref{kondoe}) that both spin and pseudospin exchange
constants contribute on an equal footing. Unlike the isotropic
Kondo Hamiltonian for ${\cal N}=3$ discussed in Refs.
\cite{Cr-th2,Cr-th3}, the non-Fermi-liquid regime is not realized
for ${\cal N}=1$ with $H_{SW}$ (\ref{SWE}). The reason of this
difference is that starting with the Anderson Hamiltonian
(\ref{H-sys}) with finite $Q,Q'$, one inevitably arrives at the
anisotropic SW exchange Hamiltonian for any $\cal N$. As a result,
two out of three orbital channels become irrelevant. However $T_K$
is enhanced due to inclusion of orbital degrees of freedom.
Moreover, this enhancement is magnetically tunable. It follows
from (\ref{emag}) that the crossover $SU(2)\to SU(4)\to SU(2)$
occurs three times within the interval $0<\Phi<6\pi$ and each
level crossing results in enhancement of $T_K$ from (\ref{TKA}) to
(\ref{kondoe}) and back (lower panel of Fig. \ref{T-phi}). These
field induced effects may be observed by measuring the
two-terminal conductance $G_{jl}$ through TTQD (the third contact
is assumed to be passive). Calculation by means of Keldysh
technique (at $T>T_{K}$) similar to that of Ref. \cite{kng} show
sharp maxima in $G$ as a function of magnetic field, following the
maxima of $T_K$ (Fig. \ref{C-phi}).

\begin{figure}[htb]
\centering
\includegraphics[width=100mm,height=38mm,angle=0]{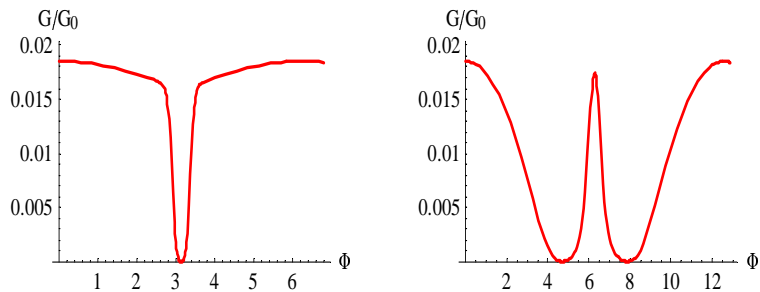}
\caption{Conductance as a function of magnetic field for
$\Phi_2=0$ (left panel) and $\Phi_1=\Phi_2=\Phi/2$ (right
panel).}\label{cond}
\end{figure}

So far we have studied the influence of the magnetic field on the
ground-state symmetry of the TTQD. In a two-lead geometry (Fig.
\ref{figTQD}b) the field $B$ affects the lead-dot hopping phases
thereby inducing the AB effect. (The necessary condition for AB
effect is that the electron coherence length in the source should
exceed the size of the electrode "tip".) The symmetry of the
device is thereby reduced since it looses two out of three mirror
reflection axes. The orbital doublet $E$ splits into two states,
but still, the ground state is $|DA \rangle$. In a generic
situation, the total magnetic flux is the sum of two components
$\Phi=\Phi_1+ \Phi_2$. In the chosen gauge, the hopping integrals
in Eqs. (\ref{H-dot}), (\ref{H-tun-2t}) are modified as,
$$W\to W\exp (i \Phi_1/3),~~V_{1,2}\to V_s\exp[\pm
i (\Phi_1/6+\Phi_2/2)],$$ and  the exchange Hamiltonian now reads,
\begin{eqnarray}
H&=&J_s {\bf S}\cdot {\bf s}_s +J_d {\bf S}\cdot {\bf
s}_d
+J_{sd}{\bf S}\cdot ({\bf s}_{sd} + {\bf s}_{ds}). \label{HAB}
\end{eqnarray}
Here
\begin{eqnarray}
 J_s&=&\frac{4V^2_s}{3}
   \left(\frac{1+\cos\left(\frac{\Phi_1}{3}+\Phi_2\right)}{\epsilon_F
   -\epsilon} +\frac{1}{\epsilon+Q -\epsilon_F}\right),\nonumber\\
J_d&=&\frac{2V^2_d}{3}
   \left(\frac{1}{\epsilon_F
   -\epsilon} +\frac{1}{\epsilon+Q
   -\epsilon_F}\right),\label{J-doubl}\\
J_{sd}&=&\frac{4V_sV_d}{3}
   \left(\frac{1}{\epsilon_F
   -\epsilon} +\frac{1}{\epsilon+{Q}' -\epsilon_F}\right)
   \cos\left(\frac{\Phi_1}{6}+\frac{\Phi_2}{2}\right)\nonumber
\end{eqnarray}
for $0\leq\Phi_1 <\pi$, and
\begin{eqnarray}
 J_s&=&-\frac{V^2_s}{3}
   \left(\frac{1-\sin\Phi_2}{\epsilon+{Q}' -\epsilon_F} -
   \frac{2}{\epsilon+Q -\epsilon_F}\right)<0,\nonumber\\
J_d&=&\frac{4V^2_d}{3}
   \left(\frac{1}{\epsilon_F
   -\epsilon} +\frac{1}{\epsilon+Q
   -\epsilon_F}\right),\label{J-pi}\\
J_{sd}&=&\frac{2V_sV_d}{3}
   \left(\frac{1}{\epsilon_F
   -\epsilon} +\frac{1}{\epsilon+{Q}'
   -\epsilon_F}\right)\sin\Phi_2,\nonumber
\end{eqnarray}
for $\Phi_1 =\pi$. Applying poor man scaling RG procedure on the
Hamiltonian (\ref{HAB}) yields the scaling equations,
\begin{eqnarray}
\frac{dj_s}{d\ln D}&=&-\left[j_s^2+j_d^2\right],\nonumber\\
\frac{dj_d}{d\ln D}&=&-\left[j_d^2+j_d^2\right],\label{sc-eqs-2term}\\
\frac{dj_{sd}}{d\ln D}&=&-j_{sd}\left[j_s+j_d\right].\nonumber
\end{eqnarray}
Eqs.(\ref{sc-eqs-2term}) give the Kondo temperature,
\begin{eqnarray}\label{T-Kondo}
T_{K}={D_0}\exp\left\{-\frac{2}{j_s+j_d+\sqrt{(j_s-j_d)^2+4j_{sd}^2}}\right\}.
\end{eqnarray}
The conductance at $T>T_{K}$ reads \cite{kng},
\begin{eqnarray}
  \frac{G}{G_0}&=&\frac{3}{4}\frac{j_{sd}^2}{(j_s+j_d)^2}\frac{1}{\ln^2
  (T/T_K)}.
  \label{cond-AB}
\end{eqnarray}
It was verified that the resulting conductance
$G(\Phi_{1},\Phi_{2})$ (\ref{cond-AB}) obeys the Byers-Yang
theorem (periodicity in each phase) and the Onsager condition
$G(\Phi_{1},\Phi_{2})=G(-\Phi_{1},-\Phi_{2})$. We choose to
display the conductance along two lines
$\Phi_{1}(\Phi),\Phi_{2}(\Phi)$ in parameter space of phases,
namely, $G(\Phi_{1}=\Phi,\Phi_{2}=0)$ and
$G(\Phi_{1}=\Phi/2,\Phi_{2}=\Phi/2)$ (figure \ref{cond} left and
right panels, respectively).
 The Byers-Yang relation
 implies respective periods of $2 \pi$ and $4 \pi$ in $\Phi$.
Experimentally, the magnetic flux is of course applied on the
whole sample as in figure 1b, and the ratio $\Phi_{1}/\Phi_{2}$ is
determined by the specific geometry. Strictly speaking then, the
conductance is not periodic in the magnetic field unless
$\Phi_{1}$ and $\Phi_{2}$ are commensurate.

The shapes of the conductance curves presented here are distinct
from those pertaining to a mesoscopic AB interferometer with a
single correlated QD and a conducting channel
\cite{Fano1,Fano4,AB} (termed as Fano-Kondo effect \cite{Fano1}).
For example, $G(\Phi)$ in Fig. 3 of Ref. \cite{Fano1} (calculated
in the strong coupling regime)  has a broad peak at $\Phi=\pi/2$
with $G(\Phi=\pi/2)=1$. On the other hand, $G(\Phi)$ displayed in
Fig. \ref{T-phi} (pertinent to Fig. \ref{figTQD}a and obtained in
the weak coupling regime), is virtually flux independent except
near the points $\Phi=(2n+1) \pi$ ($n$ integer) at which the
$SU(4)$ symmetry is realized and $G$ is sharply peaked. The phase
dependence is governed here by interference effects on the level
spectrum of the TTQD. The three dots share an electron in a
coherent state controlled by the strong correlations, and this
coherent TTQD {\it as a whole} is a vital component of the AB
interferometer. As a result, in the setup of Fig. \ref{figTQD}b,
the conductance vanishes identically on the curve
$J_{sd}(\Phi_{1},\Phi_{2})=0$. The AB oscillations arise as a
result of interference between the clockwise and anticlockwise
"effective rotations" of TTQD in the tunneling through the \{13\}
and \{23\} arms of the loop (Fig. \ref{figTQD}b), provided the
dephasing in the leads does not destroy the coherence of tunneling
through the two source channels. (Interplay between two chiral
states results in $SU(4)$ Kondo effect in carbon nanotubes in
axial magnetic field \cite{Jar,Choi}.) On the other hand, in the
calculations performed on Fano-Kondo interferometers, $G(\Phi)$
remains finite \cite{Fano1}. Incidentally, there should be a Fano
effect due to the renormalization of electron spectrum in the
leads induced by the lead-dot tunneling similar to that in
chemisorbed atoms \cite{Plih01} but that is beyond the scope of
this Thesis.

\section{Conclusions}
To conclude, we have shown that the Kondo and AB effects in TTQD
expose peculiar symmetries in the physics of strongly correlated
electrons. Dynamical symmetry is a universal tool, which allows to
derive the effective spin Hamiltonians describing low-energy
cotunneling with spin reversal through TTQD. This artificial
molecule possesses both the continuous rotation symmetry in spin
space and discrete rotation symmetry in real space. This discrete
symmetry is imposed on the whole system both in the case of
molecular trimer Cr$_3$ on the metallic surface and in the
three-terminal device - equilateral TTQD in its center (Figs.
\ref{tik-tak}, \ref{figTQD}a). The tunnel problem is mapped on the
Coqblin-Schrieffer (CS) model of magnetic impurity with "orbital"
degrees of freedom \cite{CqS,CoqC}. Like in the latter case, the
symmetry of the Kondo center is $SU(2n)$. The parameter $n$
characterizes additional orbital degeneracy in the conventional CS
model. It describes the pseudospin operator of finite
clockwise/anti-clockwise rotations of TTQD in the process of
electron cotunneling.

Abstract concepts like dynamical symmetries are shown to be
directly related to experimentally tunable parameters. The
conductance is sensitive to an external magnetic field both in
three- and two-terminal geometry even in the weak coupling regime
at $T>T_K$. In the former case, the conductance can be enhanced
due to change of the dynamical symmetry caused by field-induced
level crossing (Fig. \ref{T-phi}). In the latter case, the
conductance can be completely suppressed due to destructive AB
interference in source-drain cotunneling amplitude (Fig.
\ref{cond}). These results promise an interesting physics at the
strong coupling regime as well as in cases of doubly and triply
occupied TTQD. It would also be interesting to generalize the
present theory for quadratic QD \cite{quad}, which possesses rich
energy spectrum with multiple accidental degeneracies.

\appendix
\chapter{Diagonalization of the Trimer Hamiltonian}\label{diag}
Here we describe the diagonalization procedure of the Hamiltonian
of the isolated TQDs occupied by four and three electrons (Fig.
\ref{TQD-s}). The dot Hamiltonian has the form,
\begin{eqnarray}
H_{d}&=&\sum_{a=l,r,c}\sum_{\sigma}\epsilon_{a}d^\dagger_{a
\sigma}d_{a\sigma
}+\sum_{a}Q_an_{a\uparrow}n_{a\downarrow}
 +\sum_{a=l,r}(W_{a}d^\dagger_{c
\sigma}d_{a\sigma }+H.c.). \label{H-dot-trimer}
\end{eqnarray}
({\bf a}) Four electron occupation:\\
The Hamiltonian (\ref{H-dot-trimer}) can be diagonalized  by using
the basis of four-electron wave functions
\begin{eqnarray}
&&\left| t_a, 1 \right\rangle =
d^{\dagger}_{c\uparrow}d^{\dag}_{a\uparrow}d^{\dag}_{{\bar
a}\uparrow} d^{\dag}_{{\bar a}\downarrow}\vert 0\rangle,\ \ \
\left| t_a, {\bar 1} \right\rangle
=d^{\dagger}_{c\downarrow}d^{\dag}_{a\downarrow}d^{\dag}_{{\bar
a}\uparrow}
d^{\dag}_{{\bar a}\downarrow}\vert 0\rangle,\nonumber \\
&& \left| t_a, 0\right\rangle = \frac{1}{\sqrt{2}}\left (
d^{\dag}_{c\uparrow}d^{\dag}_{a\downarrow}
+d^{\dag}_{c\downarrow}d^{\dag}_{a\uparrow}\right )
d^{\dag}_{{\bar a}\uparrow}d^{\dag}_{{\bar a}\downarrow}\vert
0\rangle
,\nonumber \\
&&\left| s_a \right\rangle =\frac{1}{\sqrt{2}}\left (
d^{\dag}_{c\uparrow}d^{\dag}_{a\downarrow}
-d^{\dag}_{c\downarrow}d^{\dag}_{a\uparrow}\right )
d^{\dag}_{{\bar a}\uparrow}d^{\dag}_{{\bar a}\downarrow}\vert
0\rangle ,
\nonumber \\
&&|{ex}\rangle = d^{\dag}_{l\uparrow}d^{\dag}_{l\downarrow}
d^{\dag}_{r\uparrow}d^{\dag}_{r\downarrow} \vert 0\rangle,
\label{basis-f}
\end{eqnarray}
where $a =l,r$; $\overline l=r$, $\overline r=l$. The Coulomb
interaction quenches the states with two electrons in the central
dot and we do not take them into account. The states
(\ref{basis-f}) form a basis of two triplet and three singlet
states. In this basis, the Hamiltonian (\ref{H-dot-trimer}) is
decomposed into triplet and singlet matrices,
\begin{eqnarray}
H_t=\left(
\begin{array}{cc}
\tilde{\varepsilon}_l & 0 \\
0& \tilde{\varepsilon}_r
 \end{array}\right ),
\end{eqnarray}
and
\begin{eqnarray}
H_s= \left(
\begin{array}{ccc}
\tilde{\varepsilon}_l & 0 & \sqrt{2}W_l \\
0 & \tilde{\varepsilon}_r & \sqrt{2}W_r \\
\sqrt{2}W_l & \sqrt{2}W_r & \tilde{\varepsilon}_{ex} \\
\end{array}\right ),
\end{eqnarray}
where $\tilde{\varepsilon}_l={\epsilon}_c
+{\epsilon}_l+2{\epsilon}_r+Q_r$,
$\tilde{\varepsilon}_r={\epsilon}_c
+{\epsilon}_r+2{\epsilon}_l+Q_l$, and $\tilde{\varepsilon}_{ex}
=2{\epsilon}_l+2{\epsilon}_r+Q_l+Q_r.$ We are interested in the
limit $\beta_a\ll 1$
($\beta_a=W_a/(\varepsilon_a+Q_a-\epsilon_c)$). So the secular
matrix may be diagonalized in lowest order of perturbation theory
in $\beta_a$.
 The eigenfunctions corresponding to the
energy levels (\ref{En}) are,
\begin{eqnarray}
&&\vert S_a\rangle = \sqrt{1-2\beta_a^2}\left\vert
s_a\right\rangle
-\sqrt{2}\beta_a|{ex}\rangle, \nonumber \\
&&\vert T_a\rangle = \left\vert t_a\right\rangle,
\label{eg-fun4}\\
&&\vert {Ex}\rangle = \sqrt{1-2\beta_l^2-2\beta_r^2}\left\vert
{ex}\right\rangle+ \sqrt{2}\beta_l|s_l\rangle
+\sqrt{2}\beta_r|s_r\rangle.\nonumber
\end{eqnarray}

In the completely symmetric case,
$\varepsilon_l=\varepsilon_r\equiv \varepsilon,$ $Q_l=Q_r\equiv
Q,$ $W_l=W_r\equiv W$, the eigenfunctions corresponding to the
energies (\ref{degen}) are
\begin{eqnarray}
\vert S_+\rangle &=& \sqrt{1-4\beta^2}\frac{\left\vert
s_l\right\rangle+\left\vert s_r\right\rangle}{\sqrt{2}}
-2\beta|{ex}\rangle, \nonumber \\
\vert S_-\rangle &=& \frac{\left\vert s_l\right\rangle-\left\vert
s_r\right\rangle}{\sqrt{2}},
\label{degen-eg-fun}\\
\vert T_{\pm}\rangle &=& \frac{\left\vert t_l\right\rangle\pm
\left\vert
t_r\right\rangle}{\sqrt{2}},\nonumber\\
\vert {Ex}\rangle &=& \sqrt{1-4\beta}\left\vert {ex}\right\rangle+
\sqrt{2}\beta(|s_l\rangle +|s_r\rangle),\nonumber
\end{eqnarray}
where $\beta=W/(\varepsilon+Q-\epsilon_c)$.\\
({\bf b}) Three electron occupation:\\
In this case the Hamiltonian (\ref{H-dot-trimer}) can be
diagonalized by using the basis of three--electron wave functions
\begin{eqnarray}
&&| b_1,\sigma\rangle =\frac{(
[d^{+}_{c\uparrow}d^{+}_{l\downarrow}-d^{+}_{c\downarrow}
d^{+}_{l\uparrow}] d^{+}_{r\sigma}+
[d^{+}_{c\downarrow}d^{+}_{r\uparrow}-d^{+}_{c\uparrow}
d^{+}_{r\downarrow}] d^{+}_{l\sigma} ) \vert 0\rangle}{\sqrt{6}}
,\nonumber\\
&&\left| b_2, \sigma\right\rangle =-\frac{1}{\sqrt{2}}\left (
d^{+}_{l\uparrow}d^{+}_{r\downarrow}-d^{+}_{l\downarrow}
d^{+}_{r\uparrow}\right ) d^{+}_{c\sigma}\vert 0\rangle ,\nonumber \\
&&\left| q, {\pm\frac{3}{2}}\right\rangle
=d^{+}_{c\pm}d^{+}_{r\pm}d^{+}_{l\pm}\vert 0\rangle ,
 \nonumber\\
&&\left | q, {\pm\frac{1}{2}}\right\rangle =\frac{(
d^{+}_{c\pm}d^{+}_{r\pm}d^{+}_{l\mp}+
d^{+}_{c\pm}d^{+}_{r\mp}d^{+}_{l\pm}+
d^{+}_{c\mp}d^{+}_{r\pm}d^{+}_{l\pm})\vert
0\rangle }{\sqrt{3}}, \nonumber\\
&&|b_{lc},\sigma\rangle = d^{+}_{l\uparrow}d^{+}_{l\downarrow}
d^{+}_{c\sigma}\vert 0\rangle,\ \ \ \ \ \;\;\;  |b_{rc},
\sigma\rangle = d^{+}_{r\uparrow}d^{+}_{r\downarrow}
d^{+}_{c\sigma} \vert 0\rangle,\nonumber\\
&&|b_l, \sigma\rangle = d^{+}_{r\uparrow}d^{+}_{r\downarrow}
d^{+}_{l\sigma}\vert 0\rangle,\ \ \ \ \ \ \ \ |b_r, \sigma\rangle
= d^{+}_{l\uparrow}d^{+}_{l\downarrow} d^{+}_{r\sigma} \vert
0\rangle,\label{basis}
\end{eqnarray}
where $\sigma=\uparrow,\downarrow$. The three-electron states
$|\Lambda\rangle$ of the TQD are classified as a ground state
doublet $|B_1\rangle$, low-lying doublet $|B_2\rangle$ and quartet
$|Q\rangle$ excitations, and four charge-transfer excitonic
doublets $B_{ac}$ and $B_a$ ($a=l,r$). In the framework of second
order perturbation theory with respect to $\beta_{a}$ (\ref{1.9}),
the
 energy levels $E_\Lambda$ are
\begin{eqnarray}
&&E_{B_1} = {\epsilon}_c +{\epsilon}_l +{\epsilon}_r-
\frac{3}{2}\left[W_l \beta_l+W_r \beta_r\right], \nonumber \\
&&E_{B_2} = {\epsilon}_c +{\epsilon}_l+{\epsilon}_r-
\frac{1}{2}\left[W_l \beta_l+W_r \beta_r\right],
\nonumber \\
&&E_{Q} = {\epsilon}_c +{\epsilon}_l +\epsilon_r,\nonumber\\
&&E_{B_{ac}}={\epsilon}_c+2{\epsilon}_a+Q_a-W_{\bar a}\beta_{\bar a},\nonumber\\
&&E_{B_a} ={\epsilon}_a +2{\epsilon}_{\bar a}+Q_{\bar
a}+W_{a}\beta_{a}+2W_{\bar a}\beta_{\bar a}. \label{energy3}
\end{eqnarray}

 The eigenfunctions corresponding to the
energy levels (\ref{energy3}) are the following combinations,
\begin{eqnarray}
&&\vert B_1,\sigma\rangle = \gamma_1\vert b_1,\sigma\rangle
-\frac{\sqrt{6}}{2}\beta_{l}|b_{r},\sigma\rangle +
\frac{\sqrt{6}}{2}\beta_{r}|b_{l},\sigma\rangle , \nonumber\\
&&\vert B_2,\sigma\rangle = \gamma_2\vert b_2,\sigma\rangle
-\frac{\sqrt{2}}{2}\beta_{l}|b_{r},\sigma\rangle -
\frac{\sqrt{2}}{2}\beta_{r}|b_{l},\sigma\rangle , \nonumber\\
&&\vert Q,s_z\rangle =\left\vert q,s_z\right\rangle,\ \ \ \ \ \
s_z=\pm \frac{3}{2},\pm \frac{1}{2},\label{eg-func-trimer}\\
&&\vert B_{ac},\sigma\rangle =\sqrt{1-\beta_{\bar a}^2}\vert
b_{ac},\sigma\rangle
-\beta_{\bar a}|b_{\bar a},\sigma\rangle,\nonumber\\
&&\vert B_{r},\sigma\rangle = \sqrt{1-2\beta_{l}^2-\beta_r^2}\vert
b_r,\sigma\rangle+ \beta_{r}|b_{lc},\sigma\rangle
+\frac{\sqrt{2}}{2}\beta_{l}\left(\sqrt{3}|b_1,\sigma\rangle
+|b_2,\sigma\rangle\right), \nonumber\\
&&\vert B_l,\sigma\rangle = \sqrt{1-2\beta_r^2-\beta_l^2}\vert
b_l,\sigma\rangle +\beta_{l}|b_{r},
{c},\sigma\rangle
-\frac{\sqrt{2}}{2}\beta_{r}\left(\sqrt{3}|b_1,\sigma\rangle
-|b_2,\sigma\rangle\right) ,\nonumber
\end{eqnarray}
where $\gamma_1$ and $\gamma_2$ are determined by Eq.(\ref{gam}).

\chapter{Rotations in the Source-Drain and Left-Right
Spaces}\label{GR-trans}

In the generic case, the transformation which diagonalizes the
tunneling Hamiltonian (\ref{hyb}) has the form
\begin{equation}
\left(
\begin{array}{cc} c_{lek\sigma }\\ c_{lok\sigma }\\ c_{rek\sigma }\\ c_{rok\sigma }\\
\end{array} \right) =
\left(
\begin{array}{rrrr} u_l & v_l & 0 & 0 \\
                 -v_l & u_l & 0 & 0 \\
                 0 & 0 & u_r & v_r \\
                 0 & 0 & -v_r & u_r
\end{array} \right)
\left(
\begin{array}{cc} c_{lsk\sigma }\\ c_{ldk\sigma }\\ c_{rsk\sigma }\\ c_{rdk\sigma }
\end{array} \right),
\end{equation}
with $u_a=V_{as}/V_{a}$, $v_a=V_{ad}/V_{a}$; $V_{a}^2=V_{as}^2 +
V_{ad}^2$ $(a=l,r).$ In a symmetric case $V_{as}=V_{ad}=V$, this
transformation simplifies  to
\begin{eqnarray}
&&c_{aek\sigma }=2^{-1/2}\left(c_{ask\sigma }+c_{adk\sigma
}\right),\ \ \ \ \ c_{aok\sigma }=2^{-1/2}\left(-c_{ask\sigma
}+c_{adk\sigma }\right),
\end{eqnarray}
and only the even $(e)$ combination survives in the tunneling
Hamiltonian
\begin{equation}
H_{tun}=V\sum_{ak\sigma }(c_{aek\sigma }^\dag d_{a\sigma} + H.c.).
\end{equation}
So the odd combination  $(o)$ may be omitted.

If, moreover, the whole system "TQD plus leads" possesses $l-r$
symmetry, $\varepsilon_{l}=\varepsilon_r$, the second rotation in
$l-r$-space
\begin{equation}
\left(
\begin{array}{ll} c_{gk\sigma }\\ c_{uk\sigma }\\ d_{g\sigma }\\ d_{u\sigma }\\
\end{array} \right) =\frac{1}{\sqrt 2}
\left(
\begin{array}{rrrr} 1 & 1 & 0 & 0 \\
                 -1 & 1 & 0 & 0 \\
                 0 & 0 & 1 & 1 \\
                 0 & 0 & -1 & 1
\end{array} \right)
\left(
\begin{array}{ll} c_{lek\sigma }\\ c_{rek\sigma }\\ d_{l\sigma }\\ d_{r\sigma }
\end{array} \right),
\end{equation}
transforms $H_{lead}+H_{tun}$ into
\begin{eqnarray}
H_{lead}+H_{tun}=\sum_{\eta k\sigma}\Big[ \epsilon_{k\eta}n_{\eta
k\sigma} + V(c_{\eta k\sigma }^\dag d_{\eta\sigma} + H.c.)\Big],
\end{eqnarray}
with $\epsilon_{kg}=\epsilon_{k}-t_{lr}$,
$\epsilon_{ku}=\epsilon_{k}+t_{lr}$.

\chapter{Effective Spin Hamiltonian}\label{H-spin}
The spin Hamiltonian of the TQD with ${\cal N}=4$ occupation in
series geometry (Fig.\ref{TQD-s}) is derived below. The system is
described by the Hamiltonian (\ref{Hser}). The Schrieffer-Wolff
transformation \cite{SW} for the configuration of four electron
states of the TQD projects out three electron states $\vert
\lambda\rangle$ and maps the Hamiltonian (\ref{Hser}) onto an
effective spin Hamiltonian $\widetilde{H}$ acting in a subspace of
four-electron configurations $\langle\Lambda|\ldots
|\Lambda^{\prime}\rangle$,
\begin{eqnarray}\label{SW-tr}
\widetilde{H}= e^{i{\cal S}}H e^{{-i\cal S}}
= H + \sum_m \frac{(i)^m}{m!} [%
{\cal S},[{\cal S}...[{\cal S},H]]...],
\end{eqnarray}
where
\begin{equation}
{\cal S}=-i\sum_{\Lambda\lambda}\sum_{\langle k\rangle\sigma,a}
\frac{V_{a\sigma}^{\Lambda\lambda}}
{\bar{E}_{\Lambda\lambda}-\epsilon_{ka}}
X^{\Lambda\lambda}c_{ak\sigma}+H.c.
\end{equation}
Here $\langle k\rangle$ stands for the electron or hole states
whose energies are secluded
within a layer $\pm \bar{D}$ around the Fermi level. $\bar{E}%
_{\Lambda\lambda}=E_\Lambda(\bar{D})-E_\lambda(\bar{D})$ and the
notation $a=l,r$ is used.
 The effective Hamiltonian with three--electron states
$|\lambda\rangle$ frozen out can be obtained by retaining the
terms to order $O(|V|^2)$ on the right-hand side of
Eq.(\ref{SW-tr}). It has the following form,
\begin{eqnarray}
&&\widetilde{H}  =  \sum_\Lambda \bar{E}_\Lambda
X^{\Lambda\Lambda} +\sum_{\langle k\rangle \sigma,a}\epsilon_{ka}
c^{+}_{ak\sigma }c_{ak\sigma}
\label{Ham-eff} \\
&& -
\sum_{\Lambda\Lambda^{\prime}\lambda}\sum_{kk^{\prime}\sigma\sigma^{\prime}}
\sum_{a=l,r} J^{\Lambda\Lambda^{\prime}}_{kk^{\prime}a}
X^{\Lambda\Lambda^{\prime}}c^{+}_{ak\sigma}c_{ak^{\prime}\sigma^{\prime}}
  -
\sum_{\Lambda\Lambda^{\prime}\lambda}\sum_{kk^{\prime}\sigma\sigma^{\prime}}
( J^{\Lambda\Lambda^{\prime}}_{kk^{\prime}lr}
X^{\Lambda\Lambda^{\prime}}c^{+}_{rk\sigma}
c_{lk^{\prime}\sigma^{\prime}}+H.c.),\nonumber
\end{eqnarray}
where
\begin{eqnarray}
J^{\Lambda\Lambda^{\prime}}_{kk^{\prime}a} = \frac
{(V_{a\sigma}^{\lambda\Lambda})^*
V_{a\sigma^{\prime}}^{\lambda\Lambda^{\prime}}}{2} \left(
\frac{1}{\bar{E}_{\Lambda\lambda}-\epsilon_{ka}}+
\frac{1}{\bar{E}_{\Lambda^{\prime}\lambda}-\epsilon_{k^{\prime}a}}
\right),\nonumber \\
J^{\Lambda\Lambda^{\prime}}_{kk^{\prime}lr} = \frac
{(V_{l\sigma}^{\lambda\Lambda})^*
V_{r\sigma^{\prime}}^{\lambda\Lambda^{\prime}}} {2} \left(
\frac{1}{\bar{E}_{\Lambda\lambda}-\epsilon_{kl}}+
\frac{1}{\bar{E}_{\Lambda^{\prime}\lambda}-\epsilon_{k^{\prime}r}}
\right).
\end{eqnarray}
The constraint $\sum_{\Lambda}X^{\Lambda%
\Lambda}=1 $ is valid. Unlike the conventional case 
of doublet spin $1/2$ we have here an octet $\Lambda
=\{\Lambda_l,\Lambda_r\} =\{S_l,T_l,S_r,T_r\}$, and the SW
transformation {\it intermixes all these states}. The effective
spin Hamiltonian (\ref{Ham-eff}) to order $O(|V|^2)$ acquires the
form of Eq.(\ref{ex-H}).

\chapter{$\bf SO(7)$ Symmetry}\label{algebra-o7}

The operators ${\bf S}_l$, ${\bf S}_r$, ${\bf R}_l$, $\tilde{\bf
R}_1$, $\tilde{\bf R}_2$, $\tilde{\bf R}_3$ and ${A_i}$
$(i=1,2,3)$ (see Eqs.(\ref{comm1}),(\ref{so7}),(\ref{A})) obey the
commutation relations of the $o_7$ Lie algebra,
\begin{eqnarray*}
&&\lbrack S_{aj},S_{a'k}]=ie_{jkm}\delta_{aa'}S_{am},\ \ \;\;\;
[R_{lj},R_{lk}]=ie_{jkm}S_{lm},\nonumber\\
&&[R_{lj},S_{lk}]=ie_{jkm}R_{lm},\ \ \ \ \ \ \ \ \ \
[R_{lj},S_{rk}]=[\tilde{R}_{3j},S_{lk}]=0,\nonumber\\
&&\lbrack \tilde{R}_{3j},\tilde{R}_{3k}] = ie_{jkm}S_{rm},\ \ \ \
\ \ \ \ \
[\tilde{R}_{3j},S_{rk}]=ie_{jkm}\tilde{R}_{3m},\nonumber\\
&&[\tilde{R}_{1j},\tilde{R}_{1k}]=ie_{jkm}S_{rm}(1-{\delta}_{jz})
(1-{\delta}_{kz})
+\frac{i}{2}e_{jkm}S_{lm}({\delta}_{jz}+{\delta}_{kz})-
\frac{1}{2}(S_{lj}{\delta}_{kz}-S_{lk}{\delta}_{jz}),\nonumber\\
&&[\tilde{R}_{2j},\tilde{R}_{2k}]=ie_{jkm}S_{lm}(1-{\delta}_{jz})
(1-{\delta}_{kz})
+\frac{i}{2}e_{jkm}S_{rm}({\delta}_{jz}+{\delta}_{kz})-
\frac{1}{2}(S_{rj}{\delta}_{kz}-S_{rk}{\delta}_{jz}),\nonumber\\
&&[\tilde{R}_{1j},\tilde{R}_{2k}]=\frac{i}{2}e_{jkm}(S_{rm}\delta_{jz}
+S_{lm}{\delta}_{kz})
+\frac{1}{2}\left[S_{lj}{\delta}_{kz}-S_{rk}{\delta}_{jz}+
(S_{lz}-S_{rz})\delta_{jz}\delta_{kz}\right],\nonumber\\
&&\lbrack \tilde{R}_{3j},\tilde{R}_{1k}] =
ie_{jkm}R_{lm}(1-{\delta}_{jz}-\frac{{\delta}_{kz}}{2})
-\frac{{\delta}_{kz}}{2}(1-{\delta}_{jz})R_{lj},\nonumber\\
&&\lbrack \tilde{R}_{3j},\tilde{R}_{2k}] =
ie_{jkm}R_{lm}({\delta}_{jz}+\frac{{\delta}_{kz}}{2})+
\frac{{\delta}_{kz}}{2}(1-{\delta}_{jz})R_{lj},\nonumber\\
&&[\tilde{R}_{1j},R_{lk}]=ie_{jkm}\tilde{R}_{3m}({\delta}_{kz}+
\frac{{\delta}_{jz}}{2})-
\frac{{\delta}_{jz}}{2}(1-{\delta}_{kz})\tilde{R}_{3k},\nonumber\\
&&[\tilde{R}_{2j},R_{lk}]=ie_{jkm}\tilde{R}_{3m}(1-{\delta}_{kz}-
\frac{{\delta}_{jz}}{2})
+\frac{{\delta}_{jz}}{2}(1-{\delta}_{kz})\tilde{R}_{3k},\nonumber\\
&&[A_1,S_{lj}]=iA_2{\delta}_{jz}+\frac{i\sqrt{2}}{2}
(\tilde{R}_{1x}\delta_{jx}-\tilde{R}_{1y}\delta_{jy}),\nonumber\\
&&[A_2,S_{lj}]=-iA_1{\delta}_{jz}-\frac{i\sqrt{2}}{2}
(\tilde{R}_{1y}\delta_{jx}+\tilde{R}_{1x}\delta_{jy}),\nonumber\\
&&[A_1,S_{rj}]=iA_2{\delta}_{jz}-\frac{i\sqrt{2}}{2}
(\tilde{R}_{2x}\delta_{jx}-\tilde{R}_{2y}\delta_{jy}),\nonumber\\
&&[A_2,S_{rj}]=-iA_1{\delta}_{jz}+\frac{i\sqrt{2}}{2}
(\tilde{R}_{2y}\delta_{jx}+\tilde{R}_{2x}\delta_{jy}),\nonumber\\
&&[A_3,S_{lj}]=-i\tilde{R}_{2j}(1-\delta_{jz}),\ \ \ \ \ \ \ \ \ \
\ \ \ \ \ \ \
[A_3,S_{rj}]=i\tilde{R}_{1j}(1-\delta_{jz}),
\end{eqnarray*}
\begin{eqnarray}
&&[A_{1},R_{lj}]=-\frac{i\sqrt{2}}{2}(\tilde{R}_{3x}\delta_{jx}-
\tilde{R}_{3y}{\delta}_{jy}),
\ \ \ \ \ [A_{2},R_{lj}]=\frac{i\sqrt{2}}{2}
(\tilde{R}_{3y}\delta_{jx}+\tilde{R}_{3x}{\delta}_{jy}),\nonumber\\
&&[A_{1},\tilde{R}_{3j}]=\frac{i\sqrt{2}}{2}
(R_{lx}\delta_{jx}-R_{ly}{\delta}_{jy}),
\ \ \ \ \ \ \ \
[A_{2},\tilde{R}_{3j}]=-\frac{i\sqrt{2}}{2}(R_{ly}\delta_{jx}+
R_{lx}{\delta}_{jy}),\nonumber\\
&&[A_{3},R_{lj}]=-i\tilde{R}_{3z}{\delta}_{jz},\ \ \ \ \ \ \ \ \ \
\ \ \ \ \ \ \ \ \ \ \ \ \ \ \;
[A_{3},\tilde{R}_{3j}]=iR_{lz}{\delta}_{jz},\nonumber\\
&&[A_{1},\tilde{R}_{1j}]=-\frac{i\sqrt{2}}{2}
(S_{lx}\delta_{jx}-S_{ly}{\delta}_{jy}),
\ \ \ \ \ \ \ [A_{2},\tilde{R}_{1j}]=\frac{i\sqrt{2}}{2}
(S_{ly}\delta_{jx}+S_{lx}{\delta}_{jy}),\nonumber\\
&&[A_{3},\tilde{R}_{1j}]=-i(S_{rx}\delta_{jx}+S_{ry}{\delta}_{jy}),
\ \ \ \ \ \ \ \ \ \ \
[A_{1},\tilde{R}_{2j}]=\frac{i\sqrt{2}}{2}(S_{rx}\delta_{jx}-
S_{ry}{\delta}_{jy}),\nonumber\\
&&[A_{2},\tilde{R}_{2j}]=-\frac{i\sqrt{2}}{2}
(S_{ry}\delta_{jx}+S_{rx}{\delta}_{jy}),
\ \ \ \ \ \;\; [A_{3},\tilde{R}_{2j}]=i(S_{lx}\delta_{jx}+S_{ly}{\delta}_{jy}),\nonumber\\
&&[A_{1},A_{2}]=-i(S_{lz}+S_{rz}),\ \ \ \ \ \ \ \ \ \ \ \ \ \ \ \
\ \ \ \
[A_{1},A_{3}]=[A_{2},A_{3}]=0,\nonumber\\
&&\lbrack S_{aj},\tilde{R}_{\mu k}]=\tau_{jkm}^{a\mu\nu}
\tilde{R}_{\mu m}+\alpha_{jk}^{a\mu n}A_n,
\ \ \  \ \ \ \ \ \ \ \
[\tilde{R}_{3j},R_{lk}]=\beta_{jkm}^{\mu}\tilde{R}_{\mu m}+\tilde
\alpha_{jk}^{n}A_n.
\end{eqnarray}
Here $j,k,m$ are Cartesian indices, $a=l,r$; $\mu,\nu =1,2$;
$n=1,2,3$; $\tau_{jkm}^{a\mu\nu}$, $\alpha_{jk}^{a\mu n}$, $\tilde
\alpha_{jk}^{n}$ and $\beta_{jkm}^{\mu}$ are the structural
constants, $\tau_{jkm}^{l\mu\nu}=\tau_{jkm}^{r{\bar \mu}{\bar
\nu}}$, $\alpha_{jk}^{l \mu n}=-\alpha_{jk}^{r {\bar \mu} n}$
(${\bar 1}=2$, ${\bar 2}=1$). Their non-zero components are:
\begin{eqnarray*}
\tau_{xxz}^{l11}&=&\tau_{xzx}^{l11}=\tau_{yyz}^{l11}=\tau_{yzy}^{l11}=\frac{1}{2
},\\
\tau_{xzx}^{l21}&=&\tau_{yzy}^{l21}=\tau_{xxz}^{l12}=\tau_{yyz}^{l12}=-\frac{1}{
2},\\
\tau_{xyz}^{l11}&=&\tau_{xyz}^{l12}=\tau_{yzx}^{l21}=\frac{i}{2},\\
\tau_{xzy}^{l11}&=&\tau_{yxz}^{l11}=\tau_{yxz}^{l12}=\tau_{xzy}^{l21}=-\frac{i}{
2},\\
\tau_{zzz}^{l11}&=&1,\ \tau_{zzz}^{l22}=-1,\ \tau_{zxy}^{l22}=i,\
\tau_{zyx}^{l22}=-i,
\end{eqnarray*}
\vspace{-7mm}
\begin{eqnarray*}
\alpha_{xy}^{l11}&=&\alpha_{yx}^{l11}=\frac{\sqrt{2}}{2},\ \ \
\alpha_{xy}^{l12}=\alpha_{yx}^{l12}=-\frac{\sqrt{2}}{2},\\
\alpha_{xx}^{l11}&=&\alpha_{xx}^{l12}=\frac{i\sqrt{2}}{2},\ \
\alpha_{yy}^{l11}=\alpha_{yy}^{l12}=-\frac{i\sqrt{2}}{2},\\
\alpha_{xx}^{l23}&=&\alpha_{yy}^{l23}=-i\sqrt{2},
\end{eqnarray*}
\vspace{-7mm}
\begin{eqnarray*}
\beta_{xxz}^{1}&=&\beta_{yyz}^{2}=-\frac{1}{2},\ \ \
\beta_{xxz}^{2}=\beta_{yyz}^{1}=\frac{1}{2},\\
\beta_{xyz}^{1}&=&\beta_{xyz}^{2}=\frac{i}{2},\ \ \ \ \
\beta_{yxz}^{1}=\beta_{yxz}^{2}=-\frac{i}{2},\\
\beta_{xzy}^{1}&=&\beta_{zyx}^{2}=-i,\ \ \ \
\beta_{yzx}^{1}=\beta_{zxy}^{2}=i,
\end{eqnarray*}
\vspace{-7mm}
\begin{eqnarray*}
\tilde\alpha_{xx}^{1}&=&\tilde\alpha_{zz}^{3}=i\sqrt{2},\ \ \
\tilde\alpha_{xy}^{2}=\tilde\alpha_{yx}^{2}=\tilde\alpha_{yy}^{1}=-i\sqrt{2}.
\end{eqnarray*}

The following relations hold,
\begin{eqnarray}
&&{\bf S}_a\cdot {\bf R}_l={\bf S}_a\cdot \tilde{\bf R}_3=0,\ \ \
A_1A_3=A_2A_3=0,\nonumber\\
&&{\bf S}^{2}_a=2X^{\mu_a\mu_a},\;\tilde{\bf R}_1\cdot \tilde{\bf
R}^{\dag}_1+\tilde{\bf R}_2\cdot \tilde{\bf R}^{\dag}_2=
2\sum_{a=l,r}X^{\mu_a\mu_a},\nonumber\\
&&{\bf R}^{2}_l=X^{\mu_l\mu_l}+3X^{S_l
S_l},\;\;\;\;\tilde{\bf R}^{2}_3=X^{\mu_r\mu_r}+3X^{S_l S_l}, \nonumber\\
&&A_1^2+A_2^2+A_3^2=X^{\mu_l\mu_l}+X^{\mu_r\mu_r}.
\end{eqnarray}
Therefore, the vector operators ${\bf S}_l$, ${\bf S}_r$, ${\bf
R}_l$, $\tilde{\bf R}_i$ and scalar operators $A_i$ $(i=1,2,3)$
generate the algebra $o_{7}$ in a representation specified by the
Casimir operator
\begin{eqnarray}
{\bf S}_l^{2}+{\bf S}_r^{2}+{\bf R}^{2}_l+\sum_{i=1}^{2}\tilde{\bf
R}_i\cdot \tilde{\bf R}^{\dag}_i+\tilde{\bf
R}^{2}_3+\sum_{i=1}^3A_i^2=6.
\end{eqnarray}


\chapter{Young Tableaux Corresponding to Various Symmetries}\label{tab}

\begin{figure}[htb]
\centering
\includegraphics[width=30mm,height=28mm,angle=0]{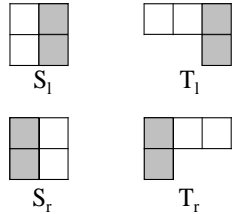}
\caption{Young tableaux corresponding to the singlet ($S_a$) and
triplet ($T_a$) four electron states of the TQD. The grey column
denote two electrons in the same dot (right or
left).}\label{scheme}
\end{figure}

A TQD with "passive" central dot and "active" side dots reminds an
artificial atom with inner core and external valence shell. The
many-electron wave functions in this nano-object may be
symmetrized in various ways, so that each spin state of $\cal N$
electrons in the TQD is characterized by its own symmetrization
scheme. One may illustrate these schemes by means of the
conventional graphic presentation of the permutation symmetry of
multi-electron system employing Young tableau
 \cite{ED}. For instance, triplet state of two electrons
which is symmetric with respect to the electron permutation is
labelled by a row of two squares, whereas the singlet one which is
antisymmetric with respect to the permutation is labelled by a
column of two squares.
 Having this in mind we can represent the singlet and triplet four
 electron states of the TQD (\ref{eg-fun4}) by the four tableaux shown in
 Fig. \ref{scheme}. The tableaux $S_l$ ($S_r$) and $T_l$ ($T_r$) correspond to
the singlet and triplet states in which the right (left) dot
contains two electrons (grey column in Fig. \ref{scheme}) whereas
electrons in the left (right) and central dots form singlet and
triplet, respectively.

\begin{figure}[htb]
\centering
\includegraphics[width=85mm,height=68mm,angle=0]{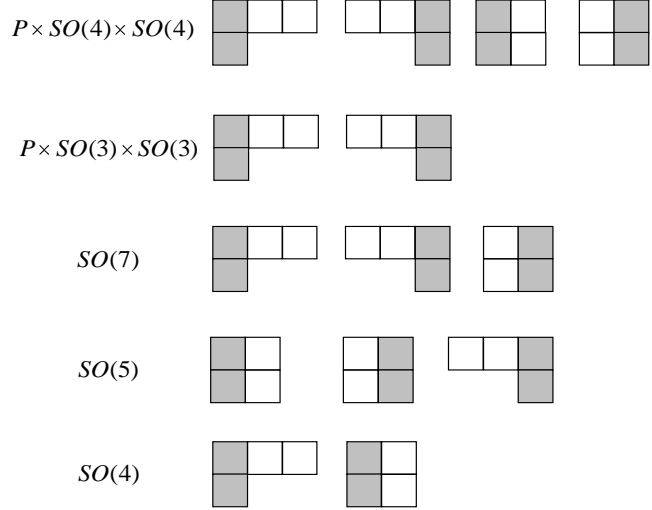}
\caption{Young tableaux corresponding to $SO(n)$
symmetries.}\label{sch-sym}
\end{figure}
The Young tableaux corresponding to various $SO(n)$ symmetries
discussed in Chapters 3 and 4 can be obtained by combining the
appropriate tableaux (Fig. \ref{sch-sym}). The highest possible
symmetry $P\times SO(4)\times SO(4)$ is represented by four
tableaux $T_l$, $T_r$, $S_l$ and $S_r$ since all singlet and
triplet states are degenerate in this case. The symmetry $P\times
SO(3)\times SO(3)$ occurs when two triplets $T_l$ and $T_r$ are
close in energy and these are represented by the couple of Young
tableaux in the second line. Following this procedure, the $SO(7)$
symmetry can be described in terms of two triplets $T_l$, $T_r$
diagrams and one singlet $S_l$ diagram. Moreover, $SO(5)$ symmetry
is represented by two singlet $S_l$, $S_r$ diagrams and one
triplet $T_l$ diagram and, finally, one triplet and one singlet
tableaux correspond to the $SO(4)$ symmetry.

\end{document}